**A comprehensive review on developments of synthetic dimensions**

Danying Yu,[a,†] Wange Song,[b,c,†] Luojia Wang,[a,†] Rohith Srikanth,[d] Sashank Kaushik Sridhar,[d] Tao Chen,[e,f,g] Chenxi Huang,[e,f] Guangzhen Li,[a] Xin Qiao,[h] Xiaoxiong Wu,[a] Zhaohui Dong,[a] Yanyan He,[a] Meng Xiao,[i] Xianfeng Chen,[a,j] Avik Dutt,[d,k,*] Bryce Gadway,[e,f,*] and Luqi Yuan[a,*]

[a]State Key Laboratory of Advanced Optical Communication Systems and Networks, School of Physics and Astronomy, Shanghai Jiao Tong University, Shanghai 200240, China

[b]National Laboratory of Solid State Microstructures, Key Laboratory of Intelligent Optical Sensing and Manipulation, Jiangsu Key Laboratory of Artificial Functional Materials, College of Engineering and Applied Sciences, Nanjing University, Nanjing 210093, China

[c]New Cornerstone Science Laboratory, Department of Physics, University of Hong Kong, Hong Kong, China

[d]Department of Mechanical Engineering, and Institute for Physical Science and Technology, University of Maryland, College Park, Maryland 20742, USA

[e]Department of Physics, University of Illinois at Urbana-Champaign, Urbana, Illinois 61801-3080, USA

[f]Department of Physics, The Pennsylvania State University, University Park, Pennsylvania 16802, USA

[g]School of Physics, Xi'an Jiao Tong University, Xi'an 710049, China

[h]College of Physics and Electronics Engineering, Northwest Normal University, Lanzhou 730070, China

[i]Key Laboratory of Artificial Micro- and Nano-structures of Ministry of Education and School of Physics and Technology, Wuhan University, Wuhan 430072, China

[j]Collaborative Innovation Center of Light Manipulations and Applications, Shandong Normal University, Jinan 250358, China

[k]National Quantum Laboratory (QLab) at Maryland, College Park, Maryland 20742, USA

*Address all correspondence to Avik Dutt, avikdutt@umd.edu; Bryce Gadway, bgadway@psu.edu; Luqi Yuan, yuanluqi@sjtu.edu.cn

†These authors contributed equally to this work.

**Abstract**

The concept of synthetic dimensions has emerged as a powerful framework in photonics and atomic physics, enabling the exploration of high-dimensional physics beyond conventional spatial constraints. Originally developed for quantum simulations in high dimensions, synthetic dimensions have since demonstrated advantages in designing novel Hamiltonians and manipulating quantum or optical states for exploring topological physics, and for applications in computing and information processing. Here we provide a comprehensive overview of progress in synthetic dimensions across photonic, atomic, and other physical platforms over the past decade. We showcase different approaches used to construct synthetic dimensions and highlight key physical phenomena enabled by the advantage of such a framework. By offering a unified perspective on developments in this field, we aim to provide insights into how synthetic dimensions can bridge fundamental physics and applied technologies, fostering interdisciplinary engagement in quantum simulation, atomic and photonic engineering, and information processing.

## I. Introduction

Quantum simulation of novel physical phenomena including topology, non-Hermitian physics, many-body interaction, etc. [1-3] is one of the important research fields in physics nowadays. There exist numerous platforms in optics and photonics, atoms, molecules as well as ions that show various demonstrations of quantum simulations that can be further extended towards quantum computations [4-7]. However, many of these platforms support the physical dynamics in low dimensions, which is constrained by the arrangement of the hopping between lattice sites and limited lattice size of the Hamiltonian model due to the complication in the spatial geometry. This has triggered the concept of synthetic dimensions in both atomic, molecular, optical and photonic communities in the past decade.

The synthetic-dimension concept, using the additional degree of freedom in a system to build an extra dimension, is of fundamental difference from the phrase of dimension in our common sense. The synthetic dimension is specifically referring to the addition of a virtual axis (onto existing spatial axes) along which a physical state can propagate or evolve, so as to increase the effective dimensionality of a Hamiltonian with an extra effective dimension. In other words, the synthetic dimension plays similar role to a spatial dimension on which the physical state can diffuse. Therefore, as people realize nowadays, there are two major categories for constructing synthetic dimensions [8-18] (see Fig. 1).

The first idea to construct the synthetic dimension is to use the discrete physical modes of light or atoms and design the connection between these modes. The discrete modes form the necessary virtual dimension while the wave packet of the system can transport along these discrete modes through the connectivity, i.e., the energy exchange between these discrete modes follows the designed connections, similar to particle movement or wave propagation in a real spatial dimension. One can then build a Hamiltonian in the basis of the virtual dimension constructed by the connection between discrete physical states, namely, the synthetic dimension. Importantly, the connectivity in the synthetic dimension can be made to have tunable and exotic attributes that are hard to realize in a real spatial dimension. From **Sec. II** to **Sec. V**, we will show recent efforts in constructing synthetic dimensions following this idea, where various degrees of freedom including frequency, orbital angular momentum, polarization, time of light as well as intrinsic atomic states and momentum states of atomic lattice are used with the proper connectivity built for constructing numerous Hamiltonian models. Relevant physical phenomena in quantum simulations are also discussed, with possible applications in both photonics and atomic physics.

The second idea to construct the synthetic dimension is to use a system parameter in a geometric structure. This idea, relating to well-studied concepts of dimensional reduction, is thus distinct from the former approach based on introducing new discrete states that are physically connected. Here, each physical state on a specific value of the parameter refers to the eigenstate of a Hamiltonian of interest at the amplitude of a variable. Usually the variable is chosen as the momentum reciprocal to a particular spatial dimension in the Hamiltonian, so varying the system parameter is equivalent to tuning the momentum for a Hamiltonian. Therefore, the system parameter can form the parameter synthetic space in this geometric structure without using the particular spatial dimension to form the desired Hamiltonian. In **Sec. VI**, we discuss how it works in detail and why it can provide an efficient way for simulating high-dimensional physics in lower-dimensional geometric structures.

Representative works are also shown to understand the essence of the parameter synthetic dimension.

The field of synthetic dimensions is growing rapidly in recent years. Given previous reviews and perspectives [8-18] and breakthrough works [19-51], the time is ripe for comprehensive review that not only summarizes the field of synthetic dimensions but also provides an introduction to people who are new to this field. This review serves this purpose by showing the development of synthetic dimensions and discussing various advantages from all examples using different synthetic dimensions, including the potential for exploring high-dimensional physics in lower-dimensional geometric structures and the designable connectivity leading to Hamiltonians supporting effective gauge fluxes, non-Hermiticity, long-range couplings, etc. Therefore, quantum simulations may be performed in high dimensions, or the simulators can be designed with simple geometric configurations but holding complicated Hamiltonians. On the other hand, physical dynamics occurring in synthetic dimensions can be used to manipulate quantum states or optical fields in novel ways, which provides unique physical or optical performance important for future applications.

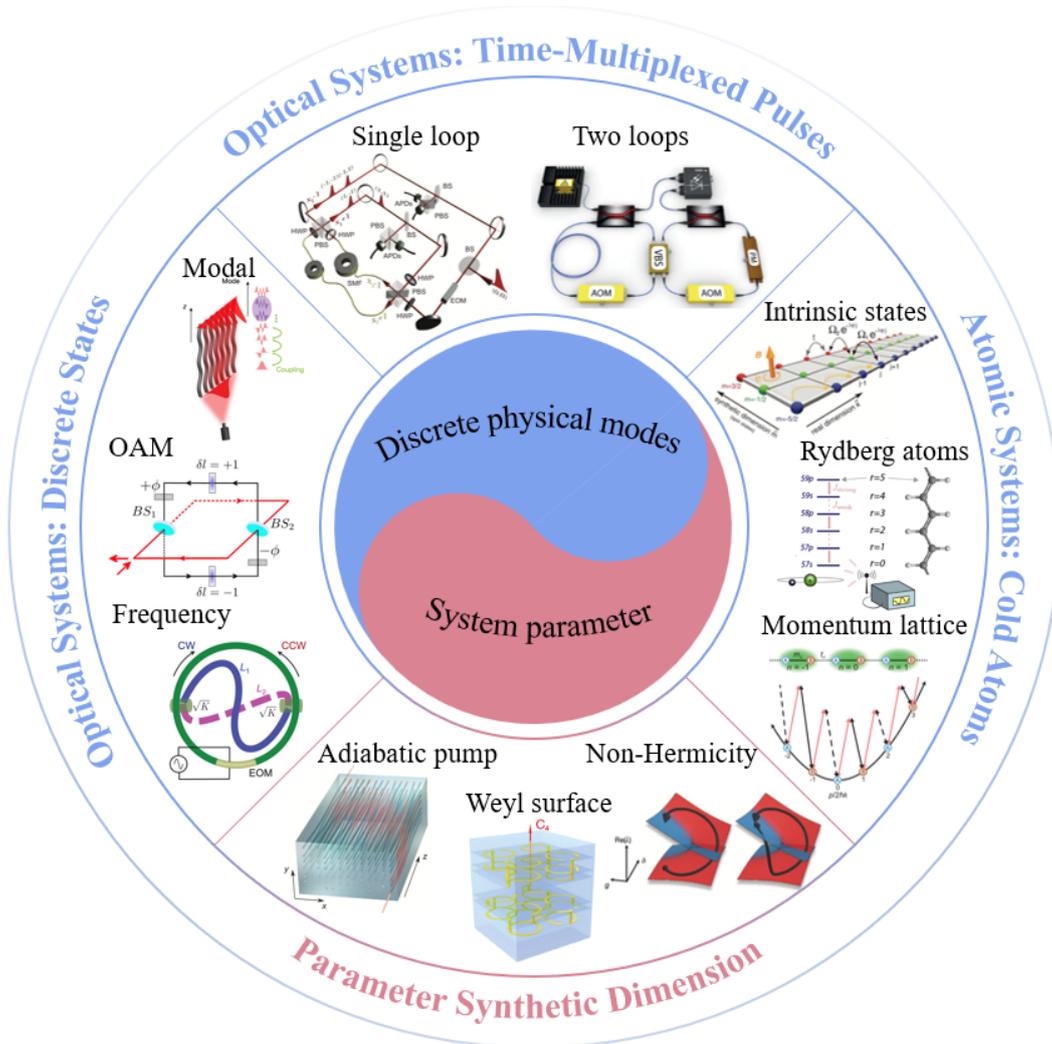

Fig. 1 Schematic diagram for the concept of the synthetic dimension. Adapted from Refs. [16, 20, 22, 24, 30, 31, 36-38, 42, 52].

## II. Optical Systems: Discrete States

As the probably simplest way to describe many interesting physical phenomena in condensed matter systems, the tight-binding model has shown the great success in describing the electronic dynamics within the one-electron approximation in the solid material [53]. An electron in a solid experiences interactions with nearby positive charges and other electrons. These interactions can be combined and treated as one average Coulomb potential for each electron as illustratively expressed in Fig. 2(a), where the potential has periodic negative dips at positions of the atomic nuclei labelled by the index $m$ [see Fig. 2(a)]. Classically the electron is trapped at each potential dip (for example, the $m$-th dip), but quantum mechanics show that the wavepacket of the electron can tunnel to nearby potential dips [$(m \pm 1)$-th]. Such a process can be described by the mathematical expression as $g_m a_{m+1}^{\dagger} a_m e^{i\phi_m}$, where $g_m$ is the tunneling strength, $a_m$ ($a_{m+1}^{\dagger}$) is the annihilation (creation) operator, and $\phi_m$ is the hopping phase. Therefore, the tight binding lattice model is built and the corresponding second quantized Hamiltonian for the electronic system is [54]

$$H_e = \sum_m V_m a_m^{\dagger} a_m + \sum_m \left( g_m a_{m+1}^{\dagger} a_m e^{i\phi_m} + \text{h.c.} \right), \tag{1}$$

where $V_m$ is the onsite potential at the $m$-th position. This model successfully reveals the property of electronic systems in a concise way, the tight binding model has widely been extended to the optical system with the aid of coupled-mode theory [55-63]. The key ingredient here is to determine appropriate discrete optical modes and then find a way to connect them in an order [8, 15]. In the spatial domain, optical modes hosted in distinct photonic structures provide discrete modes, such as waveguide modes in waveguide arrays [64, 65] and resonant modes in resonator arrays [64, 66]. With these elements, one can consider the connectivity between them from the spatial overlap between two nearby modes, which can be described by the coupled-mode theory for guided-wave optics [59, 60, 67] and coupled optical resonators [56, 58, 68], respectively. The construction of synthetic dimensions using discrete optical modes with engineered connectivity follows a similar idea, which utilizes different degrees of freedom of light including frequency, orbital angular momentum, waveguide supermodes, etc. to build the effective tight-binding model along the synthetic dimension. Nevertheless, different from the construction of lattices using photonic structures in real-space dimensions, synthetic dimensions may provide unique aspects of advantages such as complex and/or long-range hoppings, extra reconfigurable flexibility, and the potential towards higher dimensions (>3), as we discuss below.

This section is organized as follows: we show the use of the frequency degree of freedom to construct the synthetic frequency dimension in **Sec. IIA**, where we discuss theoretical approaches and experimental platforms, along with their associated physical phenomena and application perspectives. In **Sec. IIB**, we take the orbital angular momentum of light to build the synthetic dimension, with both theoretical method and experimental progress provided. We further talk about the way to use the modal states in waveguides to design the couplings along the synthetic dimension and show the resulting experimental results in **Sec. IIC**. We discuss other degrees of freedom of light that may be used to construct the synthetic dimension in photonics in **Sec. IID**.

## A. Frequency

In this subsection, we review recent progresses in constructing the so-called synthetic frequency dimension [69, 70]. We will first outline the theoretical models and then showcase different experimental platforms that have been used to construct relevant synthetic lattices. Various physical phenomena and potential applications in the synthetic frequency dimension are also discussed.

### 1. Theoretical model

In building the synthetic frequency dimension, one focuses on the construction of a tight-binding model for photons along the frequency axis of light. Here, we introduce the most attractive synthetic model in recent times, where one or more ring resonators are used to implement the frequency lattice. A ring resonator supports a set of discrete resonant modes at different frequencies, which can mimic the model of electrons on a discrete lattice of sites under the average Coulomb potential [see Fig. 2(b)]. These frequency modes are isolated and disconnected from each other if no frequency conversion happens [71]. To induce the hopping effect or the connectivity between frequency modes, frequency conversion between such resonant frequency modes is required. The nearest-neighbor hopping or the connectivity between two nearby frequency modes can then form a one-dimensional (1D) synthetic lattice in the frequency dimension, which is the essential conceit behind the synthetic frequency dimension [see Fig. 2(c)].

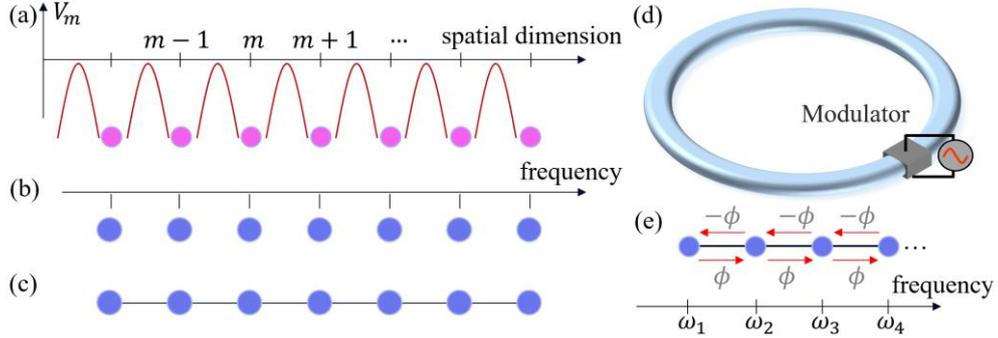

Fig. 2 The construction of the synthetic frequency dimension. (a) Electrons in the average Coulomb potential of an ionic crystal. (b) The discrete frequency modes. (c) The one-dimensional frequency lattice constructed by connecting discrete frequency modes. (d) Ring resonator under dynamic modulation. (e) The frequency conversion between resonant modes insides the ring induced by dynamic modulation.

There are several ways to achieve the frequency conversion. We take the dynamic phase modulation applied in a ring resonator as an example to demonstrate the underlying mechanism [72]. We consider the model of the modulated ring resonator as shown in Fig. 2(d), where the ring supports a set of resonant frequency modes at

$$\omega_n = \omega_0 + n\Omega_R, \tag{2}$$

where $\omega_0$ is the reference frequency, $\Omega_R = 2\pi v_g/L$ is the free spectral range (FSR) of the ring, $v_g$ is the group velocity of light propagating inside the ring resonator, $L$ is the length of the ring, and $n$ is an integer. The electric field of light that propagates inside the ring can be expanded in the basis of the resonant modes [73]

$$E(t, r_\perp, z) = \sum_n \mathcal{E}_n(t, z) E_n(r_\perp) e^{i\omega_n t}, \tag{3}$$

where $z$ is the azimuthal position along the propagation direction, $r_\perp$ is the position perpendicular to the propagation direction, $E_n(r_\perp)$ is the modal profile and $\mathcal{E}_n(t, z)$ is the modal amplitude component at the frequency $\omega_n$. The phase modulation dynamically causes a variation of the permittivity of the material

$$\delta\epsilon(t) = \delta(\boldsymbol{r}) \cos(\Omega_{\mathrm{M}} t + \phi_{\mathrm{M}}), \tag{4}$$

where $\delta(\boldsymbol{r})$ is the modulation profile generated by the modulator, $\Omega_{\mathrm{M}}$ is the modulation frequency, and $\phi_{\mathrm{M}}$ is the modulation phase. The time-periodic variation of the permittivity changes the phase of the light when the light passes through the modulation part at the location labeled by $z_{\mathrm{M}}$, which can be described by [64]

$$E(t^+, r_\perp, z_{\mathrm{M}}) = E(t^-, r_\perp, z_{\mathrm{M}}) e^{i\alpha \cos(\Omega_{\mathrm{M}} t + \phi_{\mathrm{M}})}, \tag{5}$$

where $t^\pm = t + 0^\pm$ and $\alpha$ is the modulation amplitude. Under the slowly varying envelope approximation, Eq. (5) can be expressed using the Jacobi-Anger expansion in terms of modal amplitude $\mathcal{E}_n(t, z_{\mathrm{M}})$ as

$$\mathcal{E}_n(t^+, z_{\mathrm{M}}) = J_0(\alpha) \mathcal{E}_n(t^-, z_{\mathrm{M}})$$
$$+ \sum_{q>0} i^q J_q(\alpha) \left[ \mathcal{E}_{n+q}(t^-, z_{\mathrm{M}}) e^{-iq(\Delta \cdot t^- + \phi_{\mathrm{M}})} + \mathcal{E}_{n-q}(t^-, z_{\mathrm{M}}) e^{iq(\Delta \cdot t^- + \phi_{\mathrm{M}})} \right], \tag{6}$$

where $\Delta = \Omega_{\mathrm{M}} - \Omega_{\mathrm{R}}$ is the frequency detuning, and $J_q$ is the $q$-th order Bessel function of the first kind. To see how such an operation leads to an analog of the tight binding model, we assume that the modulation is weak. In this weak modulation limit $\alpha \to 0$, $J_q$ is very close to zero for $|q| > 1$, and $J_0(\alpha) \approx 1$, $J_1(\alpha) \approx \alpha/2$. The variation of the optical field is then expressed as

$$\Delta \mathcal{E}_n(t, z_{\mathrm{M}}) = i \frac{\alpha}{2} \left[ \mathcal{E}_{n+1}(t, z_{\mathrm{M}}) e^{-i(\Delta \cdot t + \phi_{\mathrm{M}})} + \mathcal{E}_{n-1}(t, z_{\mathrm{M}}) e^{i(\Delta \cdot t + \phi_{\mathrm{M}})} \right]. \tag{7}$$

The accumulation of the optical field variation after propagating over one loop inside the ring is written by

$$\mathcal{E}_n(t + T_{\mathrm{R}}) - \mathcal{E}_n(t) = \Delta \mathcal{E}_n(t), \tag{8}$$

where $T_{\mathrm{R}} = 2\pi/\Omega_{\mathrm{R}}$ is the round trip propagation time. We define $t = \tau + T \cdot T_{\mathrm{R}}$, where $\tau$ is the fast time variable in one round trip and $T$ is the discrete round trip number [74]. Substituting the expression of $t$ into Eq. (8), we obtain

$$\mathcal{E}_n(\tau + (T+1) \cdot T_{\mathrm{R}}) - \mathcal{E}_n(\tau + T \cdot T_{\mathrm{R}}) = \Delta \mathcal{E}_n(\tau + T \cdot T_{\mathrm{R}}). \tag{9}$$

If we further regard $\partial T = T_{\mathrm{R}}$ as the time interval, the variation of optical field can be expressed in the form of a differential equation

$$\frac{\mathcal{E}_n(\tau + (T+1) \cdot T_{\mathrm{R}}) - \mathcal{E}_n(\tau + T \cdot T_{\mathrm{R}})}{T_{\mathrm{R}}} = \frac{\partial \mathcal{E}_n(\tau + T \cdot T_{\mathrm{R}})}{\partial T}. \tag{10}$$

Finally, by substituting Eq. (7) into Eq. (10), one can obtain the relationship between variation of optical field at $\omega_n$ and two nearby modes at $\omega_{n\pm1}$ in the following equation

$$T_{\mathrm{R}} \frac{\partial \mathcal{E}_n}{\partial T} = i \frac{\alpha}{2} \left[ \mathcal{E}_{n+1}(t, z_{\mathrm{M}}) e^{-i(\Delta \cdot t + \phi)} + \mathcal{E}_{n-1}(t, z_{\mathrm{M}}) e^{i(\Delta \cdot t + \phi)} \right]. \tag{11}$$

Equation (11) can be regarded as the equation of motion of the photon wave packet component at each resonant mode, dictated by the corresponding Hamiltonian

$$H = g \sum_n \left( e^{i(\Delta \cdot t + \phi_{\mathrm{M}})} c_{n+1}^\dagger c_n + e^{-i(\Delta \cdot t + \phi_{\mathrm{M}})} c_n^\dagger c_{n+1} \right), \tag{12}$$

where $g = -\alpha/2T_R$ is the coupling coefficient. We therefore see that the modulated ring model in Fig. 2(d) results in a lattice Hamiltonian along the frequency axis of light [see Fig. 2(e)], which indeed provides photonic analogue of the tight binding model of the electron. More importantly, such lattice model lies on a frequency dimension, which is virtual compared to its spatial counterparts [75-77].

The 1D Hamiltonian in Eq. (12) contains many parameters or "knobs" to design the property of the lattice in the synthetic frequency dimension using, for example, the detuning, modulation phase, and even the modulation profile. First, we introduce the way for constructing the high dimension in the synthetic frequency dimension by adding multiple modulation frequencies in the modulation profile. In particular, the dynamic modulation can not only provide nearest-neighbor coupling between the resonant frequency modes in the ring resonator, but also create long-range coupling by implementing an additional modulation frequency that is an integer multiple of the FSR of the ring ($N\Omega_R$). We assume both modulations are resonant for simplicity here. The transmission coefficient that incorporates such a two-modulation-frequency scheme becomes [78]

$$T = e^{i[\alpha\cos(\Omega_R t)+\alpha'\cos(N\Omega_R t)]},\qquad(13)$$

where $\alpha'$ is the corresponding modulation amplitude of the additional modulation frequency. The transmission coefficient in Eq. (13) can lead to the system Hamiltonian [78]

$$H = \sum_n g\left(c_{n+1}^\dagger c_n + c_n^\dagger c_{n+1}\right) + g'\left(c_{n+N}^\dagger c_n + c_n^\dagger c_{n+N}\right),\qquad(14)$$

where $g'$ is the coupling coefficient relevant to $\alpha'$. One notices that the Hamiltonian in Eq. (14) describes the physical dynamics that the field component of light at the frequency $\omega_n$ can hop to components at modes with $\omega_{n\pm1}$ and $\omega_{n\pm N}$ simultaneously. This hopping picture corresponds to an analog to a lattice in two dimensions where the photon can move along two directions. Such fact can be seen if we divide the integer $n$ into two integers, $n = q_x + q_y N$, where $q_x = \mathrm{mod}(n-n_0, N) + 1$, $q_y = (n-n_0-q_x+1)/N$, $n_0$ is a reference index, and the 1D Hamiltonian in Eq. (14) is then written as

$$H = \sum_{q_x,q_y}\left(g c_{q_x+1,q_y}^\dagger c_{q_x,q_y} + g' c_{q_x,q_y+1}^\dagger c_{q_x,q_y} + \mathrm{h.c.}\right) + \sum_{q_y}(g c_{q_x=1,q_y+1}^\dagger c_{q_x=N,q_y} + \mathrm{h.c.}).$$
$$(15)$$

Equation (15) describes a two-dimensional (2D) lattice with a twisted boundary [78]. Higher-dimensional lattices are possible to be studied in the synthetic frequency dimension by incorporating more long-range modulation frequencies [78-80], and multi-dimensional information can be solely encoded in the frequency variable [79, 81].

Another aspect that one may find from the Hamiltonian in Eq. (12) is its capability for introducing the gauge degree of freedom for photons. It is well known that the gauge potential plays a vital role to control the motion of the charged particles through electric fields and magnetic fields. In fact, the quantum mechanical way to account for the effect of electromagnetic fields on the electron is to couple it to the gauge potential. In contrast to electrons, photons, as neutral particles, cannot be directly controlled through the electric and magnetic field. Nevertheless, time modulation gives a possible solution for inducing the effective gauge potential that can couple to photons, inspired by similar ideas introduced in atomic systems [22, 23, 82]. We again take a look at the Hamiltonian in Eq. (12) and note that the hopping phase reads as $\Phi = \Delta \cdot t + \phi$, which may be used to construct the effective gauge potential. To see this fact, we consider a 2D lattice with the Hamiltonian [8]

$$H = g \sum_{\langle m,n \rangle} \left( e^{-i\Phi_{mn}(t)} c_m^\dagger c_n + e^{i\Phi_{mn}(t)} c_n^\dagger c_m \right), \tag{16}$$

where $\langle m, n \rangle$ labels the nearest-neighbor sites, $\Phi_{mn}(t)$ is the hopping phase between site $m$ and $n$. The gauge potential and the hopping phase have the following relationship [83]

$$\int_m^n \boldsymbol{A} \cdot d\boldsymbol{r} = \Phi_{nm}(t), \tag{17}$$

where $\boldsymbol{A}$ is the effective gauge potential. The effective magnetic field $\boldsymbol{B}$ can be obtained from the effective gauge potential $\boldsymbol{A}$ [84]

$$\boldsymbol{B} = \frac{1}{S} \oint_{\text{plaquette}} \boldsymbol{A} \cdot d\boldsymbol{r}, \tag{18}$$

where $S$ is the area of the plaquette. For example, the position-dependent hopping phase $\Phi_{nm} = n\phi$ along the $\hat{l}_{nm}$ direction generates the effective magnetic field $\boldsymbol{B} = \phi\hat{z}$, where $\hat{z}$ is the unit vector perpendicular to the plane of the lattice. The effective electric field $\boldsymbol{E}$ for photons can also be constructed from the effective gauge potential [85]

$$\boldsymbol{E} = -\frac{\partial \boldsymbol{A}}{\partial t} = -\hat{l}_{nm} \frac{\partial \Phi_{nm}(t)}{\partial t}. \tag{19}$$

Therefore, by carefully design hopping phase $\Phi$ between different frequency modes in the synthetic frequency dimension, one can manipulate the frequency detuning $\Delta$ to construct the effective electric field along the frequency axis of light [72, 86, 87] and design distribution of constant phase parts to generate effective magnetic field in a space including the synthetic frequency dimension [69, 88].

The discussions on synthetic frequency dimension are focused on the Hermitian lattices. It turns out that the synthetic frequency dimension can be used to design and explore non-Hermitian physics in a quite straightforward way. The idea for exploring non-Hermitian physics in the synthetic frequency dimension has been started from introducing the amplitude modulation in ring resonators in a theoretical proposal to produce skew-Hermitian or anti-Hermitian couplings [89], and later has been implemented in experiments including both phase and amplitude modulations [39]. To briefly introduce the process to construct a general non-Hermitian model using both the phase modulator and amplitude modulator, we consider the corresponding transmission [39]

$$T = e^{i\alpha \cos(\Omega_M t) - \beta \sin(\Omega_M t)}, \tag{20}$$

where $\alpha$ is the modulation amplitude from the phase modulation while $\beta$ is the modulation amplitude from the amplitude modulation. Such a transmission coefficient can be used to derive a Hamiltonian in the weak modulation limit as

$$H = \sum_n \kappa_+ c_{n+1}^\dagger c_n + \kappa_- c_n^\dagger c_{n+1}, \tag{21}$$

with $\kappa_\pm = -\alpha/2T_R \mp \beta/2T_R$. One can see that the Hamiltonian in Eq. (21) is non-Hermitian for $\beta \neq 0$. Therefore, based on the choice of both the phase modulation and the amplitude modulation parts to design specific non-Hermitian Hamiltonians, Wang et al. generated arbitrary 1-band topological windings [39] and topological 2-band complex-energy braiding [40]. These experimental demonstrations highlight the versatility of the ring resonator system in building synthetic frequency dimension models that support arbitrarily complex band structures.

The approach to analyze the synthetic frequency lattice based on the tight binding Hamiltonian provides a simple approach to solve many problems. However, such a model is limited to the weak coupling regime. It also cannot account for the variation in the propagation of light at different points within the circumference of the ring. To provide a better theoretical method, let us get back to Maxwell's equations:

$$\nabla \times \boldsymbol{E} = -\frac{\partial}{\partial t}\mu\boldsymbol{H},\tag{22}$$

$$\nabla \times \boldsymbol{H} = \frac{\partial}{\partial t}\varepsilon\boldsymbol{E},\tag{23}$$

$$\nabla \cdot \boldsymbol{E} = 0,\tag{24}$$

$$\nabla \cdot \boldsymbol{H} = 0,\tag{25}$$

where $\boldsymbol{E}$, $\boldsymbol{H}$, $\varepsilon$, $\mu$ are electric field, magnetic field intensity, permittivity, and permeability of the medium. Putting operator $\nabla \times$ on the two sides of Eq. (22), one can obtain

$$\nabla \times (\nabla \times \boldsymbol{E}) = -\frac{\partial}{\partial t}(\mu\nabla \times \boldsymbol{H}).\tag{26}$$

Substituting Eq. (23) into Eq. (26), one gets

$$\nabla \times (\nabla \times \boldsymbol{E}) = -\frac{\partial^2}{\partial t^2}(\mu\varepsilon\boldsymbol{E}).\tag{27}$$

The left side of Eq. (27) is

$$\nabla \times (\nabla \times \boldsymbol{E}) = \nabla(\nabla \cdot \boldsymbol{E}) - \nabla^2\boldsymbol{E} = -\nabla^2\boldsymbol{E}.\tag{28}$$

Substituting Eq. (28) into Eq. (27), one can obtain Maxwell's wave equation for the electric field

$$\nabla^2\boldsymbol{E} - \mu\varepsilon\frac{\partial^2}{\partial t^2}\boldsymbol{E} = 0.\tag{29}$$

In photonic structures such as waveguides, the optical field $\boldsymbol{E}$ characterized in one dimension under the slowly varying envelope approximation can be expanded as

$$E = \mathcal{E}(t,z)e^{-i\omega t + ikz},\tag{30}$$

where $z$ is the propagation direction, $\omega$ is the frequency of the light, and $k$ is the wavevector. Substituting Eq. (30) into Eq. (29), one gets the wave equation for the slowly varying envelope of the modal amplitude

$$\frac{\partial}{\partial z}\mathcal{E}(t,z) + \frac{1}{v_g}\frac{\partial}{\partial t}\mathcal{E}(t,z) = 0,\tag{31}$$

where $v_g = 1/\sqrt{\mu\varepsilon}$ is the group velocity of light in the medium.

The wave equation in Eq. (31) can hold the universal solution $\mathcal{E}(t - z/v_g, z)$, meaning that the field does not change its amplitude while it propagates inside the waveguide. One can then derive the transfer-matrix method which has been successfully applied to calculate the band structure of systems such as the three-dimensional (3D) screw dislocation [90] and photonic Weyl points in synthetic frequency dimensions [91]. In Ref. [91], the band structures calculated from the tight-binding model and the transfer-matrix method have been compared. Results show that the overall shapes of the band structures from the two models are quite similar, but there are subtle differences indicating that the transfer-matrix method may present more detail of the system than the tight-binding model. In the following, we introduce the transfer-matrix method.

The light propagation inside ring resonator follows the same way in the waveguide,

except periodic condition imposed by the length of the ring $\mathcal{E}(t, z + L) = \mathcal{E}(t, z)$. The ring resonator supports a set of resonant frequency modes $\omega_n$ due to the boundary condition. The modal amplitude with frequency $\omega_n$ at position $j$ and at time $t_i$ is $\mathcal{E}_{j,n}(t_i)$. The transfer process in the ring is

$$\mathcal{E}_{j+1,n}(t_{i+1}) = e^{i\delta\omega\delta t}e^{ik_n\delta z}\mathcal{E}_{j,n}(t_i), \tag{32}$$

where $\delta z = v_g\delta t$, $\delta t = 2\pi/(J\Omega_R)$, $J$ is the total discrete position number in the ring (the total propagation steps), $\delta\omega$ is the frequency detuning, and $k\delta z$ satisfies the resonant condition $e^{ik_nL} = e^{i2\pi n}$. Next, we consider the dynamic electro-optic modulation (EOM) process inside the ring, which provides the connectivity between nearest-neighbor frequency modes. We assume that the EOM is placed at the position with $j = 1$ and gets modulated by the external signal $\alpha\cos(\Omega_M t)$, where $\Omega_M$ is the modulation frequency and is set to be equal to the FSR of the ring ($\Omega_M = \Omega_R$). The modulation process leads to

$$\mathcal{E}_{j=1,n}(t^+) = e^{i\alpha\cos(\Omega_R t)}\mathcal{E}_{j=1,n}(t^-), \tag{33}$$

where the exponent part in Eq. (33) can be expressed using the Jacobi-Anger expansion [91]

$$e^{i\alpha\cos(\Omega_R t)} = \sum_{q=-\infty}^{+\infty} i^q J_q(\alpha)e^{iq\Omega_R t}. \tag{34}$$

Substituting Eq. (34) into Eq. (33), one can obtain the transfer relation between different frequency modes

$$\mathcal{E}_{j=1,n}(t^+) = \sum_{q=-\infty}^{+\infty} \mathcal{E}_{j=1,n+q}(t^-)\, i^q J_q(\alpha). \tag{35}$$

All optical components include every resonant frequency mode at each position $j$

$$\Psi(t_i) = \left[\underbrace{\mathcal{E}_{1,1}(t_i), \mathcal{E}_{2,1}(t_i), \ldots \mathcal{E}_{J,1}(t_i)}_{n=1}, \underbrace{\mathcal{E}_{1,2}(t_i), \ldots \mathcal{E}_{J,2}(t_i)}_{n=2}, \ldots\right]^T. \tag{36}$$

The transfer process in Eq. (32) and the modulation process in Eq. (35) can then be generalized into an evolution function

$$\Psi(t_{i+1}) = e^{i\delta\omega\delta t}H\Psi(t_i). \tag{37}$$

where $H$ is the matrix form of transfer relationship between optical components. In order to obtain the band structure between $\delta\omega$ and reciprocal lattice vector ($k_f$), one need to apply the steady-state condition, assume the optical field does not vary with time, i.e., $\Psi(t_i) = \Psi(t_{i+1})$, and then perform Fourier transformation $\mathcal{E}_{j,k_f} = \sum_n \mathcal{E}_{j,n}e^{-ik_f n\Omega_R}$. The eigenfunction equation is

$$e^{-i\delta\omega\delta t}\Psi_{k_f} = H_{k_f}\Psi_{k_f}. \tag{38}$$

As a simple example, we consider a single ring case with $J = 2$. The corresponding Hamiltonian is

$$H_{k_f} = \begin{pmatrix} 0, & e^{i\pi} \\ e^{i\pi}e^{i\alpha\cos(k_f\Omega_R)}, & 0 \end{pmatrix}. \tag{39}$$

The wavefunction is written as $\Psi_{k_f} = \left(\mathcal{E}_{1,k_f}, \mathcal{E}_{2,k_f}\right)$. The two eigen-values for Eq. (39) read as

$$\begin{cases} Eig_1 = e^{i\pi}e^{i\frac{\alpha}{2}\cos k_f\Omega_{\mathrm{R}}} \\ Eig_2 = e^{i\frac{\alpha}{2}\cos k_f\Omega_{\mathrm{R}}} \end{cases}. \tag{40}$$

Taking natural logarithm of eigen-values in Eq. (40) and using $\delta\omega_{1(2)}\delta t = i \cdot \ln[Eig_{1(2)}]$, one can then obtain $\delta\omega$

$$\begin{cases} \dfrac{\delta\omega_1}{\Omega_{\mathrm{R}}} = -\left[1 + \dfrac{\alpha}{2\pi}\cos(k_f\Omega_{\mathrm{R}})\right] \\ \dfrac{\delta\omega_2}{\Omega_{\mathrm{R}}} = -\dfrac{\alpha}{2\pi}\cos(k_f\Omega_{\mathrm{R}}) \end{cases}. \tag{41}$$

One can use band structures in Eq. (41) within $\delta\omega \in [-\Omega_{\mathrm{R}}/2, \Omega_{\mathrm{R}}/2]$ to analyze the property of the system.

For the ring coupled with external waveguide, we can write the input-output formalism [78]

$$\mathcal{E}_{j',n}(t^+) = \sqrt{1-\gamma^2}\mathcal{E}_{j',n}(t^-) - i\gamma\mathcal{E}_n^{\mathrm{in}}(t^-), \tag{42}$$

$$\mathcal{E}_n^{\mathrm{out}}(t^+) = \sqrt{1-\gamma^2}\mathcal{E}_n^{\mathrm{in}}(t^-) - i\gamma\mathcal{E}_{j',n}(t^-), \tag{43}$$

where $\mathcal{E}_n^{\mathrm{in}}$ ($\mathcal{E}_n^{\mathrm{out}}$) is the amplitude of the input (output) light with frequency $\omega_n$ in the external waveguide, $\gamma$ is the coupling strength, and we assume here the external waveguide is coupled at the position $j'$. Equation (42) - (43) can also be rewritten as a scattering matrix

$$S = \begin{pmatrix} \sqrt{1-\gamma^2}, -i\gamma \\ -i\gamma, \sqrt{1-\gamma^2} \end{pmatrix}. \tag{44}$$

The scattering matrix in Eq. (44) not only can describe the coupling between the ring and the external waveguide, but also can describe the coupling between two rings. By adding the scatter matrix properly into Eq. (37), one may compute the resulting band structure of a specific model in the synthetic frequency dimension [92].

Last but not the least, the wave equation for the optical field in Eq. (31) can also be used to perform the direct simulation, together with the modulation process in Eq. (35) and the input-output process or the coupling between rings in Eq. (44). Such simulation procedure can be used to study the exact output spectrum of the field from rings and also show the dynamics of a model in the synthetic frequency dimension [90-92].

## 2. Experimental platform

After many theoretical proposals developed to realize various simulations and functionalities with the synthetic frequency dimension, experimental efforts have been made in a number of photonic platforms. One such critical platform is a fiber-based ring resonator system incorporating electro-optic modulators [37, 39, 40, 79, 81, 87, 93-99]. The length of the fiber ring typically is $\sim 10$ m in experiments [see Fig. 3(a)], which corresponds to a FSR of the order $\sim 20$ MHz. For the input and output processes, an external waveguide accompanied with a directional coupler is required to couple light into or out of the main cavity. The frequency of the input laser can be flexibly adjusted to selectively excite the desired eigen-energy [37]. In addition, the off-resonant excitation can also be used to achieve a direct band structure measurement by analyzing the output optical field information [37, 39, 40, 81, 93, 95, 96, 98, 99]. The ring resonator system enables the flexible ability to explore photonic analogs of condensed matter physics effects. For example, the naturally existing clockwise and counter-clockwise modes in the ring can be used to mimic a pseudo spin degree of freedom [37]. The modulation phase controls the photon's gauge degree of freedom and can be directly engineered to mimic the magnetic field acting on electrons [96, 99]. The modulation frequency can be tailored to mimic an electric field by controlling the frequency detuning between the modulation frequency and the free spectral range of the ring ($\Delta = \Omega_M - \Omega_R$) [94]. Additionally, adding an amplitude modulator makes it possible to realize non-Hermitian coupling [39, 40, 81]. Apart from this, many schemes have been proposed to design boundaries for frequency lattices, that further builds upon the spatial lattice analogy [96, 100].

The properties of the synthetic frequency lattice can be obtained from the optical field information inside the ring. There is a convenient way to directly measure the optical field information in the ring, by coupling an external waveguide to the main ring, where the field can leak out of the ring. This enables the dynamic evolution of the field inside the ring to be visualized. Furthermore, it has been observed that the reciprocal lattice vector ($k_f$) of the frequency lattice is the time ($\tau$) within each round trip. By reconstructing the experimental time-resolved transmission with a simultaneous sweep of off-resonant excitation detuning ($\delta\omega = \omega_{in} - \omega_n$), the band structure of the corresponding frequency lattice in a single ring can be obtained. The transmission of the output has a relationship between $\delta\omega$ and $\tau$ as follows:

$$T_{out}(\tau = k_f, \delta\omega) = \frac{\gamma_{loss}^2}{\left[\delta\omega - g\cos(\Omega_R k_f - \Delta \cdot \tau - \phi_M)\right]^2 + \gamma_{loss}^2}, \qquad (45)$$

where $\gamma_{loss}$ is the loss in the ring and $\phi_M$ is the modulation phase. For a simple demonstration, we set $\Delta = 0$ and $\phi_M = 0$. The transmission in Eq. (45) can reach its maximum at $\delta\omega = g\cos(\Omega_R k_f)$, and thus presents a cosine-like pattern, which is the band structure for a 1D synthetic frequency lattice [93, 96]. With this useful approach, one may observe band structures for different models constructed in synthetic frequency dimensions.

Besides the flexible fiber-based ring resonator system that constructs the synthetic frequency dimension, research is also being conducted with integrated on-chip nanophotonic resonators which are scaled down versions of fiber based rings together with state-of-art integrated lithium niobate electro-optic modulators [see Fig. 3(b)] [101]. This system may not only provide a way to build synthetic frequency lattices but also show potential for multifunctional on-chip devices [102-104]. In one outstanding example, Hu et al. experimentally realized a synthetic frequency dimension with an on-chip electro-optic frequency comb, demonstrating a scheme to construct higher-dimensional synthetic lattices in a single ring resonator [105]. The length of the on-chip ring is of the order of ~10 mm, which corresponds to a FSR of the order of ~10 GHz. Consequently, the modulation frequency is also of the order of ~10 GHz. Balčytis et al. realized the synthetic frequency lattice on a silicon CMOS (complementary metal-oxide semiconductor) platform, which enables research in synthetic frequency dimensions on mature technology [106]. Moreover, Zhao et al. achieved a single layer of a matrix-vector multiplier in synthetic frequency dimensions through integrated cavity acousto-optics, which holds promise for scalable analog optical computing [107]. Miniaturization of the on-chip ring provides a condition to manipulate photons in the quantum regime. Javid et al. combined electro-optic modulation and spontaneous parametric down-conversion to study on-chip entangled quantum states in synthetic frequency dimensions [108], highlighting the ability to perform large-scale analog quantum simulations and computations within the time-frequency domain. Microwave photons ($\sim 155 \cdot 2\pi$ MHz) in synthetic frequency dimensions have also been demonstrated in a superconducting resonator [109] and a chain of modulated cavities [110], showcasing the applicability of the concept to different frequency regimes. More recent work in parametrically driven superconducting resonators has realized few-mode lattices of the Creutz ladder and the bosonic Kitaev-Majorana chain in microwave frequency dimensions [111, 112].

The above introduced schemes to construct the synthetic frequency lattice typically require the waveguide to form a loop/ring, where the resonant frequency modes naturally serve as the lattice sites. However, one can build a synthetic frequency lattice without using a ring, by taking an open waveguide and applying traveling wave modulation [80, 113-121]. In such platforms as shown in Fig. 3(c), the travelling-wave modulation induces a variation of the material's refractive index as

$$n_t = \alpha \cos(\Omega_M t - q_M z + \phi_M),\tag{46}$$

where $\Omega_M$, $q_M$, $\phi_M$, $\alpha$ are the modulation frequency, modulation wave vector, modulation phase, and modulation amplitude, respectively. The frequency of the input laser is $\omega_0$, which corresponds to the site ($\omega_0, q_0$) along the dispersion curve of the waveguide. Once the optical field inside the waveguide is under travelling-wave modulation, its frequency and wave vector experience a discrete variation ($\omega_0 \pm n\Omega_M, q_0 \pm nq_M$) with $n$ being an integer. This forms a discrete lattice along the frequency dimension [see Fig. 3(c)]. With the phase-matching condition being satisfied, the amplitude of the $n$-th mode $a_n(z)$ can be obtained by solving the coupled-mode equations

$$i\frac{\partial a_n(z)}{\partial z} = C\left[e^{i\phi_M}a_{n-1}(z) + e^{-i\phi_M}a_{n+1}(z)\right],\tag{47}$$

where $C = \Delta nk_0/2$ is the coupling coefficient and $k_0$ is the vacuum wavevector. The Bloch mode in the frequency lattice has the form $a_n = a_0 \exp(ik_\omega \cdot n\Omega_M)\exp(ik_z z)$, where $k_z$ is the propagation constant along the $z$ direction and $k_\omega$ is the Bloch wavevector. The dispersion relation for the frequency lattice is then $k_z = -2C\cos(k_\omega\Omega_M - \phi_M)$. One can incorporate a wave vector mismatch $\delta q$ of the travelling-wave modulation to induce an effective force $F = -\delta q$ along the frequency dimension. Although such platforms do not necessitate alignment of resonances in the ring, careful design of the travelling-wave modulation is required. Moreover, it is also challenging to extend the spatial dimension by coupling multiple waveguides or realize specific band excitations through choice of the input frequency of light. Nevertheless, this platform can also be potentially scaled down with integrated photonics to form functional and tunable on-chip devices.

The travelling-wave modulation process performed in the waveguide above is tunable, but this is also possible in a waveguide with $\chi^{(3)}$ nonlinearity. Cross-phase modulation of a probe signal and a sinusoidally-shaped pump in a highly nonlinear fiber was used in experiments to observe spectral Bloch oscillations [122-124], and the quantum Hall effects have been studied in one-dimensional four-waveguide array including the synthetic frequency dimension [125]. On the other hand, the nonlinear four wave mixing process has been explored to construct the frequency-domain photonic lattices [126-129]. A pair of pumps with their frequencies separated by the spectral distance $\Omega_R$ can lead to a hopping of the signal field from its original frequency to frequencies separated by $\Omega_R$ through four-wave mixing [see Fig. 3(d)]. Multiple pairs of pumps with frequency differences of multiples of $\Omega_R$, i.e., $n\Omega_R$ can then be applied to introduce long-range coupling between frequency components spanning the interval $n\Omega_R$. The Hamiltonian for the signal carrying different frequency modes that propagates along the nonlinear waveguide under the phase-matching condition is

$$H = \sum_m \sum_n C_n a_m^\dagger a_{m+n} + \text{h.c.},\tag{48}$$

where $m$ is the index labeling the frequency component and $C_n$ is the corresponding coupling coefficient, determined by four-wave mixing

$$C_n = 2\chi P \sum_m A_m A_{m-n}^*,\tag{49}$$

where $P$ is the average pump power, $\chi$ is the effective nonlinearity, and $A_m$ is the complex amplitude of one of the pump fields. The experimental realization of the spectral photonic lattice used a co-propagating signal and multiple pumps in a nonlinear medium for the tunable, long-range, and complex coupling coefficients in the synthetic frequency lattice [126]. Wang et al. further utilized the nonlinear four-wave-mixing platform to study topological effects in multidimensional synthetic chiral-tube lattices [128], illustrating the capability to study topological phenomena using nonlinear optical effects.

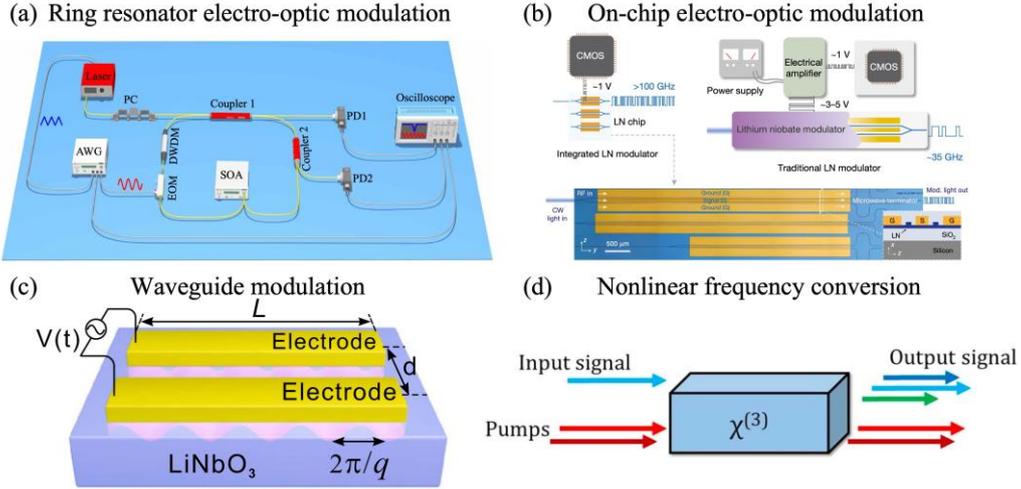

Fig. 3 Experimental platforms to realize synthetic frequency dimensions. (a) The electro-optically modulated fiber ring. Adapted from Ref. [87]. (b) The electro-optically modulated on-chip waveguide. Adapted from Ref. [101]. (c) The traveling wave modulated waveguide. Adapted from Ref. [115]. (d) Four-wave mixing in a nonlinear medium. Adapted from Ref. [126].

Besides the four introduced typical platforms to build the synthetic frequency dimensions in photonics, there are also other potential platforms that may be used. For instance, an optical microdisk-resonator supports resonant modes spaced with an interval $\Omega_{\text{disk}} = c/(n_{\text{disk}}R)$, where $c$ is the light speed in the vacuum, $n_{\text{disk}}$ is the effective refractive index of the microdisk, and $R$ is the radius of the microdisk. A continuous-wave pump laser can excite the microdisk utilizing Kerr nonlinearity to introduce the coupling between different resonant modes. By doing so, one can construct a synthetic frequency lattice therein and similarly use the phase of the pump laser to engineer an effective gauge potential [130]. A Raman medium undergoing molecular modulation also shows the potential to construct synthetic frequency dimensions [131]. This leads to a naturally asymmetric synthetic lattice model but also discusses a different direction of research at the interface of topological photonics and nonlinear optics [132].

### 3. Physical phenomena

Here, we give some outstanding examples of exploring various physical phenomena from different designs of photonic lattices with synthetic frequency dimensions. The most striking feature of studying physics using the synthetic frequency dimension is that the corresponding band structure of the synthetic lattice model can be directly visualized by measuring the transmission [see Eq. (45)] through the ring resonator in experiments. Dutt et al. performed the first experiment in a fiber-based ring resonator to measure the band structure of a 1D lattice [93]. The modulation frequency is set to the FSR of the ring and the measured band structure presents a cosine-like pattern, a typical band shape associated with a 1D tight-binding lattice [see Fig. 4(a)]. They further included the long-range coupling by adding a signal with the modulation frequency $2\Omega_R$. They introduced an effective gauge flux by tuning modulation phases, creating a nonreciprocal band structure with broken time-reversal symmetry. Later on, Li et al. measured the so-called dynamic band structure by using non-zero frequency detuning $\Delta \neq 0$ in the modulation signal, tracking the dynamic motion of the band structure with the form [87]

$$\delta\omega(t) = g\cos(\Omega_R k_f - \Delta \cdot t). \tag{50}$$

The frequency detuning $\Delta$ here plays the role of an effective electric force along the synthetic frequency dimension, as summarized in Fig. 4(b). In contrast to the static band structure which is time-independent, the dynamic band structure varies with time. Therefore, in experiments one can slowly scan the injected frequency of the input field to exactly capture the moving band structure at each time slice. Doing so provides a novel way to capture more delicate information of a dynamic system and shows the potential to design on-chip functionality with time-dependent Hamiltonians [133].

Besides the above simple models, synthetic frequency dimensions also give an approach to explore nontrivial physics in photonics. For instance, it has been well known that the realization of the quantum Hall effect in condensed-matter physics requires a strong magnetic field and a low temperature [134]. However, in photonics, these two conditions may be relaxed. Fang et al. theoretically pointed out that the effective magnetic field for photons can be created by designing non-reciprocal hopping phases [84, 135]. The theoretical proposal was then realized in synthetic frequency dimensions with a similar idea for constructing the effective magnetic flux. In the experiment, Dutt et al. utilized the clockwise- and counterclockwise-propagating modes in a single ring resonator [37] to mimic the pseudo-spin degree of freedom and used auxiliary waveguides to couple these two modes. Utilizing the individual frequency axes of each pseudo spin and electro-optic modulation inside the ring, they built a two-legged ladder lattice model with an effective magnetic flux therein.

Quantum Hall associated physical phenomena including spin-momentum locking and topological chiral one-way edge currents were observed in this single photonic cavity experiment [see Fig. 4(c)]. Ye et al. then used two coupled rings to build such two-legged ladder models but introduced gain/loss, where they observed the breakdown of chirality [99]. The design of synthetic frequency dimensions using two rings can provide more flexibility in construction of lattice models. For instance, strongly coupled rings can induce resonance splitting resulting in a photonic molecule [136-138]. Based on the symmetry along the frequency axis of light, Li et al. constructed a Su-Schrieffer-Heeger (SSH) model [98]. Different from the SSH counterparts in real space [76, 139-141], the synthetic frequency SSH lattice supports band structures including wave-function interference of the eigenstate information, which exhibit different patterns for the trivial and nontrivial cases, as shown in Fig. 4(d). One can then use this unique feature to achieve direct extraction of the topological Zak phase in experiments. Another way to obtain the topological invariant is using the mean-chiral displacement in the driven-dissipative lattice [142]. As a direct step forward from this experiment, Qiao et al. explored phenomena associated with the ultrastrong coupling regime in this synthetic SSH configuration, where the edge states of the 1D Floquet SSH lattice were observed at 0 and $\pi$ energy bandgaps [143]. Research on the photonic molecule platform can also be extended to study the wave-function tomography of the generalized SSH model including long-range couplings and the Haldane model [144, 145]. On the other hand, in a proposed system including an auxiliary ring, Yu et al. showed the direct extraction of the topological invariant in a nontrivial Hamiltonian in quench dynamics, from the collected output optical field solely in the time dimension [74]. Two rings of different lengths have also been used to construct the stub lattice (1D Lieb lattice), where gapped flat band, mode localization effect, and flat-to-nonflat band transition were observed in experiments [95].

Although non-Hermiticity can be introduced into a photonic model by tuning the gain and loss distribution [146-149], it has been observed that non-Hermiticity in synthetic frequency dimension can also be realized through incorporating both amplitude modulation and phase modulation in the ring resonator. Wang et al. constructed a synthetic frequency lattice in one ring resonator imposing both amplitude modulation and phase modulation on the ring to form a 1D Hamiltonian [39]

$$H = \sum_{m,n} (\kappa_{+m} a_{n+m}^\dagger a_n + \kappa_{-m} a_n^\dagger a_{n+m}),\tag{51}$$

where the coupling coefficient $\kappa_{\pm m} = C_m \exp(\pm i\alpha_m) \pm \Delta_m \exp(\pm i\beta_m)$, $C_m$ ($\Delta_m$) is the phase (amplitude) modulation strength, $\alpha_m$ ($\beta_m$) is the modulation phase induced by the phase (amplitude) modulation process, and $m$ is the coupling order. Nonzero amplitude modulation with $\Delta_m \neq 0$ induces non-Hermiticity in the Hamiltonian $H \neq H^\dagger$. The expression of the detected signal $I(k_f, \delta\omega)$ is

$$I(k_f, \delta\omega) \propto \frac{1}{\left[\text{Re}\left(E(k_f)\right) - \delta\omega\right]^2 + \left[\text{Im}\left(E(k_f)\right)\right]^2},\tag{52}$$

where $\text{Re}(E(k_f))$ ($\text{Im}(E(k_f))$) is the real (imaginary) part of the eigen-energy of the Hamiltonian in Eq. (51). For a given $k_f$, the detected signal $I$ has a Lorentzian function shape along the detuning axis $\delta\omega$. Therefore, through fitting one can obtain $\text{Re}(E(k_f))$ and $\text{Im}(E(k_f))$ respectively in experiments [see Fig. 4(e)]. By plotting the extracted $k_f$- dependent $\mathrm{Re}\big(E(k_f)\big)$ and $\mathrm{Im}\big(E(k_f)\big)$ on the complex plane, one can visualize winding features if the corresponding Hamiltonian is nontrivial. Wang et al. further used two ring resonators to study the topological complex-energy braiding. The Hamiltonian of their system is [40]

$$H = \sum_n g a_n^\dagger b_n + g b_n^\dagger a_n - i\gamma a_n^\dagger a_n + \kappa a_{n+1}^\dagger a_n + \kappa a_n^\dagger a_{n+1} + (C-\Delta)a_{n+m}^\dagger a_n$$
$$+ (C+\Delta)a_n^\dagger a_{n+m}, \tag{53}$$

where $a_n^\dagger$ $(a_n)$ and $b_n^\dagger$ $(b_n)$ are the creation (annihilation) operators in ring A and ring B, $\gamma$ is the loss in ring A, and $\kappa$, $g$, $C \pm \Delta$ are the coupling coefficients respectively. The real and imaginary parts of eigenenergies of the Hamiltonian in Eq. (53) on the complex 3D space $[\mathrm{Re}(E), \mathrm{Im}(E), k_f]$ represent the unlink, unknot, hopf link, and trefoil diagrams when specific parameters are chosen, as shown in Fig. 4(f).

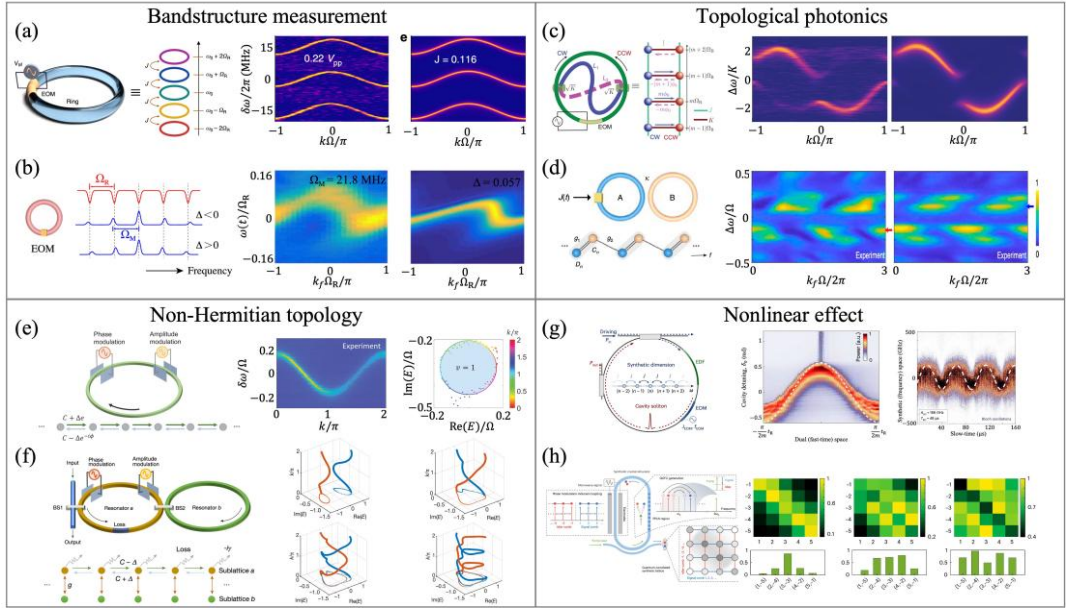

Fig. 4 (a) The static band structure of a 1D frequency lattice. Adapted from Ref. [93]. (b) The dynamic band structure of a 1D frequency lattice. Adapted from Ref. [87]. (c) The chiral band structure of the quantum hall ladder. Adapted from Ref. [37]. (d) The SSH model in the synthetic frequency dimension. Adapted from Ref. [98]. (e) The topological winding of a non-Hermitian band. Adapted from Ref. [39]. (f) The topological complex-energy braiding of non-Hermitian bands. Adapted from Ref. [40]. (g) The dissipative solitons in a synthetic dimension. Adapted from Ref. [150]. (h) The quantum correlation in synthetic space. Adapted from Ref. [108].

The synthetic frequency lattice may further be enriched by including nonlinearity into the lattice model [151-153]. Tusnin et al. theoretically explored the nonlinear effects induced by $\chi^{(2)}$ and $\chi^{(3)}$ susceptibilities in the ring resonator [154], and later Englebert et al. experimentally investigated dissipative solitons in synthetic frequency dimensions, where long-living Bloch oscillations are guaranteed by the Kerr-type nonlinearity and dissipative structures (solitons), as shown in Fig. 4(g) [150]. Javid et al. utilized electro-optic modulation and spontaneous parametric down-conversion to construct a 2D synthetic frequency lattice, where nonlinearity in the form of parametric down-conversion plays the role of expansion of the frequency dimension [108]. In this example, the quantum correlation in the synthetic space has been shown experimentally [see Fig. 4(h)].

All previous examples focus on the model in one dimension, i.e., a singular frequency dimension. It is useful to extend the model to a higher-dimensional space. To achieve higher dimensions, one can build a synthetic space by combining the synthetic frequency dimension and a geometric dimension [155]. For example, Yu et al. theoretically studied the Lieb lattice including the frequency axis of light using a 1D array consisting of two types of rings and explored the isolated photonic flatband [156]. The four-dimensional Hall effect has been theoretically proposed in a 3D geometric structure by adding frequency as the additional fourth dimension [70]. Moreover, Lin et al. constructed a 3D screw dislocation model with 2D ring arrays [90]. On the other hand, one can also combine synthetic frequency dimensions with other synthetic dimensions (which we shall introduce in later sections) to create a higher-dimensional synthetic space. For example, Yuan et al. theoretically investigated the photonic gauge potential in synthetic frequency and orbital angular momentum dimensions [157]. Furthermore, synthetic frequency dimensions can also be combined with the time dimension in a nonintuitive way to build 2D models [117, 158]. Last but not the least, it has also been theoretically proposed [78] and experimentally demonstrated [79, 81] that one can fold the frequency axis of light by introducing long-range coupling between resonant modes at multiples of the FSR to construct 2D or 3D models. In particular, traveling-wave modulation in waveguides using incommensurable frequencies was proposed to construct higher dimensional lattices [80].

## 4. Application perspective

The concept of synthetic frequency dimensions not only can be used to build different lattice models for simulating various physical phenomena, but also the resulting dynamics from the evolution of the Hamiltonian or the steady-state output optical field distribution may find many applications. Here, we briefly list several aspects of the synthetic frequency dimension applications.

One of the earliest goals for developing synthetic frequency dimensions was for quantum simulations, especially for nontrivial physics, as well as for exploring many-body physics [69, 159]. To serve such a purpose, the local interaction between photonic modes is required. In this sense, large nonlinearity is usually desired for achieving the photon-photon interaction in synthetic frequency lattices. Previous experiments have demonstrated dissipative solitons in a synthetic frequency lattice built in a Kerr-type nonlinear material, showing the existence of classical light-light interactions [150]. Moreover, nonlinearity in the form of spontaneous parametric down-conversion in on-chip platforms shows the potential to execute photon-photon interactions from the generated quantum entangled photon pairs [108]. By introducing nonlinearity into the synthetic frequency lattice, it has also been theoretically pointed out that many-body physics may be explored, which could lead to frequency entanglement for photons [159, 160]. Moreover, it has also been observed that if an atom, or quantum emitter, is added into a ring resonator that supports synthetic frequency dimensions, an artificial giant atom is induced that can interact with the photon via multiple quantum channels, due to the fact that all modes in the synthetic frequency lattice can interact with the atom even with frequency detunings [161-163]. The research that follows this line can lead to quantum optics in the synthetic frequency dimension, which could later contribute to quantum simulations of many-body physics with photonic technologies.

Another direct output from experiments with synthetic frequency dimensions is frequency comb generation, previously overlooked in this context. It has been recently observed that one may use physical dynamics in the synthetic frequency lattice to control comb generation [164]. Hechelmann et al. have taken the synthetic frequency space to demonstrate quantum walk comb generation in a fast gain laser, which produces a low-noise, nearly flat broadband comb [48]. Pontula et al. theoretically studied the mechanism to shape above-threshold frequency combs that exhibit tunable bright squeezing and quantum correlations in a multimode nonlinear cavity through dissipation and Bloch mode engineering [165]. Li et al. used two rings with different lengths to build a synthetic moiré frequency superlattice and demonstrated frequency comb generation with mode spacing reduction [166]. Different from conventional comb generation experiments, these works specifically use physical phenomena in the synthetic lattice to design a platform to generate tunable frequency combs with desired spectra.

Synthetic frequency dimensions may also find use in the field of quantum information and optical computing, where the encoding, manipulation, and decoding of information along the synthetic axis of light becomes an important ingredient. Lu et al. has reviewed emerging developments towards frequency-bin quantum information processing and networking, showing its unique advantages for multiplexing, interconnects, and high-dimensional communications [167]. Other than that, implementation of optical computing in synthetic frequency dimensions offers an integrated photonic architecture to achieve essential applications spanning quantum information processing, classical signal processing, and neural networks [168, 169]. For example, Buddhiraju et al. proposed a scheme to realize arbitrary linear transformations for photons in the frequency synthetic dimension, which uses ring resonators to implement tunable couplings between multiple frequency modes in a single waveguide [170]. Fan et al. theoretically and experimentally demonstrated convolution processing in photonic synthetic frequency dimensions with one ring resonator incorporating a phase and an amplitude modulator, which lays the foundation to perform artificial-intelligence applications on compact devices [97, 171].

In addition to this, many physical phenomena including unidirectional frequency conversion [69], non-Hermitian effects [39], and higher dimensional topology [90] have been well studied theoretically and experimentally in synthetic frequency dimensions, which may trigger future applications on integrated platforms with different functionalities. For example, Yu et al. theoretically showed the integration of multiple functions in the synthetic frequency dimension on an array of ring resonators [88]. With respect to non-Hermiticity, the anti-PT symmetry and the PT symmetry in synthetic frequency dimensions can lead to pulse-shortening in mode-locked lasers [89, 172, 173]. Moreover, the theoretical design of the 2D non-Hermitian skin effect in a synthetic photonic lattice would enable programmable light propagation and frequency conversion in an array of ring resonators [174]. Recently, Dikopoltsev et al. experimentally demonstrated the quench dynamics of Wannier-Stark states in the active synthetic frequency space [175], highlighting the ability to generate coherent multi-frequency sources.

Current prosperous research in synthetic frequency dimensions shows it is a powerful tool for not only quantum simulations but also photonic emulator designs based on the photon's frequency degree of freedom. More possibilities can be further explored in this space, including non-Abelian physics [176-178], mirror-induced reflection [179], time reflection and refraction [180], topological spin pumps [181], synthetic floquet lattices [182-184], and mode-locked topological insulator lasers [185]. In particular, recent work theoretically predicted that homogeneous non-Abelian lattice gauge potentials may induce Dirac cones [176], which was also confirmed by the experiment [177]. Therefore, synthetic frequency dimensions show tremendous potential for fundamental exploration as well as on-chip applications.

## B. Orbital Angular Momentum

As another very important degree of freedom for photons, the orbital angular momentum (OAM), carrying the spatial information of light, attracts fundamental research interest in the structured light manipulation [186-190] and has broad applications from quantum information [191, 192] to optical and photonic communications [191, 193]. It has been observed that OAM can also be used to construct a synthetic dimension in photonics. In this section, we will introduce the theoretical description of a synthetic OAM lattice model, and recent experimental progress.

### 1. Theoretical model

Theoretical works [194, 195] indicate that a main cavity coupled with an auxiliary cavity can be used to construct the synthetic OAM dimension [see Fig. 5(a)], where light with the OAM information passes through spatial light modulators multiple times to form a lattice structure. Coupling between sites is achieved by converting a portion of the energy from its original OAM mode into a mode with a lower (higher) OAM number $l$ [see Fig. 5(b)]. The corresponding tight-binding model reads as [52, 194, 195]

$$H = \kappa \sum_l \left( e^{i\varphi} c_l^\dagger c_{l-1} + e^{-i\varphi} c_l^\dagger c_{l+1} \right), \tag{54}$$

where $\kappa$ is coupling coefficient between two nearby OAM modes, $\varphi$ is the hopping phase that is determined by the propagation length inside the auxiliary cavity, and $c_l^\dagger$ ($c_l$) is the creation (annihilation) operator for the $l$-th OAM mode. Although the model in Eq. (54) only includes nearest-neighbor coupling, long-range couplings may also be achieved by engineering multiple auxiliary cavities accommodating suitable spatial light modulators [196].

The tight-binding model can provide a clear way to reveal the intrinsic physics in the synthetic lattice with different OAM modes. It is also important to show the ability to measure the corresponding band structure and evolution dynamics of the system. One can use the input and output formalism of light with OAM following the same procedure as that for the synthetic frequency dimension in Eq. (44) to model the coupling between different cavities and construct high-dimensional synthetic OAM lattices. Different from the synthetic Bloch momentum of the frequency dimension which is the time taken by light to circulate inside the ring for one round trip, here the Bloch momentum for the synthetic OAM lattice is the azimuthal angle $\phi$ (considering the fact that the wave function for light carrying OAM has the expression $e^{il\phi}$) [14]. Through measuring the transmitted intensity along azimuthal angle $\phi$, it has been proposed that one can reconstruct the band structure of the synthetic

OAM lattice [14, 194].

Based on the framework introduced above, a lot of physical models based on the cavity structure have been established. Zhou et al. used a single degenerate optical cavity and proposed to demonstrate 1D topological physics in the synthetic OAM lattice [see Fig. 5(c)], where they designed a pinhole structure in the cavity to form a sharp boundary along the OAM dimension so that the edge state associated with the SSH model can emerge [52]. Luo et al. used a one-dimensional array of optical cavities to study 2D topological physics [see Fig. 5(d)], with one dimension being the OAM dimension and the other being the spatial dimension [194]. By engineering a hopping phase along the OAM dimension, it is possible to realize an effective gauge field for photons. Consequently, topologically protected edge states for photons in the synthetic OAM space have thus formed, and the projected band structure versus the hopping phase displays a Hofstadter-Harper butterfly pattern [see Fig. 5(d)]. Furthermore, Sun et al. used a 2D geometric structure to construct a 3D synthetic lattice [see Fig. 5(e)], where the extra synthetic dimension is OAM. Particularly, they used this synthetic 3D structure to demonstrate Wely semimetal phases [197]. Besides the above introduced fundamental physical effects studied in synthetic OAM lattices, Luo et al. showed potential applications in optical communication and quantum information by theoretically proposing a topological photonic OAM switch [198]. They also illustrated synthetic OAM lattice applications involving quantum memory and optical filters through designing the hopping phase and coupling between two cavities [195].

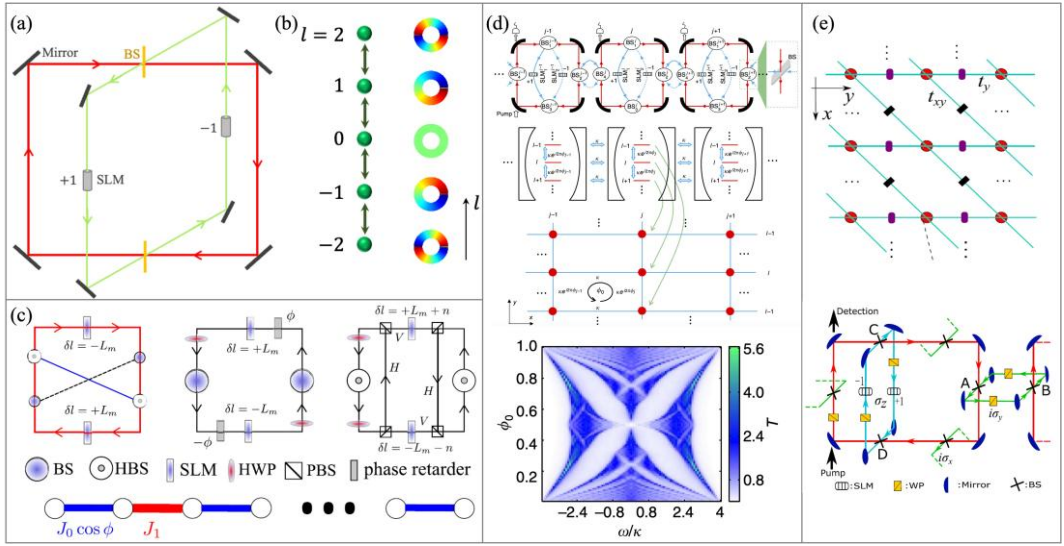

Fig. 5 (a) A main cavity with an auxiliary cavity equipped with two spatial light modulators on the two arms. (b) The coupling among OAM modes forms a tight-binding lattice. (c) The SSH lattice constructed using OAM modes in a single degenerate optical cavity. Adapted from Ref. [52]. (d) A 2D synthetic lattice with one dimension along the OAM degree of freedom and the other dimension along the spatial degree of freedom. The band structure displaying a Hofstadter-Harper butterfly pattern. Adapted from Ref. [194]. (e) The construction of a 3D synthetic lattice with one OAM dimension and two spatial dimensions. Adapted from Ref. [197].

As side notes, it has also been observed that one may build a 2D synthetic space in a single cavity using OAM and frequency dimensions [157, 199]. In these proposals, the mechanisms for constructing the synthetic OAM dimension using an auxiliary cavity with spatial light modulators and the synthetic frequency dimension using dynamic modulation are combined in the single main cavity. Therein, an effective magnetic flux is naturally induced in the 2D synthetic space due to the linear dependence between the additional propagation phase in the auxiliary cavity for each mode and its carrier frequency. Apart from forming a cavity to build the synthetic OAM dimension, it has also been observed that one may establish the OAM dimension by arranging sequences of Q-plates (consisting of anisotropic liquid crystal molecules) and wave plates without using the cavity. This structure can then be used to design quantum walks in the OAM space and study the physics therein [200].

## 2. Experimental progress

Although the original theoretical proposal for constructing the OAM dimension is based on one main cavity being coupled to an auxiliary cavity containing spatial light modulators, experimental verification is not easy with such a complicated configuration. However, Yang et al. devised a simplified design where they utilized a single Q-plate inside a cavity to vary the OAM number of the light by one. By doing so, and further including the right-handed and left-handed circular polarizations of light, they demonstrated the construction of the synthetic OAM lattice in experiment [201]. They then constructed a topological photonic model and measure the topological invariant by analyzing the band structure of the synthetic lattice therein [see Fig. 6(a)]. As a step further, the same authors also showed that this experimental platform can be used to study non-Hermitian physics once loss is added into the system. In experiments, they explored exceptional points (EPs) in the synthetic OAM dimension [see Fig. 6(b)] [202], pointing to potential applications in EP-based sensing. Liao et al. used a similar experimental setup and realized a sharp boundary in a synthetic OAM lattice by drilling a pinhole in the cavity. Unique features associated with the boundary state, dynamic moving of the edge modes, as well as the spectrum discretization are observed in experiments [203].

In a different experimental platform, Cardano et al. demonstrated a discrete quantum walk in a lattice along the synthetic OAM dimension and showed that the OAM degree of freedom can be utilized as a versatile photonic platform to perform quantum simulation tasks [204]. Furthermore, they used probability distribution moments to reveal the topological quantum transition [205] and the mean chiral displacement method to measure Zak phases and topological invariants in the synthetic OAM lattices [see Fig. 6(c)] [206]. Apart from the 1D synthetic OAM space, Wang et al. also constructed a 2D quantum walk with one spatial dimensional and one OAM dimension in experiment [see left panel in Fig. 6(d)], where they observed topologically protected bound states but with vanishing Chern numbers [see right panel in Fig. 6(d)] [200].

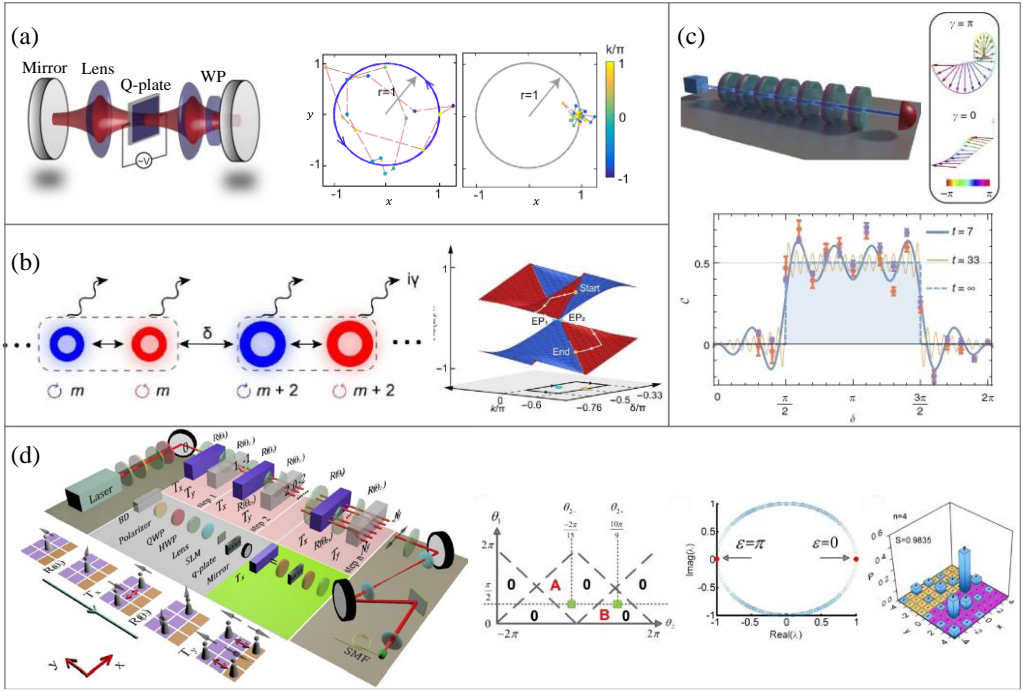

Fig. 6 (a) Left panel: The cavity model used to construct the synthetic OAM lattice. Right panel: The topological invariant obtained from the experimental data. Adapted from Ref. [201]. (b) Left panel: A synthetic OAM lattice with loss included. Right panel: The exceptional topological band. Adapted from Ref. [202]. (c) Detection of the topological invariant in a quantum walk along the synthetic OAM dimension. Adapted from Ref. [206]. (d) Left panel: The construction of the 2D synthetic lattice with $x$ dimension being the OAM state and $y$ dimension being the spatial position. Right panel: The topologically protected bound state with vanishing Chern numbers. Adapted from Ref. [200].

## C. Modal Dimension

We have seen that the basic strategy to construct a synthetic dimension is to find a set of discrete states, for example, using the degrees of freedom from light, and then introducing the appropriate coupling mechanism to connect these states to form a regular structure. It has also been noted that one may use the spatial dimension to construct the synthetic dimension [36]. The key idea is to use the modal dimension of supermodes propagating in a spatially-distributed array of waveguides and connect these discrete modal states through the evanescent-wave coupling via the design of waveguide spacings. This idea is interesting as there is no additional dimension added into the physical space if we convert one spatial dimension to one synthetic modal dimension. Nevertheless, the advantage is obvious once it is noted that the coupling between modal states can be engineered in a complex way and long-range coupling can also be induced.

### 1. Theoretical model

Here, we briefly summarize the method for constructing the synthetic modal dimension from the spatially-distributed array of waveguides. We first review a so-called $J_x$ photonic lattice [207, 208], where the interchannel coupling of the waveguide array can map to the matrix elements of $J_x$ and the eigenenergies of the $J_x$ lattice form an equally spaced ladder illustrated in Fig. 7(a). According to the theory of angular momentum in the quantum mechanics [209], the *x*-component of OAM has the relationship between ladder operators

$$J_x = \frac{J_+ + J_-}{2}. \tag{55}$$

The expressions for the ladder operators are

$$J_+|j, m\rangle = \sqrt{(j-m)(j+m+1)}\hbar|j, m+1\rangle, \tag{56}$$

$$J_-|j, m\rangle = \sqrt{(j+m)(j-m+1)}\hbar|j, m-1\rangle, \tag{57}$$

where $j$ is integer or a half-integer, $m\hbar$ is the eigenvalue of the operator $J_z$. We set $\hbar = 1$ in the following for the simplicity. Based on Eqs. (56) - (57), we can get the matrix form of $J_x$

$$(J_x)_{p,q} = \frac{1}{2}\left[\sqrt{(j-p)(j+p+1)}\delta_{q,p+1} + \sqrt{(j+p)(j-p+1)}\delta_{q,p-1}\right]. \tag{58}$$

The dimension of the $J_x$ matrix is $N = 2j + 1$, and the indices $p$, $q$ range from $-j$ to $j$. This matrix satisfies $(J_x)_{p,q} = (J_x)_{q,p}$, which means that if we map this matrix to the 1D lattice the coupling is reciprocal. In photonics, one can use waveguide arrays to construct a 1D lattice with the Hamiltonian elements equal to matrix $J_x$. Therein, the coupling strength between *n*-th waveguide and (*n*+1)-th waveguide is $0.5\sqrt{(N-n)n}$ [210]. Therefore, to achieve such coupling strength arrangement in the lattice model, one should engineer the positions of the waveguides being not equidistantly distributed and hence the coupling strength is maximum in the middle area [see Fig. 7(b)]. By further calculating the eigenvalues of $J_x$ matrix in Eq. (58), one can get the eigenvalues of the system as $-j$, $-j + 1$,…, $j - 1$, $j$, which are equally spaced. We label these eigenvalues with notation $l$, i.e., the *l*-th eigenvalue. The eigenstate of the *l*-th eigenvalue is [207]

$$u_n^{(l)} = 2^{-\frac{1}{2}(N+1)+n}\sqrt{\frac{(n-1)!\,(N-n)!}{(l-1)!\,(N-l)!}}\,P_{n-1}^{(l-n,N-l-n+1)}(0), \tag{59}$$

where $P_{n-1}^{(l-n,N-l-n+1)}(0)$ is the Jacobi polynomial of order $(n-1)$, and the range of $n$ is $n \in [1, N]$. Fig. 7 (a) shows a schematic eigenstate distribution, which gives the modal states in waveguide arrays. These eigenstate distribution with different eigenvalues is discrete and cannot transform to each other if no perturbation is added, which provides the discrete states that are needed for constructing the synthetic dimension.

We next discuss the way to introduce the connectivity between these discrete eigenstates by designing the oscillation of the array of waveguides [see Fig. 7(b)], where the oscillation of waveguides is restricted within the *z-y* plane and light propagates along the *z* direction. The variation amplitude of the oscillation is

$$R\sin(\Omega z + \phi), \tag{60}$$

where $R$ is an amplitude, $\Omega$ is the frequency that is equal to eigenvalue interval, and $\phi$ is the hopping phase. The equation to describe the coupling between oscillatory waveguides is

$$i\frac{\partial}{\partial z}c_n(z) = p_n e^{i\varphi_n}c_{n+1} + p_{n-1}e^{i\varphi_{n-1}}c_{n-1}, \tag{61}$$

where $c_n$ is the amplitude at *n*-th waveguide, $p_n$ is the coupling strength between the *n*-th and (*n*+1)-th waveguides, $\varphi_n = d_{y,n}k_0\Omega R\cos(\Omega z + \phi)$, $d_{y,n}$ is the distance between *n*-th waveguide and (*n*+1)-th waveguide that is tailored according to the coupling strength between the two waveguides, and $k_0 = 2\pi n_0/\lambda$ is wavenumber in the waveguide. A static waveguide array ($R = 0$) supports the discrete modal states without the connectivity. It has been found that for small $R$, the oscillation gives perturbation, which introduces the connectivity between modal states and hence constructs the synthetic modal dimension in such a curved array of waveguides [36]. The construction of the synthetic modal dimension uses the array of waveguides along the y direction. Copying such array of waveguides over the x-axis can build the 2D lattice with one spatial dimension and one modal dimension [see Fig. 7(c)]. Experimental efforts have used such a construction to demonstrate the first photonic topological insulator in 1 real-space + 1 synthetic-space dimension [36].

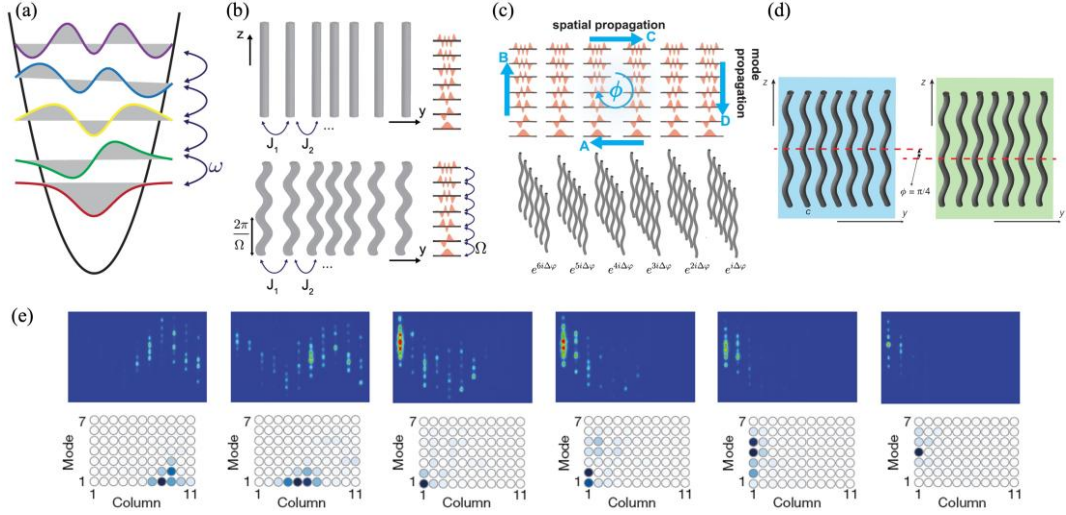

Fig. 7 (a) The schematic eigenstate distribution of $J_x$ lattice that is used to construct the modal dimension. (b) Upper panel: The unequally spaced arrangement of 1D waveguides that can form the equally spaced (in propagation constant) modal distribution in (a). Lower panel: The helical waveguide with its oscillation period consistent with the eigenvalue interval can create the hopping between nearby modes. (c) Upper panel: The synthetic 2D lattice with one spatial dimension and one modal dimension. Lower panel: The arrangement of arrays of waveguides in 2D space. (d) The schematic phase offset between nearby arrays of waveguide. (e) Upper panel: The experimental results of light transporting in the real spatial dimension. Lower panel: The transport of light in the synthetic modal dimension. (a)-(c) are adapted from Ref. [12]. (d)-(e) are adapted from Ref. [36].

## 2. Experimental progress

Lustig et al. first experimentally realized photonic topological insulator in synthetic spatial-modal space [36] based on the framework introduced in the previous part. The effective magnetic field induced topological edge state for photon is realized through designing appropriate hopping phase $\phi$ in each column [see Fig. 7(d)]. The experimental results show that the light propagates clockwise along the edge of the synthetic lattice. Wherein the light with the lowest modal mode moves towards the left direction along the spatial dimension, and then the model mode of the light starts to convert towards higher modal numbers when light reaches the spatial boundary [see the lower panel in Fig. 7(e)]. Such topologically protected edge state cannot be seen in the original waveguide arrays in 2D spatial space [see the upper panel in Fig. 7(e)], as the edge state is embedded in the bulk in real space, which is potential to find possible applications in topological insulator lasers. In this setup, the anticlockwise edge state can also be achieved by exciting the lower bandgap. In contrast, once the effective magnetic field is tuned to zero, the topological edge state disappears and the light in the synthetic lattice penetrates into the bulk [36].

Lustig et al. further studied 3D topological insulator based on the synthetic space with two spatial dimensions (x-y plane) and one modal dimension [43]. In the spatial dimension, they used helical waveguides to construct a Floquet lattice, which has the periodicity along with the propagation direction [see Fig. 8(a)]. This 2D lattice is topological nontrivial through adjusting the relative initial phase offset between nearest helical waveguides. The synthetic modal dimension is then introduced to the system through replacing each site with three waveguides, whose refractive indexes are different to each other [see Fig. 8(b)]. The set of these three waveguides can form a synthetic modal dimension with limited lattice site number. Nevertheless, such system can still show experimental feature that is associated to the 3D screw dislocation, which is then used to reveal the topological transport in the strong photonic topological insulators [see Fig. 8(c)] and paves the way towards the potential application based on 3D topology.

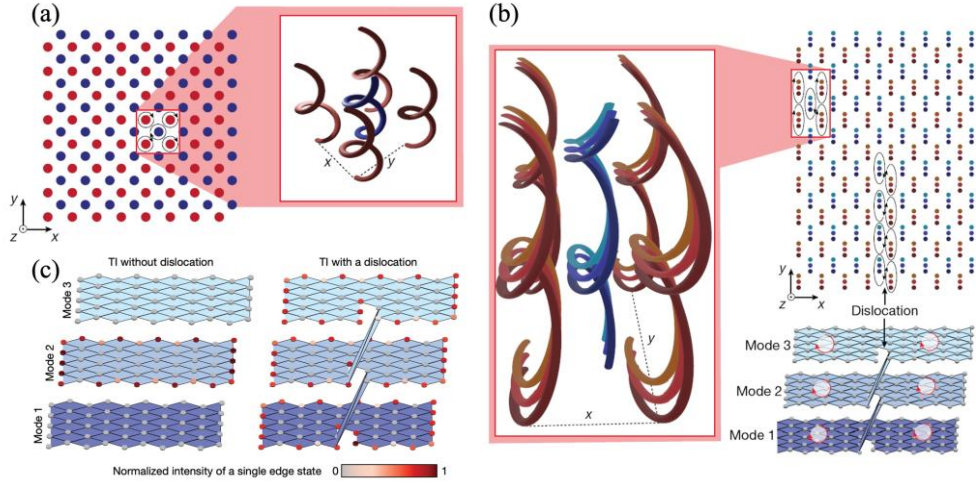

Fig. 8 (a) The 2D lattice in the spatial space, where each site is a helical waveguide. (b) The synthetic 3D lattice with two spatial dimensions and one modal dimension. (c) The schematic illustration of the screw dislocation in synthetic space including the modal dimension. (a)-(c) are adapted from Ref. [43].

## D. Other Degrees of Freedom

In the framework of optical systems, there are other degrees of freedom forming discrete states that can be used to construct a synthetic dimension. The time degree of freedom of light is a key way to construct synthetic lattices, which we will focus in the next section. Here, we briefly discuss another degree of freedom of light, namely the polarization degree of freedom of photon pairs [211], and show how one can use the polarization of light to construct a synthetic dimension [212].

Light has a polarization degree of freedom, perpendicular to its propagation direction. The difference between the polarization states of light is observed when it propagates through birefringent materials. A direct laser-written waveguide in fused silica is birefringent [213]. One can use the two modes of polarization to describe light propagation in this birefringent waveguide, where one mode is parallel to the slow principal axis and the other mode is parallel to the fast principal axis. The refractive indices for these two modes are different, which lead to different corresponding on-site potentials. One can use the following Hamiltonian to describe this system as

$$H = \beta_s a_s^\dagger a_s + \beta_f a_f^\dagger a_f, \tag{62}$$

where $a_s(a_s^\dagger)$ and $a_f(a_f^\dagger)$ are the annihilation (creation) operators for the modes on the slow principal axis and the fast principal axis, with propagation constants (as the on-site potential) $\beta_s$ and $\beta_f$ respectively.

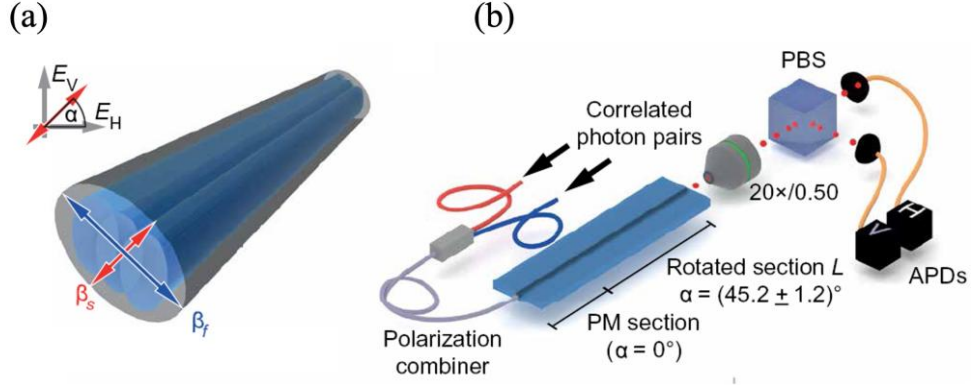

Fig. 9 (a) Direct laser-written waveguides in fused silica. (b) Experimental setup for using the polarization degree of freedom to verify the Hong-Ou-Mandel effect. (a)-(b) are adapted from Ref. [212].

The two modes here form a synthetic space of the two polarizations of light in the waveguide. However, the polarization states are discrete, so if the synthetic dimension is desired to be constructed, one needs to introduce connectivity between them. To do so, one can rotate the waveguide using the following procedure [214]. First, one can assume that the fast principal axis is along the horizontal direction (H) and the slow principal axis is along the vertical direction (V). Next, the waveguide gets rotated clockwise with an angle $\alpha$ [see Fig. 9(a)]. The two polarizations of the light can be considered to be along the horizontal and vertical directions of the waveguide, i.e., H and V. Therefore, the Heisenberg equation of motion for this dynamic process is

$$i\frac{d}{dz}\begin{pmatrix} a_H^\dagger \\ a_V^\dagger \end{pmatrix} = \begin{pmatrix} \bar{\beta} + \Delta \cdot \cos 2\alpha , \Delta \cdot \sin 2\alpha \\ \Delta \cdot \sin 2\alpha , \bar{\beta} - \Delta \cdot \cos 2\alpha \end{pmatrix} \begin{pmatrix} a_H^\dagger \\ a_V^\dagger \end{pmatrix}, \tag{63}$$

where $a_H^\dagger$ and $a_V^\dagger$ are the creation operators for the horizontal polarization mode and the vertical polarization mode, $\bar{\beta} = (\beta_s + \beta_f)/2$ is the mean propagation constant, $\Delta = (\beta_s - \beta_f)/2$ is the birefringence strength. Interestingly, Eq. (63) is mathematically equivalent for two coupled and detuned waveguides. However, for the coupled polarization degree of freedom, the on-site potential $\Delta \cdot \cos 2\alpha$ and the coupling strength $\Delta \cdot \sin 2\alpha$ both can be controlled by the birefringence strength. When $\alpha = \pi/4$, the on-site potentials for both modes are equal and the coupling strength between two modes is maximum.

Ehrhardt et al. used the above setup to verify Hong-Ou-Mandel effect [see Fig. 9(b)] by using a correlated photon source, where they can explore quantum optical phenomena from the quantum walks of photons in the synthetic polarization dimension. In particular, they used two and even three waveguides to explore quantum interference of correlated photons on 3D graphs, which may open up avenues for experimental explorations of quantum dynamics, and for emulating many bosonic or fermionic models including the further study of the nonlinearities and nontrivial topologies with the graph isomorphism problem on optical platforms [215].

The construction of the synthetic dimension using the polarization states in the birefringent waveguides gives a continuous model in the time evolution [or the propagation distance, see Eq. (63)]. On the other hand, there are also many works using sets of quarter-wave and half-wave plates to manipulate the polarization states of light in a sequence, where a discrete quantum walk on the two polarization states can be demonstrated [206, 216-220]. Although some of these works may not claim the construction of synthetic dimensions, the connectivity is built between discrete polarization states and the unitary conversion processes occur based on a lattice framework, which makes these photonic quantum walks using the polarization degree of light be an important supplementary to the synthetic dimension in optical systems. Other than the polarization states, the transverse momentum of light can be engineered to design synthetic dimensions. For example, D'Errico et al. used the 2D momentum lattice to reveal the topological quantum walks [221] and two corelated photons in quantum walks [222]. Moreover, the geometrical angular coordinate around a ring resonator has also been used to construct the synthetic dimension, where a combination between synthetic dimensions and effective magnetic fields with local interactions was proposed [223].

### III. Optical Systems: Time-Multiplexed Pulses

The previous section uses the discrete states of light as the basic components to create the synthetic dimension. As one of the most important parameters for light, the time information has not been discussed in the previous section, despite the volume of research into temporal synthetic dimensions in the past few years [38, 44, 49, 50, 224-237] (and a long history of time-bin multiplexing for myriad other applications). Unlike synthetic dimensions employing discrete photonic states, where the models usually can be treated safely as continuous evolutions in time from a corresponding Schrödinger equation, the time dimension here is not straightforward in a similar fashion. In this section, we will see how one uses the temporal ordering of light pulses inside a ring as the basic components to construct the synthetic dimension (i.e., the synthetic time lattice built from the time-multiplexed pulses with a fast integer time variable), where the dynamics of the model evolves in discrete steps in time (a slow integer time variable) [19, 238]. Nevertheless, we shall see that with discrete-time models, one may still be able to simulate various physical phenomena, and more importantly, this kind of system provides a natural platform to emulate quantum walks.

We organize this section as follows: in the theoretical model part (**Sec. IIIA**), we introduce the model to describe the construction of the synthetic dimension using time-multiplexed pulses. In **Sec. IIIB** and **Sec. IIIC**, we introduce the novel physical phenomena in these synthetic time lattices by using single loops and two coupled loops respectively. Lastly, we briefly summarize this section in **Sec. IIID**.

### A. Theoretical Model

In this subsection, we review the basic idea in constructing the synthetic time lattice using the temporal axis of light based on a time-multiplexed network. This method is developed to facilitate the connectivity between pulses at different arrival times in fiber loop(s) by using delay lines or by coupling two loops with a length difference. We start by using delay lines in a single loop to construct the synthetic time dimension in the first part and discuss its extension by coupling two loops to form the synthetic time lattice in the second part.

#### 1. Time-multiplexed network in a single loop

We consider a main fiber loop in Fig. 10(a), where optical pulses can propagate along the main loop. A sequence of pulses circulate inside the loop with equal time interval between two nearest-neighbor pulses (labelled as $\delta t$), where the time degree of freedom of each pulse is specified by the arrival time of the pulse on a reference position inside the main loop. We can use the discrete index $n$ to label each pulse and $u_n$ to mark the amplitude of the $n$-th pulse. To induce connectivity between the nearest-neighbor pulses, two delay lines are introduced into the main loop. One should then deliberately design the lengths of two delay lines (where the longer one is $L_l$ and the shorter one is $L_s$) and the length between two joint points inside the main loop as $L_{\mathrm{arc}}$ to satisfy the condition $L_{\mathrm{arc}} - L_s = L_l - L_{\mathrm{arc}} = v_g \delta t$, where $v_g$ is the group velocity for the field in the fiber loop. Then a portion of $u_n$ in the $n$-th pulse can go ahead of (behind) the $(n$-1)-th pulse [$(n$+1)-th pulse] through the shorter (longer) delay line and contribute to $u_{n-1}$ ($u_{n+1}$), and vice versa [239]. Following such procedure, a tight-binding type connectivity between pulses is built after the fields finish each roundtrip circulation (labelled by an integer $m$), where $u_n$ in the $n$-th pulse at the $(m$+1)-th roundtrip consists of a combination of the major portion of $u_n$ and small portions of $u_{n-1}$ and $u_{n+1}$ at the $m$-th roundtrip [see the lower panel of Fig. 10(a)]. This hence constitutes the most basic construction of a synthetic time lattice in a single loop. In principle the connectivity can be manipulated arbitrarily to be long range of order $N$ by introducing suitable delay lines that satisfy $L_{\text{arc}} - L_s = L_l - L_{\text{arc}} = N v_g \delta t$. Moreover, the synthetic lattice evolves with discrete time-steps, quantified by the number of round trips for the pulse circulation.

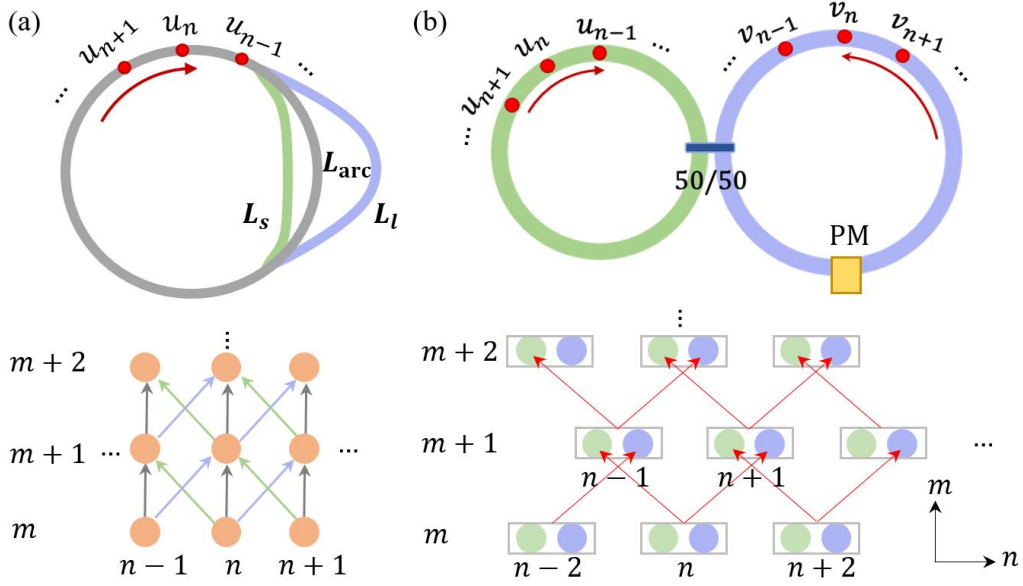

Fig. 10 (a) Synthetic time dimension created from a main loop with short and long delay lines to couple pulses. Upper panel: The main loop for time-multiplexing, which incorporates green and purple delay lines to induce the hopping between nearest-neighbor pulses. Lower panel: The lattice network in synthetic time dimension. (b) Synthetic time dimension created from two loops of unequal lengths. Upper panel: The short loop and long loop are connected by a 50/50 coupler. A phase modulator can impose phases on pulses in the long loop. Lower panel: The lattice network in synthetic time dimension, where $n$ constitutes the 1D synthetic lattice and $m$ denotes the evolution time.

## 2. Time-multiplexed network in coupled two loops

A potentially simpler design for constructing the synthetic time lattice (or time-multiplexed mesh lattice) is to use two loops with different lengths coupled by a 50/50 coupler, as shown in Fig. 10(b). In this setup, again the sequence of pulses get circulated, now in two loops, and one can use the arrival time to label pulses in each roundtrip. We assume the time for a pulse to complete one round trip in the short loop as $T_s$ and the time for that in the long loop as $T_l$. The time difference is set to be $T_l - T_s = 2\Delta T$, and the average time is $(T_l + T_s)/2 = T$, so $T_s$ and $T_l$ can be re-written as $T_s = T - \Delta T$ and $T_l = T + \Delta T$. We denote a pulse at the $n$-th arrival time in the short loop as $u_n^m$ and a pulse at the $n$-th arrival time in the long loop as $v_n^m$ for the $m$-th round trip. Hence the time that a pulse arrives at the coupler is defined as $t = mT + n\Delta T$. Now two pulses $u_n^m$ and $v_n^m$ arriving at the coupler together can contribute to the pulse circulating in the short loop for the next roundtrip with the time $\Delta T$ ahead (i.e., the pulse spends time $T_s$ for a roundtrip), which gives $u_{n-1}^{m+1}$. This process can be expressed as [8, 19]

$$u_{n-1}^{m+1} = \frac{1}{\sqrt{2}}(u_n^m + iv_n^m). \tag{64}$$

Similarly, the same two pulses can contribute to the pulse circulating in the long loop instead, which makes $\Delta T$ behind as the pulse spends time $T_l$ for the next roundtrip, which leads to $v_{n+1}^{m+1}$ by

$$v_{n+1}^{m+1} = \frac{1}{\sqrt{2}}(iu_n^m + v_n^m)e^{i\phi(n)}. \tag{65}$$

Here $\phi(n)$ is the modulation phase that can be imprinted from the phase modulator in the long loop. The lower panel in Fig. 10(b) shows the hopping diagram described in Eqs. (64)-(65), forming the discrete *m-n* mesh grid, which creates the synthetic time lattice model.

## B. Physics in a Single Loop

There are many interesting phenomena that can be observed in the time-multiplexed network in a single loop, using the concept of the synthetic time lattice. As examples, we discuss achievements in photonic emulation of the Ising machine, topological physics observed with dissipative photonics, efforts towards quantum computation, and realization of quantum walks.

### 1. Ising machine

The Ising model, which models magnetic interactions in spin-chains, has been proposed to tackle Quadratic Unconstrained Binary Optimization (QUBO) problems by mapping the binary variables of interest to individual spins on a chain, and finding the ground state of the corresponding Ising Hamiltonian. This method can have potential applications in drug discovery [240] and artificial intelligence [241]. Classical annealing methods have already shown great promise for utilizing Ising machines as hardware solvers [242]. However, quantum annealing, which takes advantage of quantum tunneling and superposition shows great potential to speed up the optimization process [242]. Marandi et. al. therefore have used the synthetic time dimension framework for construction of the coherent Ising machine in experiment [243], which we briefly introduce here.

The standard Hamiltonian of the Ising model is [243]

$$H = -\sum_{ij}^{N} J_{ij}\sigma_i\sigma_j, \tag{66}$$

where $\sigma_i$ represents the *i*-th *z* projection of the spin, and $J_{ij}$ is the coupling between spin $\sigma_i$ and $\sigma_j$. The photonic emulation of the Ising model is applied in a time-multiplexed network as shown in Fig. 11(a), where four sequenced pulses with the same time intervals are circulating inside the fiber loop and represent four spins [see Fig. 11(b)]. The binary phase of the degenerate optical parametric oscillator（OPO）above threshold [244] mimics the spin up and spin down states, and they are coupled to each other below threshold to realize a superposition of different spin configurations. The couplings $J_{ij}$ between each spin [see Fig. 11(c)] are tailored by controlling the three delay lines. More flexible connectivities beyond nearest neighbor have been obtained using a measurement-feedback architecture for nearly arbitrary amplitudes of the $J_{ij}$ coupling. When the frustrated Ising spin model is considered, the couplings are set to be out-of-phase, $J_{ij} < 0$. Ground energy of Hamiltonian in Eq. (66) corresponds to a set of eigen state distributions, which are given by the alignment of the spin patterns, and the Ising Hamiltonian's energy spectrum is thus mapped onto the loss of each coupled OPO state. In order to obtain the ground energy, one can gradually increase the OPO gain through increasing the pump field, until the OPO gain balances out minimum loss, which corresponds to the minimum energy of the Ising problem [see Fig. 11(d)], and collapses the superposition of degenerate OPOs to a spin configuration corresponding to this ground state. Hence, at this stage, only the ground states from the Ising Hamiltonian in Eq. (66) can stably exist in the system and then get observed.

## 2. Dissipative photonics

Dissipation is ubiquitous in all areas of physics [245, 246], and is typically a deleterious effect in the study of particle transport dynamics, but can be a resource in nonlinear physics for creating optical solitons. Besides this, dissipation can also be used to realize exceptional points in non-Hermitian Hamiltonians, offering possible improvements for optical sensing [27]. Recently, Leefmans et al. revealed that dissipation can also be utilized to engineer the topological nature of a system [229], by using a time-multiplexed photonic network and introducing dissipative coupling in a synthetic time lattice.

In the experimental setup [see Fig. 11(e)], a single loop, where a sequence of pulses circulate inside, is used to construct a lattice model in the synthetic time dimension. The delay lines are also introduced in the loop contributing to a dissipative coupling between lattice sites. The modulators are added in the delay lines, which are set to design the coupling strengths and phases. In such a configuration, one can construct the dissipative SSH lattice model, where the Hamiltonian is purely imaginary, so the topological invariants now are calculated from topologically non-trivial imaginary bands. In experiments, one can observe the diffusion of the pulses where the energy couples into the initially unoccupied lattice sites to form a topologically trivial structure [see Fig. 11(f)]. On the other hand, if the topologically nontrivial model is constructed, the distribution of the pulses in the synthetic time lattice is consistent with the theoretical topologically-protected edge-state distribution [see Fig. 11(g)].

The synthetic time lattice configured here can also be expanded to two dimensions by incorporating multiple delay lines. For example, the same loop is coupled with four delay lines, two of which are introduced additionally to provide the long-range coupling, in which way the lattice in the time dimension can be folded to effectively a two-dimensional lattice. In the experiment, a $4 \times 10$ Harper–Hofstadter (HH) lattice is then constructed in the synthetic space together with the effective magnetic field induced by setting suitable modulation phases on each pulse travelling in the delay lines. The corresponding topologically-protected edge state is then observed in the experiments, where the distribution of pulses shows localization behavior along the boundary of the synthetic lattice [see Fig. 11(h)]. The demonstration of topological phenomena with purely dissipative Hamiltonians opens a novel direction to understanding topological phases, which has led to further studies including the creation of a temporally mode-locked laser [247] and the generalization of the topological laser proposal in a dissipative time-frequency synthetic space [158].

## 3. Quantum computing

Quantum computation has been proposed as an alternative computational paradigm with potential applications in combinatorial optimization, factoring large integers, and quantum simulation [248-251]. Photonics can provide alternative ways for the quantum computation process [5, 6]. However, operations on optical qubits need to be conducted mid-flight, which typically requires sequential optical components. These components make photonic quantum circuits requiring massive resource overheads, especially to achieve fault-tolerance [252], and hard to be integrated in photonic chips [253]. Bartlett et al. proposed a scalable architecture for a deterministic photonic quantum computer by using the synthetic time dimension [226], with the construction steps briefly summarized below:

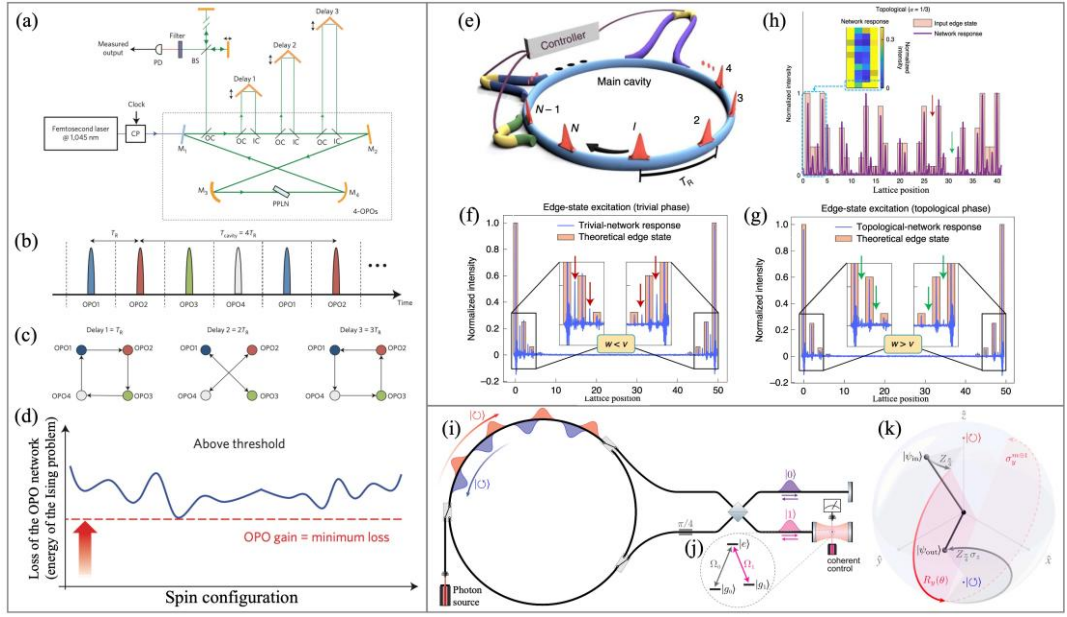

Fig. 11 (a) A single loop system with three delay lines to design the Ising model. (b) The output pulse train, where the pulses are labeled by OPO1 to OPO4. (c) The couplings between pulse slots provided by three delay lines. (d) The diagram illustrating the search for the ground state, where the OPO gain reaching the minimum energy of the Ising problem triggers a collapse of the OPOs to the ground state distribution of the spins. (a)-(d) are adapted from Ref. [243]. (e) A main cavity with delay lines is used to construct the synthetic time lattice with dissipative coupling. The optical field distribution in the synthetic lattice of the 1D SSH model (f)-(g) and 2D HH model (h). (e)-(h) are adapted from Ref. [229]. (i) The CW and CCW modes in the storage ring constitute the photonic qubits. The optical switches can guide the photonic qubits to scatter with an atom which is controlled by a laser. (j) The atom has a Λ-shaped three-level energy structure. (k) The transformation of the photonic qubit depicted by the Bloch sphere. (i)-(k) are adapted from Ref. [226].

Single photons propagating in the clockwise (CW) and counterclockwise (CCW) modes of the loop constitute the photonic qubit basis $|\circlearrowright\rangle$ and $|\circlearrowleft\rangle$. The scattering unit consisting of $\pi/4$ phase shifter, 50:50 beam splitter, an atom, and a cavity is coupled to the loop through two optical switches, as illustrated in Fig. 11(i). The atom has a Λ-shape three-level energy structure, where two ground states $|g_0\rangle$ and $|g_1\rangle$ represent the atomic qubit basis, with one of the Λ-level transitions resonant with the cavity [see Fig. 11(j)]. In this configuration, when the photonic qubit scatters with the atom, the photonic qubit and the atomic qubit are then entangled. After the scattering process is completed, a rotation operation between states $|g_0\rangle$ and $|g_1\rangle$ is applied by using external coherent lasers, and the measurement on atomic state in the $|g_0\rangle$ and $|g_1\rangle$ basis is also performed. Due to the entanglement between the atomic and photonic qubits, the operation and the measurement on the atom can be teleported to the photonic qubit. An arbitrary single-qubit gate can be designed by configuring a three-step rotation scheme, which forms an arbitrary rotation in the Bloch sphere, providing a direct visualization of the subsequent transformation on the photonic qubits [see Fig. 11(k)]. More importantly, a two-qubit controlled Z gate can be deterministically implemented by directing the two pulses corresponding to the qubits successively on the cavity-coupled atom [254]. The prepared photonic qubit can then be readout by using the swap operation, thus satisfying all of the criteria for a deterministic photonic quantum computation in the synthetic time dimension [226]. This scheme has the potential to be significantly more scalable by re-using the same atom or quantum emitter in the cavity in a time-multiplexed fashion, obviating the need for multiple such identical atoms or emitters which itself is experimentally challenging. Other quantum computing schemes re-using a single quantum emitter through time-multiplexed pulses for cluster state generation have also been proposed [255, 256].

## 4. Quantum walk

The classical random walk puts forward the idea that the shift of a walker is dependent on a coin toss. The quantum walk as the quantum counterpart of the classical random walk has aroused broad interest, as it provides insights into quantum algorithm to speed up search processes [257-261]. Apart from quantum walks based on the previously introduced OAM framework, the synthetic time dimension provides another experimental candidate to simulate discrete-time quantum walks in photonics [20, 211, 224, 234, 236, 262-265]. A typical quantum walker can be described by the product of the position Hilbert space $\mathcal{H}_x$ and the coin Hilbert space $\mathcal{H}_c$, where the evolution of the wave function under the quantum walk is [262]

$$|\Psi(t_1)\rangle = \hat{S}\hat{C}|\Psi(t_0)\rangle, \tag{67}$$

where $\hat{C}$ is the coin operation and $\hat{S}$ is the step operation. For the coin operator, the polarization degree of freedom of light is usually a good candidate to perform the coin toss procedure. The position operator $\hat{S}$ that moves the quantum walker according to the result of the coin toss with the expression [262]

$$\hat{S} = \sum_{x\in\mathbb{Z}}(|x+1,H\rangle\langle x,H| + |x-1,V\rangle\langle x,V|), \tag{68}$$

where $H$ is the horizontal polarization and $V$ is the vertical polarization, is emulated in the spatial dimension as before. The use of the synthetic time lattice is to replace such movement by the conversion between pulses at different arrival times. There are several important experiments that have achieved this, which we summarize in the following.

Nitsche et al. demonstrated dynamic control of a 1D discrete-time quantum walk by actively adjusting the coin operator, where tunable graph structures can now be observed [262]. Barkhofen et al. configured a topological structure by designing time-reversal symmetry, particle-hole symmetry, and chiral symmetry into the synthetic time lattice, and topological invariants were measured and edge states were observed during the discrete-time quantum walks [263]. Lorz et al. studied a four-dimensional coin operator by adding clockwise and counter-clockwise propagating modes in the loop, where multiple wavefronts are observed in the quantum walk process [264]. Chalabi et al. used two beam splitters in the loop model to construct a 2D synthetic lattice [265], where the effective gauge field in the discrete-time multiplexed quantum walk can induce band gaps and give rise to suppressed diffusion. Besides this, they also showed a particular walk along the domain boundary when two topological edge states with opposite velocities are excited. Moreover, Chalabi et al. engineered synthetic electric fields into the 2D quantum walk, where Bloch oscillations, revival of the optical field, and waveguide photons in nonuniform electric fields were experimentally observed in the time-multiplexed network [224]. The same group has more recently demonstrated dynamic control of the non-Hermitian skin effect in 2D quantum walks [237]. Schreiber et al. used the time-multiplexed network to construct a 2D quantum walk, where two-particle dynamics and strong nonlinearities were experimentally obtained [20].

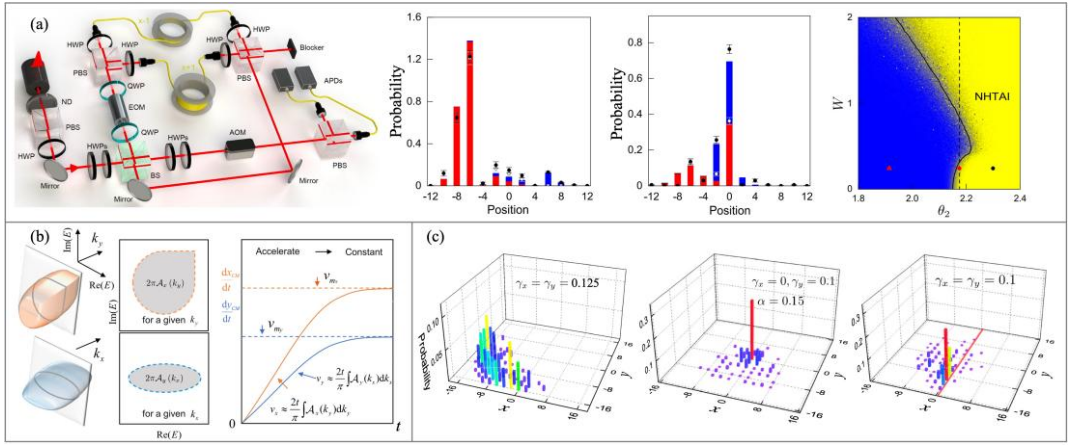

Fig. 12 (a) The non-Hermitian skin effect, Anderson localization, and disorder-induced topological phase transitions in the synthetic time lattice. Adapted from Ref. [231]. (b) The correspondence between self-acceleration and the spectral geometry encircled by the complex eigenenergy. Adapted from Ref. [236]. (c) The non-Hermitian skin effect, the magnetic suppression phenomena, and the Floquet topological edge modes in the time-multiplexed network including the non-Hermicity and the magnetic flux. Adapted from Ref. [234].

Non-Hermicity and Anderson localization can also be explored in the time-multiplexed network by incorporating wave plates to introduce polarization-dependent on-site loss and incorporating an electro-optical modulator to induce walker-position-dependent disorder, respectively. Lin et al. thus demonstrated the interplay of non-Hermiticity, disorder, and topology, where they observed the non-Hermitian skin effect, Anderson localization, and disorder-induced topological phase transitions [see Fig. 12(a)] [231]. The non-Hermicity can be further combined with the quasiperiodicity by constructing an Aubry-André-Harper (AAH) model in the synthetic time dimension. Lin et al. used the polarization- and position-dependent phase operators to fulfill the quasiperiodicity condition, and notably they experimentally observed the mobility edge as well as the concurrence of global delocalization-localization transitions, the spectral topological transition, and the PT-symmetry breaking transition [230]. In the non-Hermitian framework, an interesting correspondence between non-Hermitian spectral topology and transient self-acceleration was also experimentally unveiled by Xue et al., where they showed that the 1D self-acceleration is proportional to the spectral area encircled by the complex eigenenergy while the 2D self-acceleration is proportional to the volume encircled by the complex eigenenergy [see Fig. 12(b)] [236]. The non-Hermicity together with the synthetic magnetic flux can be utilized to engineer the directional flow of light in the 2D square lattice, where Lin et al. used the time-multiplexed 2D quantum walk to reveal the novel phenomena, including the adjustable orientation of the non-Hermitian skin effect by controlling the photon-loss parameters, the suppression of the bulk flow by applying the magnetic confinement, and the Floquet topological edge modes under the impacts from non-Hermiticity and the magnetic flux [see Fig. 12(c)] [234].

## C. Physics in Two Loops

In the previous part, we discussed interesting physics explored in the synthetic time lattice using a single loop. Although the models constructed in a single loop are straightforward to be understood using the tight-binding picture as illustrated in Fig. 10(a), the two-loop configuration in Fig. 10(b) has been developed with more broad aspects due to its relatively simpler setup in experiments. Here, we show some outstanding instances for studying various physical models in two loops.

### 1. Topology

Photonic topological effect can be used to control the flow of the light [266-269], where different platforms use ferrite rods [266], delay lines [267], etc. to create the effective magnetic field for photons to break the time-reversal symmetry of the system and then create topologically-protected edge states. The time-multiplexed network formed by two loops can provide reconfigurable and scalable synthetic time lattices, which manifests a powerful tool to simulate photonic topological effect [38, 44, 225, 228, 233, 235, 270-273], where the phase modulator in the loop provides the effective magnetic field for photons.

Weidemann et al. constructed the prominent 1D topological structure, the SSH model, in the synthetic time dimension based on the two loop setting, where they displayed the topological funneling of light with the incorporation of the non-Hermitian effect [see Fig. 13(a)] [38]. With the two-loop setup, they designed the modulation phase to induce effective vector potentials for the construction of the AAH model in the synthetic time dimension, which results in the Floquet Hofstadter butterfly pattern in the energy spectrum [see Fig. 13(b)] [44]. The topological triple phase transition is also observed in this experiment, where the bulk metal-insulator phase transition, topologically non-trivial phase for the non-Hermitian Floquet model, and parity-time (PT) symmetry breaking simultaneously emerge in experiments by controlling one single parameter [see Fig. 13(c)] [44]. A method to quantify the topological invariant has also been developed in this platform. For example, Longhi et al. demonstrated the dynamical topological winding of the synthetic lattice by adopting mean survival time of a pulse to characterize the topological winding of the non-Hermitian system [272]. Wimmer et al. also used the anomalous displacement of wavepacket to obtain the Berry curvature in the mesh lattice based on the two-loop setting [270]. In addition, Ye et al. used scalar and vector gauge potentials to experimentally design the electric and magnetic Aharonov-Bohm effect in the synthetic time dimension, which shows the potential application in the quantum information processing [233].

A single-shot technique to measure the band structure from the synthetic time lattice has also been developed for observing topological effects, where band gaps close and reopen by controlling the modulation phase [228]. The emergent topology in the band closing and reopening process can then be used to realize a time-reversal operation by utilizing the exchange of eigenstates. Based on this mechanism, Wimmer et al. realized the restoration of a pulse sequence in the photonic time lattice [271]. Adiyatullin et al. used heterodyne measurement to directly visualize the bulk winding bands as well as the bands with anomalous edge states in the synthetic time lattice, where they found Bloch sub-oscillations within the bulk and edge states at the interface [235]. The robustness and vulnerability of the topological anomalous Floquet interface state are also experimentally examined by Bisianov et al. [225].

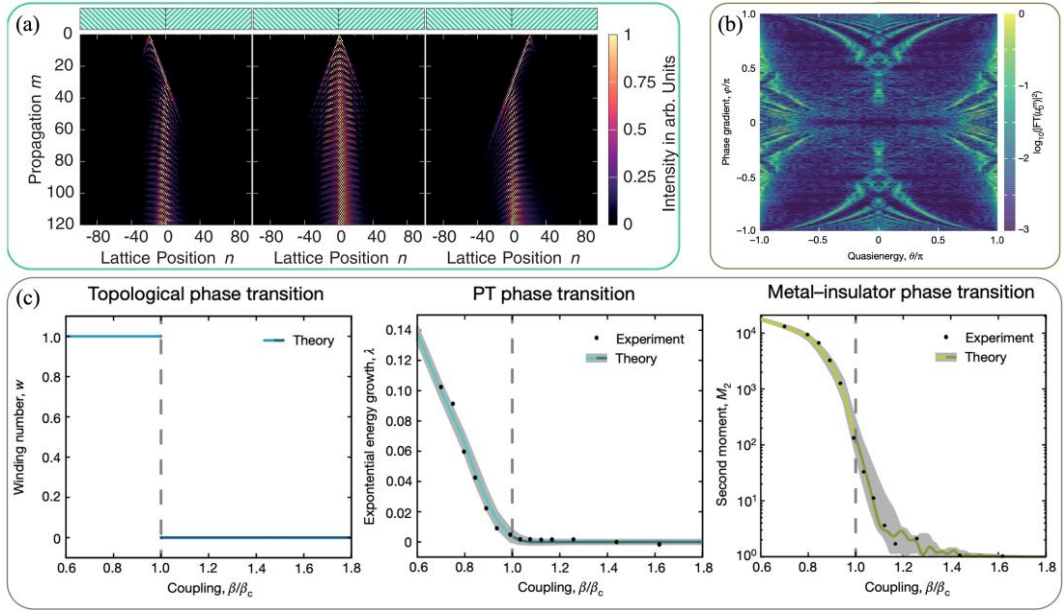

Fig. 13 (a) Topological funneling of light in the synthetic time lattice. Adapt from Ref. [38]. (b) The Floquet Hofstadter butterfly pattern of the AAH model in the synthetic time lattice. (c) The topological triple phase transition by controlling one single parameter, which is the coupling coefficient $\beta$. (b)-(c) are adapted from [44].

## 2. Non-Hermiticity

Non-Hermiticity has been widely explored by the photonics community [28, 274, 275]. The introduction of non-Hermicity in a time-multiplexed network reveals numerous novel physical effects [19, 38, 44, 50, 227, 232, 238, 272, 276-279], which promotes potential applications in optical sensing and light harvesting. Wimmer et al. utilized the phase gradient generated from the phase modulator to form a linear potential for Bloch oscillations and utilized acousto-optical modulators to generate the effective gain and loss for the system, where they discovered secondary emissions during the Bloch revivals. They also demonstrated reconstruction of the non-Hermitian band structure by using Bloch oscillations [277]. Regensburger et al. explored the synthetic photonic lattice with parity-time symmetry, where power unfolding, secondary emissions, and unidirectional invisibility are experimentally obtained [19].

Steinfurth et al. utilized amplitude modulators to induce controllable gain and loss of the system, which plays the role as the imaginary part of the lattice potential. In addition, they utilized phase modulators to induce the hopping phase of the light in the mesh lattice in the time dimension, which plays the role of the real part of the underlaying lattice potential [232]. By designing suitable parameters in experiments, they realized shape-preserving beam transmission and non-Hermitian-induced transparency [see Fig. 14(a)] [232].

The specific design of the intensity modulations and phase modulations can also be used to explore Dirac mass with optical gain and loss included [50]. Yu et al. used the synthetic time lattice to experimentally reveal phenomena related to Dirac masses. In their design, by carefully tuning the gain/loss parameter $g$, they build a massive Dirac cone, following the effective Dirac Hamiltonian [50]

$$H_{\text{eff}} = M(g)\sigma_1 + \frac{1}{2}v_D k\sigma_2 - v_D(\phi - \pi)\sigma_3, \tag{69}$$

where $\sigma_{1,2,3}$ are the Pauli matrices, $k$ is the quasimomentum, $\phi$ is a tunable phase parameter, $M(g) = \cosh(g - 1)$ is the real Dirac mass, and $v_D$ is the Dirac velocity. The scalar potential barrier formed by appropriately engineered phase modulation is added into the synthetic time lattice, and Klein tunneling of massive Dirac quasiparticles is experimentally observed. Furthermore, with proper design of the gain and loss of the system, the sign flip of the mass is also engineered to realize control of time reflection and refraction [see Fig. 14(b)] [50].

The dissipation induced non-Hermicity in time-multiplexed networks in two loops can be further adapted to reveal novel Anderson localization phenomena, where the disorders are introduced from the modulators [227]. The resulting stochastic dissipation leads to breakdown of the Hermitian Anderson model, where dynamical delocalization and spectral localization are concurrently observed in experiment [see Fig. 14(c)] [227].

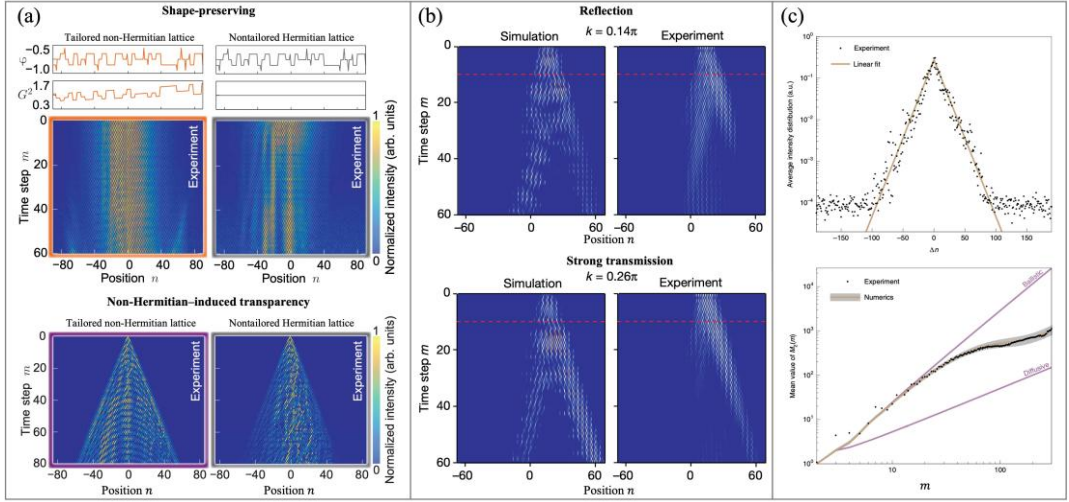

Fig. 14 (a) The shape-preserving beam transmission and non-Hermitian-induced transparency phenomena in the synthetic time lattice. Adapted from Ref. [232]. (b) The time reflection and refraction phenomena at the interface formed by the mass-flipping temporal boundary. Adapted from Ref. [50]. (c) The exponential localization at time step $m$=110 and the mean value of second moment in the non-Hermitian Anderson model. Adapted from Ref. [227].

## 3. Thermodynamic processes in photonics

Thermodynamics, with temperature as an important physical quantity, is significant in many physical processes including superconductivity [280, 281], photovoltaic module electrical performance [282], and soil carbon decomposition [283]. The photonic synthetic time lattice is capable of simulating phenomenon involving thermalization [49], where Joule expansion effects and isentropic expansions-compressions can be experimentally simulated at a negative temperature. The underlying mapping to the thermodynamic process under negative temperature is summarized as follows.

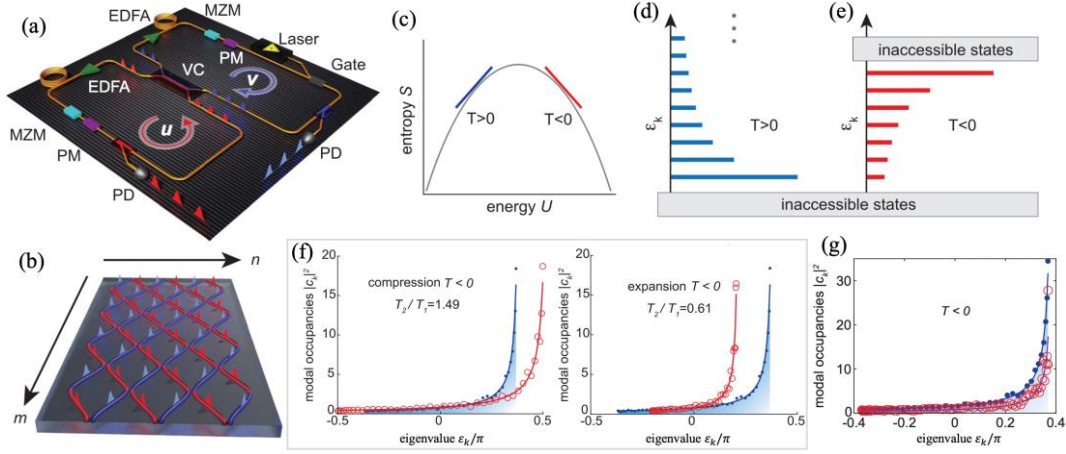

Fig. 15 (a) The optical platform including two fiber rings and various components. (b) The time-synthetic mesh lattice constructed by the two-loop setup. (c) Entropy-energy diagram based on the framework of the thermodynamics. The eigen-spectrum under the positive temperature (d) and negative temperature (e). (f) The isentropic compression and expansion under the negative temperature condition. (g) The optical Joule expansion in the negative temperature regime. (a)-(g) are adapted from Ref. [49].

We again consider two coupled loops with a slight length difference to construct the synthetic time lattice [see Fig. 15(a)-(b)]. A segment of nonlinear fiber supporting photon-photon interaction (four-wave mixing) is used to simulate a thermalization process with the evolution in this time-multiplexed mesh lattice [284, 285]. According to the framework of thermodynamics, the temperature $T$ can be described by entropy $S$ and internal energy $U$, e.g., $1/T = \partial S/\partial U$ [see Fig. 15(c)]. Novel phenomena associated with the temperature can be explored, for example, as a system with the positive temperature favors lower energy states while the microcanonical system with a negative temperature favors high energy levels [see Fig. 15(d)-(e)]. In the experiment, the band structure of the corresponding synthetic time lattice is designed as [49]

$$\cos \varepsilon_k = -C \cos\left(\frac{2k\pi}{M+1}\right) - (1-C)\cos(\varphi_0),$$ (70)

where $C$ is the coupling coefficient between two loops, $M$ is the total lattice sites, $k \in [1, 2, \ldots M]$ is an integer that mimics the Bloch momentum, and $\varphi_0$ is the constant phase to control the band gaps. The internal energy $U$ and the total optical power $P$ can be obtained according to its eigenvalue and corresponding eigenstate, which are expressed as

$$\begin{cases} U = \sum_{k=1}^{M} \varepsilon_k |c_k|^2 \\ P = \sum_{k=1}^{M} |c_k|^2 \end{cases}.$$ (71)

The entropy can then be obtained by calculating the eigenstate distribution, $S = \sum_{k=1}^{M} \ln|c_k|^2$. When one initially excites the negative part of the band structure ($\varepsilon_k < 0$), a negative value of $U$ can be acquired ($U = -22$). On the other hand, the positive value of $U$ ($U = 25$) can be prepared by initially exciting the positive part of the band structure ($\varepsilon_k > 0$). Referring to the entropy-energy diagram in Fig. 15(c), the choice of $U$ can therefore realize the positive and negative temperature condition in experiments. The optical thermalization in the experimental observation matches the theoretical prediction in Fig. 15(d)-(e). The coupling strength between synthetic lattice sites adjusted by the coupler between two loops can also be utilized to simulate the isentropic compressions and expansion with negative temperature conditions. Counterintuitive phenomena are then simulated in the experiments, where the system shows a cooling tendency during the compression under the negative temperature condition [see Fig. 15(f)]. Furthermore, by expanding the lattice range, the synthetic mesh lattice is also able to experimentally simulate the Joule photon-gas expansion. Novel phenomena related to the Joule irreversible expansion are revealed under negative temperature, e.g., the temperature rise of thermalized photon gas after suffering a sudden expansion [see Fig. 15(g)] [49].

## 4. Quantum walk

The two-loop setup for the synthetic time lattice is also capable of simulating quantum walks in experiments. The coupler between two loops can play the role of the coin, and the coupler ratio acts as the probability to get two states of the coin, while the pulse propagation in short (long) loop corresponds to the left (right) movement of the walker [286]. Regensburger et al. experimentally used a two-loop setup to reveal the ballistic spreading of discrete-time quantum walks [238]. Numerous phenomena are shown in the synthetic time dimension based on the two-loop setting, for example, the non-Hermitian potential induced invisibility [287] and unconventional anomalous topology emerging during the discrete quantum walk [273].

Random walks present completely distinct behaviors for quantum and classical particles, and Longhi et al. used traps in the synthetic lattice to distinguish the difference between these two realms [see Fig. 16(a)] [288]. With a similar construction to build a synthetic time lattice in two-loop resonators [see Fig. 10(b)], the discrete-time equations for the light dynamic in the loops can be described as [288]

$$\begin{cases} u_n^{m+1} = [\cos(\beta) u_{n+1}^m + i \sin(\beta) v_{n+1}^m] \exp(-i\phi_n^m - \gamma_n) \\ v_n^{m+1} = [\cos(\beta) v_{n-1}^m + i \sin(\beta) u_{n-1}^m] \end{cases}, \qquad (72)$$

where $\beta$ defines the coupling strength of the fiber coupler, $\phi_n^m$ is the modulation phase induced by the phase modulator, and $\gamma_n$ is the loss induced by the amplitude modulator that can mimic the trap at site $n$. The design for the quantum random walk can be realized by setting $\phi_n^m = 0$, which indicates a coherent walk, while for the classical random walk, the modulation phase is randomly distributed in the range $(-\pi, \pi)$, which indicates an incoherent walk. Four traps are distributed in the 1D synthetic time lattice at different sites with loss rates being $\gamma_n/J = 1, 0.4, 0.5,$ and $0.6$, where $J = 0.5 \cos\beta$. The simulation results show different behavior of a photon in the two situations. For the classical random walker, the photon is annihilated by the dissipative traps and the walk process is terminated [see Fig. 16(b)], while for the quantum random walk, the photon can tunnel out of the traps and continue the walk process [see Fig. 16(c)].

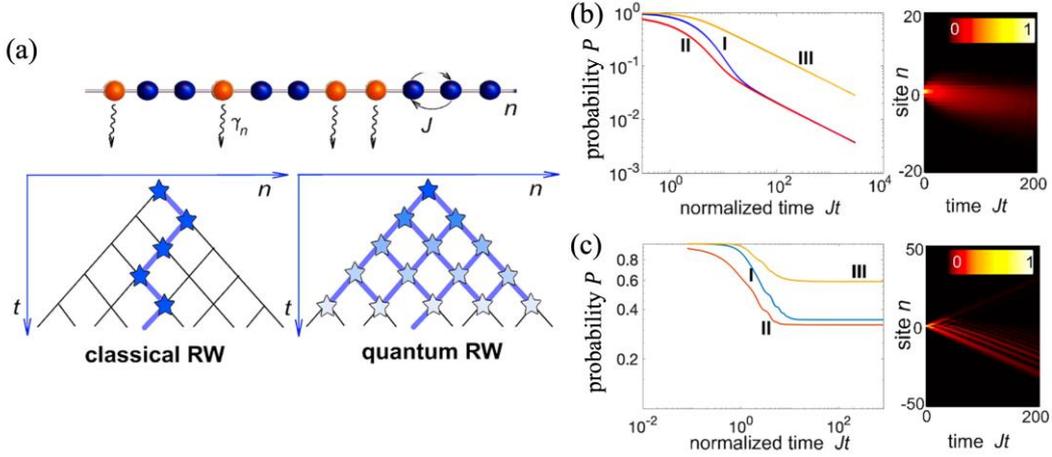

Fig. 16 (a) The classical random walk and the quantum random walk in the 1D lattice with annihilation traps. The survival probability for the classical random walk (b) and the quantum random walk (c). (a)-(c) are adapted from [288].

### D. Summary of the Time-Multiplexed Pulses

In this section, we give a comprehensive summary on using time-multiplexed pulses to construct the concept of a synthetic time dimension. The arrival time of pulses in optical loop(s) is used to label each synthetic lattice site while delay lines added in the single loop or two loops with a length difference are used to introduce connectivity between different sites, so the tight-binding lattice model is built along the temporal positions inside each round trip while the model evolves along the multiple round trips. In the single loop framework, we review frontier research progress, such as the Ising model, dissipative photonics, quantum computation, and quantum walks. On the other hand, in the two-loop framework, we discuss some remarkable physical effects therein, e.g., topology, non-Hermitian effects, thermodynamics, and quantum walks vs classical walks. However, there are many other works that we do not get into the details of here, such as Landau-Zener tunneling [289], temporal refraction [290], Goos-Hänchen shift [291], sampling problem [292], quantum state processing [293], Kapitza light guiding [294] and the quantum walk in the synthetic time dimension based on a sequence of birefringent $\alpha$-barium borate crystals [295]. Furthermore, localization features of physical models have been further studied in the synthetic time dimension [296], such as dynamic localization [297-299], Anderson localization [231, 300, 301] and quasicrystals [230]. Hydrodynamic phenomena [302] and superfluidity of light [303] have been studied and trigger further interest to explore the dynamics of fluids of light based on a time-multiplexed mesh lattice. Nonlinearity in the synthetic time lattice [304, 305] has also been explored to seek potential applications in soliton-based gas sensing [306] and optical neural networks [239, 307]. Some counterintuitive phenomena have been experimentally simulated by taking advantage of this unique platform. For example, Wimmer et al. demonstrated the optical diametric drive acceleration by breaking the action-reaction symmetry [308]. In general, the synthetic dimension based on time-multiplexed pulses can fulfill promising functionalities covering fundamental physics and possible applications [237].

## IV. Atomic Systems: Cold Atoms

The synthetic dimension is widely studied in the atomic systems where atomic states are used to build the extra dimensionality and the connectivity is induced by external lasers and electromagnetic fields (rf and microwave). The field of atomic synthetic dimensions includes many efforts and approaches spanning low-energy atomic states, laser-excited metastable states and Rydberg states, discrete states of atomic motion, and even synthetic dimensions by parametric encoding. In this section, we discuss the former three topics which mainly focus on the use of a discrete degree of freedom in the cold atom platform. We review the basic conceit underlying each of these approaches, discuss considerations relevant to the study of interactions and many-body physics, and review key progress and results from the past decade. Related efforts on superradiant states of neutral atoms, ultracold molecules, and parametric synthetic spaces will also be touched on here and in later sections.

### A. Model

#### 1. Intrinsic atomic states

We take the $^{87}$Rb model as an example from Ref. [82] to briefly introduce the idea of using the intrinsic atomic states to build the synthetic dimension. In many ways, the picture of using laser-coupled or Raman-coupled atomic states to encode a synthetic dimension was a natural extension of previous work on spin-orbit coupled quantum gases [309-313]. As considered in Ref. [82] and illustrated in Fig. 17(a), atomic moving in a hybrid 2D lattice involving real-space and synthetic-space motion may be realized by first confining atoms in a 1D optical lattice, shown along the $x$-axis with a lattice constant of $a = \lambda/2$ and for a lattice laser wavelength of $\lambda = 1064$ nm. A moderate lattice depth of $V_{\text{lat}} = 5E_L$ (where $E_L = \hbar^2 k_L^2 / 2M$ is the lattice recoil energy, with $k_L = 2\pi/\lambda$ the recoil momentum and $M$ the atomic mass) is chosen to validate the tight-binding approximation while still enabling appreciable hopping in real space, characterized by a nearest-neighbor tunneling amplitude of $t = 0.065E_L$. Considering the ground hyperfine ($F = 1$) manifold, a small bias magnetic field with strength $B_0$ defines a set of three magnetic sublevels $|m_F = 0, \pm 1\rangle$ [see Fig. 17(b)] that are Zeeman-split in energy by $\hbar\omega = g_F\mu_B B_0$, where $\mu_B$ is the Bohr magneton and $g_F$ is the Landé $g$ factor. The MHz-level Zeeman-splitting for Gauss-level $B_0$ fields would far exceed any other energy scales ($t$, thermal energies, or interaction energies) in the system, but connectivity between the bare atomic states can be realized by Raman transitions via a pair of coherent laser fields, where the laser fields can be expressed via an effective magnetic field [82]

$$\Omega_T = \delta\boldsymbol{e}_z + \Omega_R\big[\cos(2k_R x)\boldsymbol{e}_x - \sin(2k_R x)\boldsymbol{e}_y\big], \tag{73}$$

where $\delta$ is the detuning, $k_R = 2\pi\cos(\theta)/\lambda_R$ is the Raman recoil wave vector, $\theta$ is the angle deviating from $x$-axis, $\lambda_R$ is the wavelength of the Raman lasers, $\Omega_R$ is the Raman coupling strength and $\hbar = 1$ is taken. The effective atom-light Hamiltonian is [82]

$$H_{\text{al}} = \Omega_T \cdot \boldsymbol{F} = \delta F_z + \big(F_+ e^{ik_R x} + F_- e^{-ik_R x}\big)\Omega_R/2, \tag{74}$$

where $F_{\pm} = F_x \pm iF_y$ can lift/lower the atomic state $|m\rangle$, e.g., $F_+|m\rangle = g_{F,m}|m+1\rangle$, where $m = -F, \ldots, F$ is the label of the sublevels and $g_{F,m} = \sqrt{F(F+1) - m(m+1)}$. The $x$-dependent hopping phase can then generate Peierls phases when atoms move along the $m$ direction. The combination of the 1D optical lattice in the $x$ direction and the synthetic dimension along intrinsic atomic states can yield a hybrid 2D lattice [see Fig. 17(c)], and the corresponding 2D Hamiltonian with $\delta = 0$ is [82]

$$H = \sum_{n,m} (-t a_{n+1,m}^\dagger + \Omega_{m-1} e^{-i2k_R a n} a_{n,m-1}^\dagger) a_{n,m} + \text{h. c.}, \qquad (75)$$

where $n$ is the spatial index, $\Omega_m = \Omega_R g_{F,m}/2$ is the corresponding synthetic hopping strength, $a_{n,m}^\dagger$ and $a_{n,m}$ are the creation and annihilation operators. One convenient feature of the synthetic dimension is that "site-resolved" detection along that axis can be achieve relatively simply, using techniques such as the Stern-Gerlach separation of magnetic sublevels.

Thus, the spin-orbit-coupled atomic gas in a superimposed stationary lattice can naturally be interpreted as a Harper-Hofstadter-like Hamiltonian in a hybrid real/internal space, where $\theta$ and the ratio $\lambda/\lambda_R$ defines the gauge field (magnetic flux per lattice plaquette). This approach is naturally extensible to large-spin atoms [314]. Importantly, however, interactions along the synthetic dimension are sensitive to the details of specific scattering lengths for same-spin and mixed-spin configurations. While the interactions are easy to interpret (but nearly all-to-all) for $^{87}$Rb and alkaline earth (and alkaline earth-like) atoms with SU($N$) symmetry, they can be quite random [315] and potentially exothermic [316] for generic multi-level atoms. To note, the use of metastable "clock" states as a small (two-site) synthetic dimension has also been explored [317-320], but interactions are typically quite complicated in this approach as well [321, 322].

## 2. Rydberg atoms

Because of the complicated nature of interactions (either all-to-all or potentially inelastic) in neutral atom synthetic dimensions experiments, it was motivated to consider the engineering of synthetic dimensions in systems that are violently inelastic when in contact [323, 324] but that can support strong and coherent interactions even when separated in real-space [325, 326]. While originally envisioned in the context of ultracold polar molecules [325, 326], recent development has mainly occurred in the entirely analogous synthetic dimension of dipolar-interacting Rydberg levels of excited atoms [16]. To note, both of these systems (molecules and Rydberg atoms) host many stable or quasi-stable internal states that can be coupled to one another by relatively strong electric dipole-allowed microwave transitions. Both the Rydberg array and the polar molecule array platform had demonstrated coherent dipole-dipole interactions [327, 328] at the time that the ideas for synthetic dimensions with dipolar spin arrays [325, 326] were developed, and earlier work had demonstrated tight-binding-like dynamics [329, 330] in both systems as well.

We emphasize here that for the synthetic dimensions of (lattice-pinned) molecules and dipolar Rydberg atoms, one can simply think of these systems as dipolar spin models with rather exotic external (transverse and longitudinal) fields relating to the microwaves that induce direct state-to-state transitions to mimic "hopping" in the synthetic dimension. Unlike in the hybrid 2D lattice (real and synthetic dimensions) of spin-orbit-coupled gases described earlier, the excitations in Rydberg synthetic dimension experiments only experience single-particle hopping in the synthetic space, with no hopping of the actual particles in real space (ignoring the cases of itinerant molecules or Rydberg-dressed atoms).

Here, we focus on the case of Rydberg synthetic dimensions, which utilize the valence electron of Rydberg atoms as the constituent particles. In contrast to conventional low-lying atomic states, the Rydberg states refer to the valence electron of atoms being excited to high principal quantum numbers. These states feature a large distance between the electron and the positive atomic core [see Fig. 17(d)] [331, 332], imbuing these states with exaggerated properties (large polarizability, large dipole-dipole interactions) and a lifetime that far exceeds typical laser-excited levels. The arrangement of the valence electron and the atomic core is similar to the hydrogen atom, and the modified Rydberg formula to characterize the energy of such Rydberg levels is

$$E_{n,l} = -\frac{R_y}{(n - \delta_l)^2},$$ (76)

where $R_y \approx 13.6$ eV is the Rydberg constant, $n$ is the principal quantum number, $l$ is the orbital angular momentum of the state, and $\delta_l$ is the quantum defect that accounts for the Hartree interaction shift of the valence electron with the core electrons [333, 334]. In addition to their principal quantum number $n$, orbital angular momentum $l$ and its projection $m_l$, the Rydberg levels are characterized by their total angular momentum $J$ and its projection $m_J$, and all together the levels are split by fine-structure as well as Zeeman splittings in the presence of a magnetic bias field (considering the case of zero electric field). Spontaneous decay and blackbody-induced decay and transitions can be mostly ignored on short (few microsecond) timescales, and dipole-allowed microwave transitions between Rydberg levels can be used to explore fundamental transport phenomena in synthetic tight-binding models [335, 336]. In a finite bias field, all state-to-state transitions are associated with a unique microwave frequency, such that all hopping terms in the synthetic dimension can be individually controlled due to their spectroscopic isolation.

We briefly review the general approach to Rydberg synthetic dimensions, as explored in Refs. [332, 335-339]. A comb of coherent microwaves that address (either resonantly or with controllable detunings) a number of state-to-state transitions are utilized to engineer the effective connectivity along the synthetic dimension (see, e.g., Fig. 17(e) for the case of an SSH model). Because each transition is separately controlled, the generic form of the single-particle Hamiltonian along the synthetic dimension is

$$H = \sum_i (-\hbar J_{i,i+1} |i\rangle\langle i+1| + \text{h.c.}) + \sum_i \hbar \delta_i |i\rangle\langle i|,$$ (77)

for a simple one-dimensional geometry of state connectivities (where the $J_{i,i+1}$ terms are the tunable inter-state coupling amplitudes, which can have controllable complex phases, $\delta_i$ are effective site energies introduced via transition detunings, and $\hbar$ is Planck's constant). Because of the large state space available and the ability to drive multi-photon processes, there are many opportunities to engineer longer-range hopping terms and even higher-dimensional lattices within the Rydberg synthetic dimension. The physics of the engineered synthetic lattice, and in particular the phenomena that arise due to dipolar interactions between Rydberg atoms, can be explored either through spectroscopy of the microwave-dressed states as in Ref. [336] or by studying the microwave-driven population dynamics within the set of Rydberg levels (after first populating a particular Rydberg level by laser excitation) as in Refs. [332, 335, 337-339].

One of the most attractive aspects of the Rydberg synthetic dimensions platform is the existence of strong and stable dipole-dipole interactions that enable the exploration of many-body phenomena. These experiments also benefit from the rapidly expanding capabilities related to the preparation, control, and detection in atom array experiments. At present, however, one of the practical difficulties associated with Rydberg synthetic dimensions is the inability to perform high-fidelity simultaneous detection of many Rydberg levels in an atom-resolved fashion. High-fidelity and atom-resolved detection of individual Rydberg levels has been achieved through standard approaches in tweezer arrays [335, 338, 339] and, complementing this, fully state-resolved detection via sequential field-induced ionization has been achieved for single atoms [332, 336, 337]. But the combination of these capabilities for full and high-fidelity state readout is still outstanding.

### 3. Momentum lattice

In the previous two examples, the synthetic dimensions occur in an internal degree of freedom that is almost entirely distinct from (and in addition to) the real-space dynamics, given the huge energy separation between the internal and external degrees of freedom. Those approaches can thus be used to explore higher-dimensional physics by combining real-space and synthetic lattices. A bit different from those approaches, one can also consider to use momentum states or generic motional states of atoms as a discrete degree of freedom to form an effective tight-binding model. In this case (similar to the works in photonics [36]), the synthetic dimension replaces a real dimension, but one still benefits from the ability to induce the connectivity between states in the synthetic dimension through laser-driven transitions (so as to engineer complex hopping terms, e.g.). One approach to engineer synthetic dimensions based on laser-coupled momentum states of neutral atoms was considered by Gadway et al., where the presence of the quadratic free-particle dispersion of massive atoms leads to a unique transition frequency for two-photon Bragg transitions (and Raman-Bragg transitions) between states in the synthetic dimension [340]. To note, the idea to utilize discrete momentum states of atoms as a landscape for exploring tight-binding transport phenomena was highly influenced by the body of research on kicked rotor experiments with cold atoms [341-347] and the use of the quadratic dispersion to enable the control of complex hopping was highly influenced by contemporary work on Raman- or modulation-assisted hopping in biased real-space optical lattices [348-350].

Roughly speaking, the momentum state (or space) lattice (MSL) technique can simply be thought of as connecting a set of discrete plane-wave momentum states of ultracold atoms (as derived from a Bose-Einstein condensate) through a set of coherent two-photon Bragg transitions [351, 352]. Because the atomic waves have mass and the free-particle dispersion is quadratic, the energy difference between adjacent pairs (or nearest neighbors) of momentum states is unique, reflecting the Doppler shift of the Bragg transition. Unlike in kicked rotor experiments, which apply short (in time) but strong lattice pulses to atoms, one can operate at weak Bragg coupling and utilize this momentum-dependence of the Bragg transition frequency to achieve a unique control over all of the site-to-site connectivity terms. This is similar to the case of Rydberg and molecule synthetic dimensions, which were heavily inspired by the momentum lattice technique.

We now review the formalism behind the MSL technique. We consider a pair of counter-propagating laser beams applied to atomic waves having mass $M$, where the left-traveling field contains a set of discrete frequency components and the right-traveling field contains a single frequency component [see Fig. 17(f)]. For a system of two-level atoms with ground state $|g\rangle$ and excited state $|e\rangle$ [see Fig. 17(g)], the interaction process between light and atoms gives the following Hamiltonian after considering the dipole approximation [340]

$$H = \frac{\boldsymbol{p}^2}{2M} + \hbar\omega_e|e\rangle\langle e| + \hbar\omega_g|g\rangle\langle g| - \boldsymbol{d}\cdot\boldsymbol{E}, \tag{78}$$

where $\boldsymbol{d} = -|e|\boldsymbol{r}$ is the dipole operator with $\boldsymbol{r}$ being the distance vector, $\boldsymbol{p}$ is the free-particle momentum, $\hbar\omega_g(\hbar\omega_e)$ is the ground (excited) state energy, and $\boldsymbol{E}$ is the electric field of the driving lasers. The right-traveling field and the left-traveling field have the expressions as [340]

$$\boldsymbol{E}_+(x,t) = \boldsymbol{E}_+ \cos(\boldsymbol{k}_+ \cdot \boldsymbol{x} - \omega_+ t + \phi_+), \tag{79}$$

$$\boldsymbol{E}_-(x,t) = \sum_j \boldsymbol{E}_j \cos(\boldsymbol{k}_j \cdot \boldsymbol{x} - \omega_j t + \phi_j), \tag{80}$$

where $\boldsymbol{x}$ is the (one-dimensional) position, $\omega_+$ and $\omega_j$ are frequencies, $\boldsymbol{k}_+$ and $\boldsymbol{k}_j$ are wave vectors of the lasers, $\phi_+$ and $\phi_j$ are the phase of the driving lasers, $\boldsymbol{E}_+$ and $\boldsymbol{E}_j$ correspond to the amplitudes of the respective components. The detuning between atomic resonance ($\omega_{eg} = \omega_e - \omega_g$) and laser frequencies is $\Delta \equiv \omega_{eg} - \omega_+ \simeq \omega_{eg} - \omega_j$. The resonant Rabi coupling between ground and excited states are $\Omega_+ = -\langle e|\boldsymbol{d}\cdot\boldsymbol{E}_+|g\rangle/\hbar$ and $\Omega_j = -\langle e|\boldsymbol{d}\cdot\boldsymbol{E}_j|g\rangle/\hbar$. The Hamiltonian can then be re-written as [340]

$$H = \frac{\boldsymbol{p}^2}{2M} + \hbar\omega_e|e\rangle\langle e| + \hbar\omega_g|g\rangle\langle g|$$
$$+\hbar\left[\Omega_+ \cos(kx - \omega_+ t + \phi_+) + \sum_j \Omega_j \cos(-kx - \omega_j t + \phi_j)\right](|e\rangle\langle g| + |g\rangle\langle e|). \tag{81}$$

By assuming that the single-photon detuning $\Delta$ far exceeds all the single-photon Rabi coupling terms ($\Omega_+$ and $\Omega_j$), tracing out the excited state, and making a rotating wave approximation, the Hamiltonian in Eq. (81) becomes [340]

$$H_{\text{eff}} = \sum_n 4n^2 E_r|n\rangle\langle n| + \frac{\hbar}{4\Delta}\sum_n \Omega_+\Omega_n e^{-i(\omega_+-\omega_n)t}e^{i(\phi_+-\phi_n)}|n+1\rangle\langle n|$$
$$+ \frac{\hbar}{4\Delta}\sum_n \Omega_+\Omega_n e^{i(\omega_+-\omega_n)t}e^{-i(\phi_+-\phi_n)}|n\rangle\langle n+1|, \tag{82}$$

where $n$ is an integer to characterize the momenta $\boldsymbol{p}_n = 2n\hbar k\hat{x}$ and $E_r = \hbar^2 k^2/2M$ is the recoil energy. The introduced states $|n\rangle$ relate to the motional, plane-wave-like states of ground state atoms that have momentum $\boldsymbol{p}_n$. The Hamiltonian can be further transformed into the interaction picture as [340]

$$H_{\text{eff}}^I = \sum_n \frac{\hbar\widetilde{\Omega}_n}{2} e^{\frac{i(2n+1)4E_r}{\hbar}t} e^{-i(\omega_+-\omega_n)t}e^{i(\phi_+-\phi_n)}|n+1\rangle\langle n|$$
$$+ \sum_n \frac{\hbar\widetilde{\Omega}_n}{2} e^{-\frac{i(2n+1)4E_r}{\hbar}t} e^{i(\omega_+-\omega_n)t}e^{-i(\phi_+-\phi_n)}|n\rangle\langle n+1|), \tag{83}$$

where $\widetilde{\Omega}_n = \frac{\Omega_n\Omega_+}{2\Delta}$ is the two-photon Rabi coupling. By setting $\omega_n = \omega_+ - (2n+1)\,4E_r/\hbar$ and neglecting non-resonant components, the Hamiltonian can be simplified to [340]

$$H_{\text{eff}}^I = \sum_n \left(\frac{\hbar\widetilde{\Omega}_n}{2} e^{i(\phi_+-\phi_n)}|n+1\rangle\langle n| + \frac{\hbar\widetilde{\Omega}_n}{2} e^{-i(\phi_+-\phi_n)}|n\rangle\langle n+1|\right). \tag{84}$$

Finally, by denoting the hopping amplitude as $t_n = \frac{\hbar \tilde{\Omega}_n}{2}$ and the hopping phase as $\varphi_n = (\phi_+ - \phi_n)$, the resulting Hamiltonian in the synthetic momentum lattice reads as [340]

$$H_{\text{eff}}^I = \sum_n \left( t_n e^{i\varphi_n} |n+1\rangle\langle n| + \text{h.c.} \right). \tag{85}$$

We can see that in the Eq. (85), the coupling strength and the phase can be individually adjusted (and the ignored detuning terms can be used to engineer arbitrary site-energy landscapes [353]), which provides high tunability in simulating various models for different aims.

The above formalism discusses the scenario of a nearest-neighbor-coupled one-dimensional MSL with one internal state. The single-particle control can be and has been expanded in many ways, including with multiple internal states (with two-state non-Abelian lattices also considered in Ref. [340] and realized in Ref. [354]), longer-range hopping [355] or multiple driving wavelengths [356] leading to the engineering of gauge fields, dissipative loss [357–359], and higher dimensional lattices [360, 361]. Perhaps most importantly, however, atomic $s$-wave collisions between identical bosons naturally give rise to a nonlinear interaction [362] that is effectively local in momentum space and which is analogous to the Kerr nonlinearity of photonics [363, 364]. This interaction term in the MSL allows for the study of nonlinear phenomena in the synthetic dimension. Experiments with bulk gases are just now pushing to the regime of strongly correlated dynamics [365], and this exciting direction has also been accessed in recent years by combining the MSL concept with high cooperativity optical cavity experiments [366, 367].

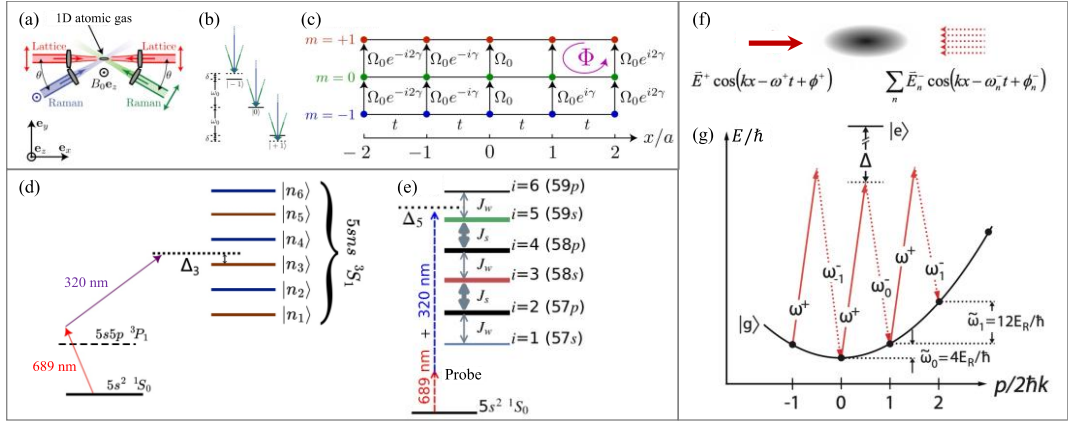

Fig. 17 (a) The proposed experimental layout to construct the synthetic dimension based on the intrinsic states of cold atom. (b) The three magnetic sublevels of the $F = 1$ ground state. (c) The 2D lattice with combination of the 1D optical lattice and the synthetic spin lattice. (a)-(c) are adapted from Ref. [82]. (d) The excitation of the Rydberg states through two-photon transitions. Adapted from Ref. [332]. (e) The connection between Rydberg levels through the employment of millimeter waves on the atomic system. Adapted from Ref. [336]. (f) A schematic diagram depicting how to form a momentum state lattice for ultracold atoms by driving with a pair of counter-propagating laser beams. (g) The coupling between momentum lattice sites is achieved by two-photon Bragg transitions. (f)-(g) are adapted from Ref. [340].

## B. Physical Phenomena

### 1. Intrinsic atomic states

We first discuss some novel physical phenomena explored based on intrinsic, stable atomic states (treating Rydberg states as a separate consideration). Since the first realizations of atomic quantum gases, the internal state space had been considered as a relevant resource for quantum dynamics studies and many-body physics [368], even considered through the lens of an analog to real-space transport phenomena [369]. While there have been studies relevant to synthetic dimensions phenomena arising purely in the internal state space of ultracold atoms [32], we will mostly restrict our discussion to studies that combine real-space and internal-space control, with the atomic internal state providing an extra, synthetic dimension.

The first concrete proposal for using the internal states of cold atoms to embed an extra "synthetic" dimension was the work by Boada, et al [370], in the context of higher-dimensional Hubbard models. Because of the massive separation of frequency scales between external motional states (Hz to kHz) and internal states (MHz to GHz), one can effectively decouple the two and consider the dynamics in the internal-state space as being separate from that in real space. Later works would discuss extensions to this based on potential realizations of the four-dimensional quantum Hall effect [371]. While each of these proposed models required some rather intricate requirements on the spatial arrangement of internal-state transitions or the form of internal state-dependent interactions, these works and others [215] helped to stimulate the use of internal states as a resource for engineering Hamiltonians relevant to topological phases.

The 2014 work by Celi et al. [82] constituted a major practical breakthrough, in effect reinterpreting the phenomena of spin-orbit coupling in cold atomic gases [372] (in an underlying real-space lattice) in terms of the effective dynamics in a hybrid real-internal space and showing that this approach allowed for the generation of Harper-Hofstadter models (describing charged electrons moving in a perpendicular magnetic field) with large effective magnetic fields. This proposal helped to provide an intuitive picture to how such models are implemented in hybrid real-synthetic lattices, with the effective gauge field resulting directly from the spatial variation of the relative phase of the lasers driving state-changing Raman transitions. This pioneering work also made clear the power of the approach, from the ability to observe dynamics with "single-site" resolution in the internal space by spin-selective (Stern-Gerlach) imaging to the robustness of the implementation with respect to technical heating. This theoretical proposal was followed soon thereafter by two experimental studies that realized large effective magnetic fields in few-leg ladder systems, Refs. [22] and [23], both observing hybrid "real-synthetic"-space skipping orbits associated with the appearance of quantum Hall edge states, as depicted in Fig. 18(a). The generality of this approach was further emphasized by the theoretical proposal by Wall et al. [317], where single-photon transitions to long-lived laser-excited internal states could also be utilized to achieve effective magnetic fields in two-leg ladder-like systems. Experimental realizations similarly followed soon thereafter based on optical clock transitions in alkaline earth [319, 320] and alkaline earth-like atoms [318], where the ability to drive one-photon transitions provided additional tunability of the effective artificial gauge field [318].

Continued advances in the control and engineering of internal state synthetic dimensions continued for several years, with improved measurement techniques [373] and the realization of systems with period boundary conditions [374-377]. A significant advance came with the realization of internal-state synthetic dimensions (spin-orbit coupling) in highly magnetic atoms possessing many internal states, where the exploration of Chalopin et al. [314] with dysprosium atoms (possessing 17 low-energy internal states) allowed for the clear visualization of both bulk cyclotron orbits and chiral edge-state dynamics, as depicted in Fig. 18(b) (cf. also Refs. [378, 379]). Such studies have been extended in several directions, with periodic boundary conditions [380] and the effective realization of small four-dimensional lattices [51].

Considering the many advances in the engineering of effective magnetic fields based on atomic internal states, combined with the decades-long exploration of many-body phenomena in ultracold gases [381], it is natural to ask if one could use these systems to directly explore analogs of correlated matter (bosonic or fermionic) in large effective magnetic fields associated with the fractional quantum Hall effect. In this context, it is important to consider how interactions appear along the synthetic dimension, and how those compare to the local-in-space interactions one would associate with (screened) Coulomb interactions in, e.g., electronic matter. For species like rubidium-87 and the alkaline earth (and alkaline earth-like) atoms, the interactions are either SU($N$) symmetric or nearly so, such that the interactions appear as being very long-ranged (infinitely so) in the synthetic dimension. In the case of magnetic atoms such as dysprosium and erbium, the dependence of two-body interactions on the positions of particles in the internal space is expected to be highly random [315] and often inelastic. All told, these experimental details have complicated the pathway from realizing single-particle dynamics in large effective magnetic fields to the realization of correlated states of atoms in large effective magnetic fields. However, there has been recent success when restricting to a synthetic two-leg ladder of two internal states, where the universal Hall response of strongly interacting fermions in large magnetic fields [as depicted in Fig. 18(c)] has been explored [382]. With continued advances, the synthetic dimension based on intrinsic atomic internal states holds great promise for the quantum simulation of interacting fermionic matter in large effective magnetic fields.

While our discussion has largely focused on advances in the experimental realization and exploration of atomic synthetic dimensions, we should also stress that there have been significant theoretical studies on the kinds of phenomena that may explored in atomic synthetic dimensions, including Haldane phases in synthetic ladders [383], orbital magnetism of frustrated Creutz ladder systems [384], synthetic Hall ribbons with an unconventional squished baryon fluids phase [385, 386], and Laughlin-like states associated with chiral currents in ladder-like systems [387]. A number of theoretical studies have sought to capture the potential influence of unconventional interactions along the synthetic dimensions on the possible many-body phases, with Bilitewski et al. considering the competition between density-density interaction and density-dependent tunneling [388], Barbarino et al. considering the enhancement of chiral currents due to repulsive atom-atom interactions [389] and the emergence of fractional helical liquids from repulsive contact interactions [390], Anisimovas et al. considering the interplay of frustration and nonlocal atom-atom interactions [391], and Yan et al. demonstrating the presence of Majorana zero mode states and a topological superfluid under the influence of attractive Hubbard interactions [392]. Building on the early study of Boada et al. [215], researchers have also explored several possibilities for engineering novel geometries and topologies based on synthetic dimensions [393, 394]. Finally, there have been several theoretical studies exploring how interactions can be combined with the control over periodic boundary conditions to study fractional charge pumping and fractional quantum Hall behavior in synthetic dimensions experiments [395, 396].

In describing the above studies in the context of synthetic dimensions, many of which are related to the engineering of effective magnetic fields for neutral atoms, we should not overlook the more general body of works that explore the spin-orbit coupling of atomic systems based on two-photon Raman transitions. Indeed, a good number of the above-discussed works can equivalently be described in the context of spin-orbit coupling [397-402]. While potential distinctions could be made based on whether the synthetic dimension size is more than just two states or whether the approach involves an effective magnetic field giving rise to topological edge states, this distinction is a bit arbitrary, with some theoretical works more explicitly melding the approaches of synthetic dimensions and spin-orbit coupling [403]. One recent study by Zhang et al., which utilized two internal states to represent two layers of a twisted bilayer structure [see Fig. 18(d)] [47], is a beautiful example of how internal states can be used to introduce a synthetic dimension without the need to invoke magnetic fields or topological effects. Similar to the scheme [404] utilized in [47], it has alternatively been proposed [405] that one may implement twistronics without a physical twist by imprinting spatially dependent interlayer hopping between two internal states of atoms in a synthetic bilayer.

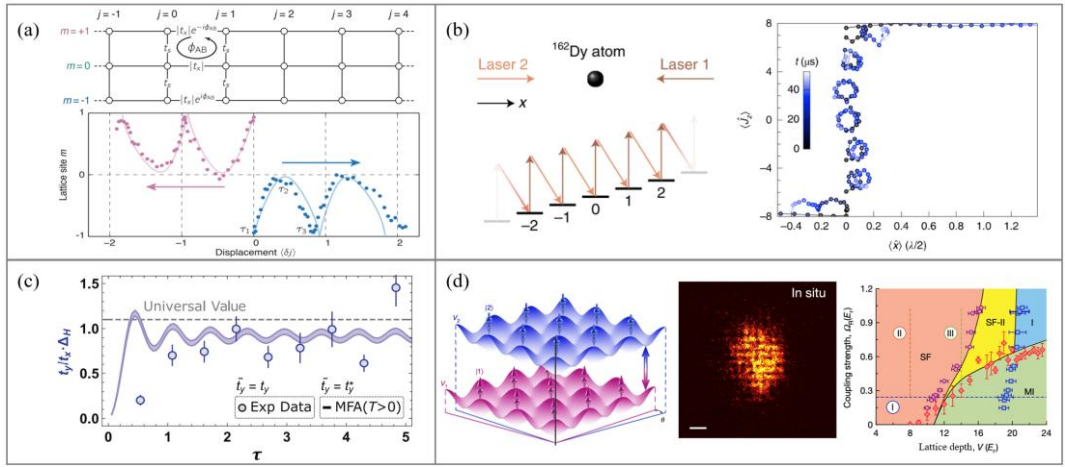

Fig. 18 Synthetic dimensions with low-lying atomic internal states. (a) In few-leg Hofstadter lattices based on Raman-coupled hyperfine states, researchers have observed the skipping orbits associated with topological boundary states. Adapted from Ref. [23] (cf. also Ref. [22]). (b) In larger-spin magnetic atoms such as dysprosium, one can similarly explore both bulk and boundary dynamics in spin-orbit-coupled gases. Adapted from Ref. [314]. (c) Control of atomic interactions in few-state systems has recently enabled the exploration of many-body effects, namely the observation of universal Hall response in two-leg flux ladders. Adapted from Ref. [382]. (d) The moiré pattern and rich phase diagram resulting from a synthetic twisted bilayer structure. Adapted from Ref. [47].

## 2. Rydberg atoms

We next summarize some of recent progresses made in the framework of Rydberg-level synthetic dimension. This platform is just a few years old, building on the proposals for synthetic dimensions in dipolar molecules [325], but has already made some fast progress in demonstrating single-particle control and the influence of interactions. In some respects, the earliest study on synthetic dimensions with Rydberg atoms pre-dated the aforementioned proposal [325], with a tight-binding lattice formed in the high angular momentum states of individual Rydberg atoms in Ref. [329] by application of a common resonant radiofrequency tone. In 2022, Kanungo et al. were the first to implement a Rydberg synthetic dimension with spectroscopic control, constructing a topological SSH tight-binding model by applying a set of distinct microwave fields to couple together a set of Rydberg levels in individual Sr Rydberg atoms. They further used frequency-resolved laser-driving to probe the system's band structure, including the zero-energy topological edge states and their robustness to disorder [see Fig. 19(a)] [336]. Recently, the study of Rydberg synthetic dimensions in Sr atoms has been extended by Lu et al. to the study of real-time dynamics following state initialization, with the bulk topology (winding number) of the SSH lattice structure extracted through the real-time quench dynamics of Rydberg state populations [see Fig. 19(b)] [332]. Through dynamics, Lu et al. have also directly observed topologically-protected edge states, striking long-range tunneling between zero-energy modes, as well as the destruction of topological protection under the introduction of disorder breaking the chiral symmetry [406]. In a separate effort using similar techniques, Trautmann et al. have recently utilized time-dependent control of the microwave driving fields to accomplish Thouless pumping of individual Rydberg excitations in the synthetic dimension [407].

The above studies in Refs. [329, 332, 336, 406, 407] were all based on individual atoms from atomic beams and bulk gases excited to Rydberg levels, allowing for high repetition rates as well as the efficient detection of many Rydberg states through sequential field ionization. However, these studies on single atoms did not allow for the exploration of how strong dipole-dipole interactions between particles (Rydberg atoms or molecules) can dramatically influence the dynamics in the synthetic dimension, as explored in Ref. [325]. The first study combining Rydberg synthetic dimensions and strong dipolar interactions was that of Chen and Huang et al. based on one-dimensional arrays of individual potassium atoms trapped in optical tweezers and excited to Rydberg states [335]. At the level of single-atom control, they introduced a tunable artificial gauge field (over a single plaquette) to the Rydberg synthetic dimension by introducing a closed path in the internal state space [see Fig. 19(c,left)]. They further explored how the dipolar interactions between atoms in the array influenced the Rydberg state population dynamics, controlling the spacing between the atoms to tune the strength of the dipole-dipole interactions. In the limit of large dipolar interactions, the interactions between atoms led to a suppression of uncorrelated transport [see Fig. 19(c,right)], with population becoming trapped in the initially prepared Rydberg level. Additional explorations of larger one-dimensional synthetic lattices by Chen et al. explored Stark localization and its breakdown due to dipolar interactions, as well as Floquet-activated hopping of atom pairs in the synthetic dimensions [338]. And recently, Chen and Huang et al. have engineered kinetically frustrated band structures - specifically a diamond lattice with flux - in the Rydberg synthetic dimension [339], studying how strong dipolar interactions could lead to band mixing and delocalization due to the breakdown of Aharonov-Bohm caging, as shown in Fig. 19(d).

Looking forward, beyond the results referenced above and other forthcoming works, the platform of Rydberg synthetic dimensions can be extended in several straightforward directions. First, such multi-level controls can be introduced into larger optical tweezer arrays, which have been created with several thousands of atoms [408]. Second, these early demonstrations can easily be extended to larger and more elaborate internal lattices involving many dozens of Rydberg levels. Third, the global microwave control can be combined with local optical addressing to expand the control over initial state preparation, Hamiltonian engineering, and state readout. Some possible science targets for these systems have been explored since the earliest proposal for dipolar synthetic dimensions, with some rather exotic quantum string and membrane phases expected to arise even for simply a uniform tight-binding model along the synthetic dimension [325, 326, 409, 410]. Even more unusual properties have been predicted to emerge in small synthetic lattices of just three or four Rydberg states, which may naturally give rise to quasiparticles with exotic exchange statistics as predicted by Wang and Hazzard [411]. These works point to the possibility of a variety of rich phenomena to be explored in Rydberg synthetic dimensions.

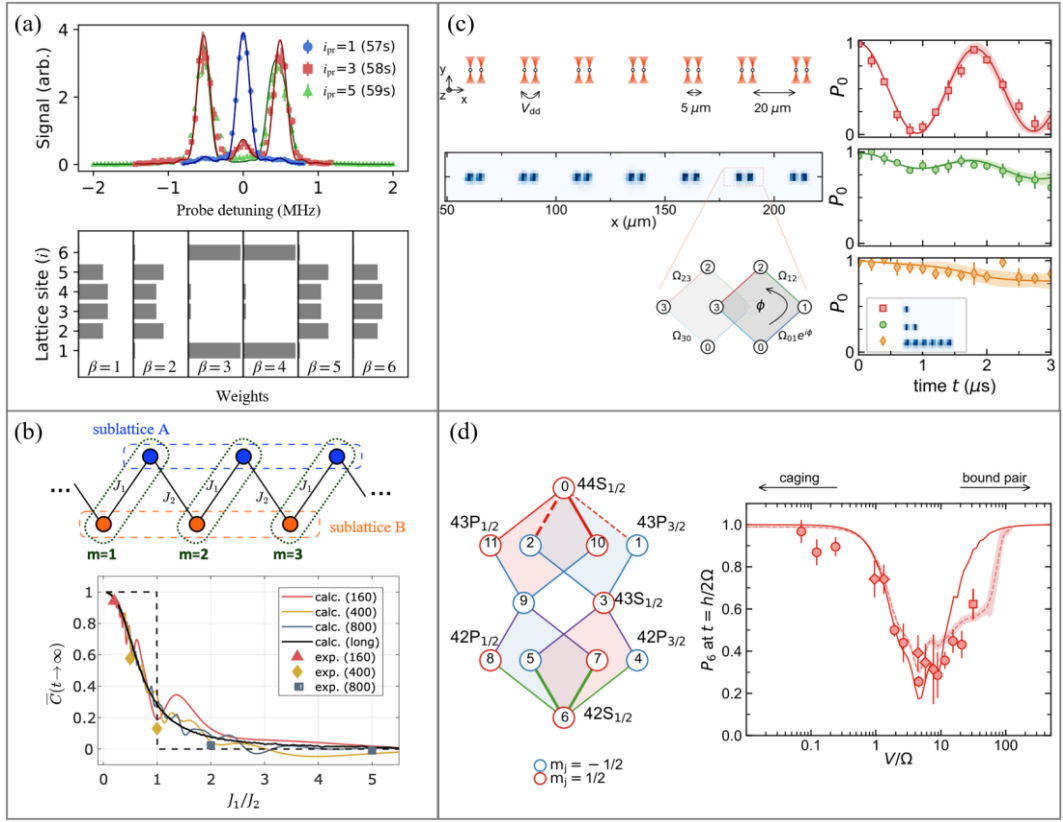

Fig. 19 Synthetic dimensions in Rydberg atoms. (a) The excitation spectra and the state decomposition weights in a SSH lattice formed by microwave-coupled Rydberg levels. Adapted from Ref. [336]. (b) The topological winding number extracted from the dynamics of Rydberg level populations in a SSH lattice. Adapted from Ref. [332]. (c) The influence of strong dipolar interactions leading to inhibited dynamics of Rydberg electrons in a synthetic flux plaquette. Adapted from Ref. [335]. (d) The observation of Aharonov-Bohm (AB) caging and its breakdown due to strong dipolar interactions in a twisted diamond lattice. Adapted from Ref. [339].

## 3. Momentum lattice

The original proposal for momentum space/state lattices (MSLs) [340] emphasized that the experimental ingredients required for their engineering had been present in many cold atom experiments dating back to (and even before) the creation of Bose-Einstein condensates. One simply had to drive the atoms with a pair of lasers and operate with weak Bragg coupling, such that local control over the hopping elements (Bragg transitions) along the synthetic "momentum lattice" could be independently controlled through their unique resonance conditions (Doppler-shifted Bragg resonances). Thus, the first implementations of momentum lattices followed soon thereafter by Meier et al. [353], demonstrating control of hopping amplitudes, complex hopping phases, and effective potential landscapes.

Restricting to only nearest-neighbor tunneling terms in one-dimensional MSLs, follow-up efforts by Meier at al. [412] and An et al. [413] would respectively demonstrate the detection of topological edge states in the SSH model and the engineering of near-arbitrary disorders (diagonal, off-diagonal, time-dependent). By combining the chiral symmetric structure of the SSH Model with tunable off-diagonal disorder, Meier et al. would soon also demonstrate the phenomenon of disorder-induced topology associated with the topological Anderson insulator phase [31], depicted in Fig. 20(a). Additional studies occurring in simple nearest-neighbor-coupled one-dimensional models included explorations of helical Floquet channels [414], studies of critical localization phenomena [415], studies of fine-tuned, self-dual quasiperiodic potentials [416], and the first experimental realizations of time reflection and refraction by Dong et al. [417], depicted in Fig. 20(b).

The ability to directly control the complex phase of all hopping elements was one of the original motivations to explore MSLs [340], but this capability is of little consequence in nearest-neighbor 1D models. The first realization of artificial gauge fields, or effective magnetic fields, in MSLs was by An et al., based upon driving two independent sets of Bragg transitions to create an effectively two-dimensional MSL [356], albeit with only two sites in one dimension. In a separate study, An et al. created a two-leg zig-zag ladder system in a one-dimensional MSL with only one set of Bragg lasers by combining nearest-neighbor and next-nearest-neighbor hopping [355]. In both of these studies [355, 356], the two-leg ladder systems could be interpreted in terms of one-dimensional lattices with sublattice structure within a unit cell, with the flux-dependent dynamics being recast in the language of spin-orbit coupling and flux-dependent control of the band dispersion, leading to flux-dependent localization transitions and other effects [355]. A much more extreme version of this flux-dependent control of band dispersions was accomplished by Li et al. with the realization of a diamond lattice with control of the flux penetrating each diamond plaquette [418]. Starting with a frustrated diamond lattice under the all-flat-bands condition of Aharonov-Bohm caging, Li et al. showed that the addition of disorder would actually induce transport via the phenomenon of inverse Anderson localization, with a transition to Anderson localization under very larger disorder [418]. Later studies by Zeng et al. [419] and Mao et al. [420] would demonstrate and explore in detail the same phenomena in the analogous Tasaki (or sawtooth) lattice, depicted in Fig. 20(c).

The possibility of introducing non-Hermitian loss in MSLs was first proposed and explored by Lapp et al., through either effective loss by weakly coupling to a reservoir of states or by weakly coupling to an auxiliary internal state that can experience scattering-induced loss from the relevant set of momentum levels [357]. These techniques would be demonstrated in a set of pioneering experiments that followed, with demonstrations of non-reciprocal transport by Gou et al. [358], quantum Zeno effects and parity-time symmetry breaking by Chen et al. [421], and the first observation of the non-Hermitian skin effect in an atomic system by Liang et al. [359].

In addition to non-Hermitian effects, which may be thought of through the lens of complex hopping phases and complex gauge fields, MSLs also offered the possibility of exploring other exotic gauge field Hamiltonians. As discussed in Ref. [340], the combination of state-preserving Bragg transitions and state-changing Raman Bragg transitions could enable the engineering of tight-binding models with non-Abelian hopping elements and gauge fields. A significant advance in MSL experiments came with the development of the Raman MSL and the exploration of chiral dynamics driven by non-Abelian gauge fields by Liang et al. [354], depicted in Fig. 20(d). In principle, the presence of many internal states of atoms opens the future possibility to explore highly complex non-Abelian gauge fields well beyond SU(2).

In the studies described above, even with the added complexity of non-Hermitian loss terms and non-Abelian gauge fields, the phenomena that could be explored were somewhat limited by the fact that the full control over MSLs was limited to one-dimensional situations (or quasi one-dimensional geometries) where one could independently control all tunneling terms. Recently, Agrawal et al. proposed a generic path to implementing highly controllable MSLs in higher dimensions by giving up complete control over all tunneling terms and instead settling for a large amount of control that repeats with some pattern in momentum space [360]. A recent study by Dong. et al. [361] has demonstrated that rich phenomena can even be explored in fully separable two-dimensional MSLs, and have paved the way for future studies of MSL dynamics in two and three dimensions.

Finally, we discuss the role of atomic interactions, which ultimately lie at the heart of utilizing atomic systems for quantum simulations, as opposed to simply achieving experimental demonstrations of phenomena that have been predicted and explored in theory. The fact that atomic interactions should be relevant to momentum space dynamics should have been clear based on early pioneering experiments with ultracold atomic gases [362, 422]. An et al. first pointed out in detail how the mode-dependence of interactions in momentum space, stemming from the exchange interactions of bosonic atoms colliding in distinct momentum states, would give rise to an effectively local nonlinearity that was highly analogous to the Kerr nonlinearity in many photonics systems [363]. The influence of these momentum-space nonlinearities were explored in a number of experiments, with the hysteresis-like response of Landau-Zener transitions by An et al. [363], the interaction-induced localization in momentum-space quantum walks by Xie et al. [423], the observation of non-exponential tunneling related to swallowtails by Guan et al. [424], the exploration of macroscopic nonlinear self-trapping and soliton-like features in MSL arrays by An et al. [364], and the influence of interactions on the mobility edges in quasiperiodic models by An et al. [416]. Significant breakthroughs were made with the realization of MSLs in an atomic species with controllable interactions, with the Feshbach control of interactions in caesium allowing for the observation of interaction-induced mobility edges by Wang et al. [425] as well as the observation of flux-dependent self-trapping in zig-zag ladders by Li et al. [426].

The future exploration of MSLs with strong interactions and quantum correlations could be achieved in several ways, either through the use of optical cavities to effect large effective nonlinearities as proposed and explored in Refs. [367, 427], through additional spatial confinement along transverse directions, or by simply looking for collective effects beyond mean-field in existing experiments, as explored recently by Williams et al. [365].

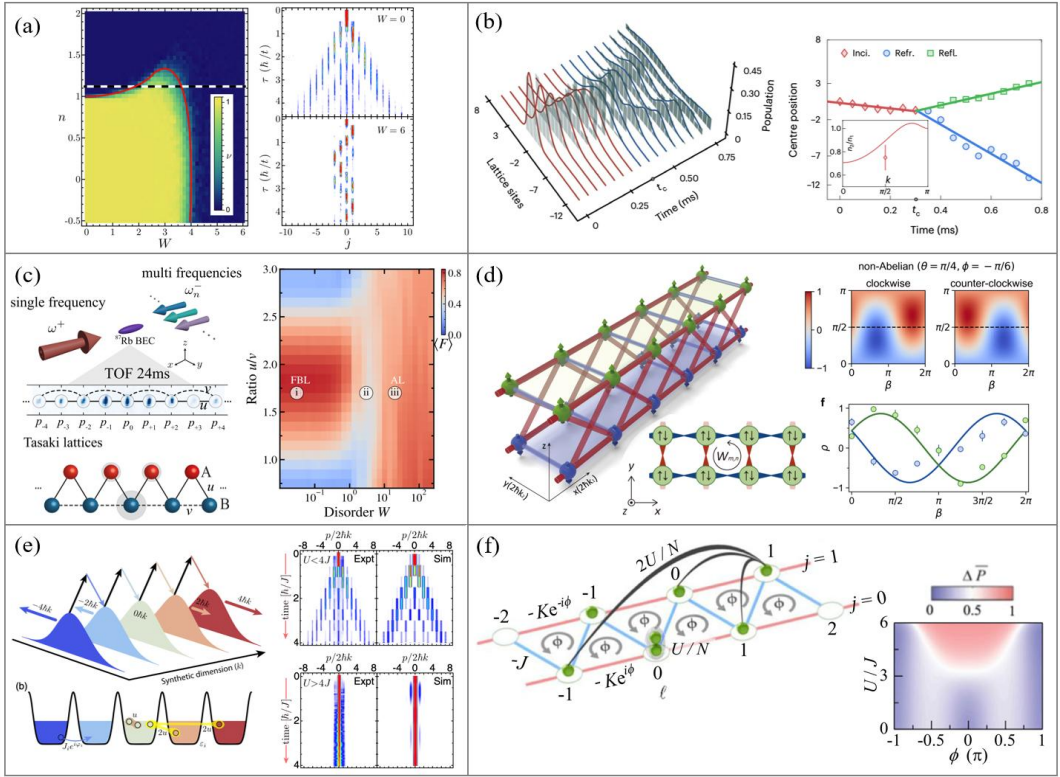

Fig. 20 Synthetic lattices based on laser-coupled atomic momentum states. (a) The realization of disorder-induced topology in momentum lattices. Adapted from Ref. [31]. (b) The observation of time reflection and refraction based on engineering a temporal boundary in momentum lattices. Adapted from Ref. [417]. (c) The transition between flat-band localization and Anderson localization in a synthetic momentum-state Tasaki lattice. Adapted from Ref. [419]. (d) The observation of chiral dynamics in a synthetic momentum lattice with non-Abelian gauge fields. Adapted from Ref. [354]. (e) The observation of nonlinear self-trapping when the mean-field interaction shift $U$ exceeds the tunneling bandwidth $4J$. Adapted from Ref. [364]. (f) The observation of flux-dependent nonlinear self-trapping in a momentum-state flux ladder. Adapted from Ref. [426].

## C. Brief Summary

We have presented three of the main ideas associated with the engineering of synthetic dimensions in cold atomic systems based on low-energy internal states (e.g., hyperfine states), Rydberg levels, and atomic momentum states. We have furthermore reviewed the state of the art of these approaches and presented some considerations related to the exploration of many-body quantum phenomena. However, given the large number of degrees of freedom in atomic and atom-like systems and the many interpretations of the term "synthetic dimensions", our discussion has been highly selective and ignores many important related developments. For example, in discussing synthetic dimensions based on atomic motional states, we focused solely on the case of laser-coupled linear momentum states of otherwise freely propagating matter waves. One could reasonably pursue the same basic approach in annular atomic fluids to couple discrete angular momentum states [428]. Such an approach would avoid the spatial separation of the coupled modes (a limiting factor in the MSL experiments), but would be restricted to much lower operating energies due to the concomitant large length scales. Alternatively, discrete motional states of trapped atoms could be utilized, analogous to work in photonics [36]. This approach follows the pioneering proposal by Price et al. on the engineering of synthetic dimensions, artificial gauge fields, and the quantum Hall effect based on harmonic oscillator states [429]. Experiments along this direction have been performed in different regimes, both with bulk bosonic gases exploring Bloch oscillations in a synthetic dimension [430] and with the engineering of chiral ladders of lattice-trapped fermions [431]. Extensions to this approach have been proposed for the exploration of quantized Hall conductance in two-terminal experiments [432]. The use of discrete motional trap states has also come of recent interest in trapped ion experiments, allowing for the engineering of synthetic dimensions and an artificial magnetic field [433]. While trapped ions have been relatively under-explored in the synthetic dimensions context, they could be of interest either for the natural coupling between discrete motional states in many-ion arrays [434] or by considering the tunability of spin-spin interactions [435] in systems with more than two internal states [436], possibly with the incorporation of transverse and longitudinal field terms analogous to the described work on Rydberg synthetic dimensions.

In thinking about many-body spin systems with internal state synthetic dimensions, we ignored the original motivating platform or polar molecule arrays [325, 326], which are analogous to the dipolar Rydberg system but also possess additional internal states (nuclear states) and are fundamentally longer-lived. While the challenges or working with molecules and the issue of state-dependent polarizabilities had initially challenged work on molecule synthetic dimensions, there has been much recent progress [437, 438], and this platform appears set to take off. Finally, we highlight some additional platforms and techniques that could be brought into the fold of synthetic dimensions. For instance, the use of multiple internal levels in artificial atoms [366, 367, 439] could be used to enrich quantum simulations in solid-state experiments. Additionally, powerful tools such as cavity quantum electrodynamics could be brought to bear on synthetic dimensions experiments, to transform the role of (or to increase the range and strength of) interactions in the many discussed platforms, as in Refs. [366, 367]. The area of studies in synthetic dimensions with atomic systems are still vividly growing and we expect to see more interesting results in the near future.

### V. Other Systems for Constructing Synthetic Dimensions Using Discrete Physical States

The concept of synthetic dimensions is developing rapidly in many subjects and areas, some of which may not be initially considered as the synthetic dimension but later are found to fit within this framework. This section will discuss some ideas, that have not been included in any of previous sections, to construct synthetic dimensions using discrete physical states either in the quantum regime (such as Fock-state lattice in **Sec. VA** and superradiance lattice in **Sec. VB**) or with classical fields (using electronic circuit in **Sec. VC** and others in **Sec. VD**)

### A. Fock-State Lattice

Fock states refer to photon number states [440-443], and the transition rate of putting one more photon into a Fock state increases with its existing occupation number as $c^{\dagger}|n\rangle = \sqrt{n+1}|n+1\rangle$. The Hilbert space of Fock states can serve as a new synthetic dimension to form Fock-state lattices [444], revealing the topology of the quantized light. Meanwhile, the dimension of the Fock-state lattice is determined by the number of the cavity modes, which shows unique advantage to study high-dimensional physics based on this framework. Cai et al. utilized Fock-state lattices to reveal various topological effects including the edge state in 1D SSH model, the Valley Hall effect in 2D honeycomb structure, and the chiral edge states on the incircle in 2D Haldane model [see Fig. 21(a)] [445]. Wang et al. theoretically explored the mesoscopic superposition states in the Fock-state lattice configured in a system composed of three cavities and a two-level atom, where the helical currents were used as an efficient scheme to generate NOON quantum states [446]. Yuan et al. based on a strained honeycomb lattice to elucidate the unification of valley and anomalous Hall effects, where the spin precession of valley Hall current is around an in-plane magnetic field while the spin precession of chiral edge current is around a magnetic field that is perpendicular to the synthetic lattice plane [see Fig. 21(b)] [447]. Deng et al. experimentally realized the construction of the Fock-state lattice in a superconducting circuit, where they displayed a set of topological physics including topological zero-energy states of the SSH model, pseudo-Landau levels, Haldane chiral edge currents, and the valley Hall effect of 2D Fock-state lattices [see Fig. 21(c)] [45]. Other than the honeycomb structure, Wang et al. theoretically constructed a two-leg ladder based on the Fock state (of phonon occupation) of a single trapped ion, where they demonstrated the topological chiral motion of wave packets on the synthetic lattice [433]. The robustness of coherent states using a mapping to an SSH edge state has also been shown, thus showing a connection between quantum optics and topological physics using the Fock-state lattice [448]. In terms of the many-body physics, Yao et al. experimentally observed the many-body localization by introducing disorder into the synthetic Fock space [449]. Classical analogue of the Fock-state lattice can be implemented by using waveguides to mimic Fock states [450], where Keil et al. theoretically and experimentally revealed the intriguing quantum correlations of the Glauber-Fock photonic lattice [451]. Yang et al. precisely controlled the coupling strength between photonic lattice sites to mimic the Fock-state lattice, where the all-band-flat phenomena were observed in the experiment [452]. Yuan et al. proposed a nonadiabatic topological transfer protocol using the gap-protected edge state of 1D Fock-state lattices [453], which has been experimentally demonstrated in photonic [454] and nanomechanic lattices [455]. Cai et al. theoretically extended the pseudo-Landau levels to a three-dimensional diamond model based on 3D Fock-state lattices [456].

We also note that the rich dynamics of excitations in Rydberg Ising simulators, namely the long-lived oscillations associated with quantum scarring and PXP model dynamics [457-459], can be considered through the lens of a quantum walk in the higher-dimensional state-space (or Fock space), with interactions giving rise to structure of the free energy landscape so as to constrain the dynamics.

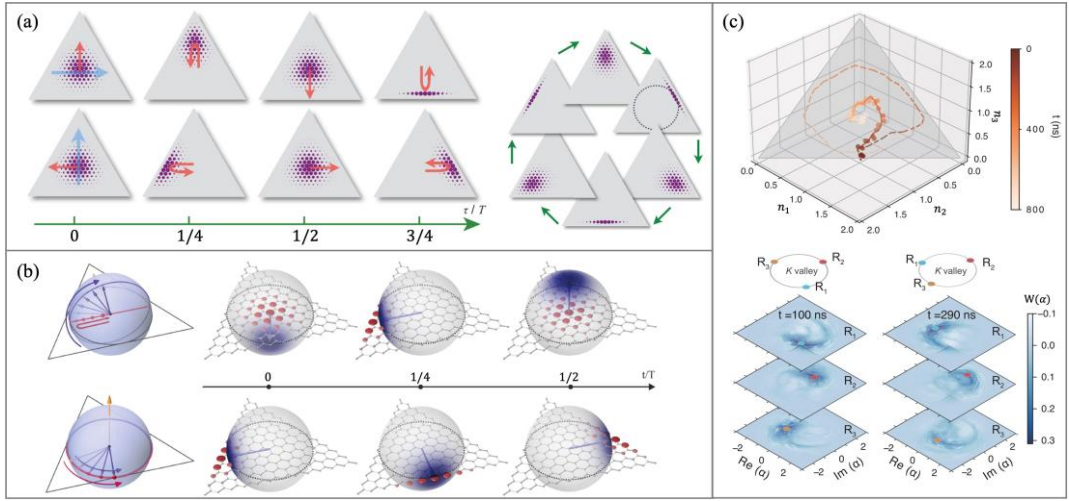

Fig. 21 (a) The valley Hall effect and the chiral edge states of the Fock-state lattice. Adapted from Ref. [445]. (b) The valley and anomalous Hall effects in a strained honeycomb lattice. Adapted from Ref. [447]. (c) The Haldane chiral edge currents and valley Hall effect in the Fock-state lattice. Adapted from Ref. [45].

## B. Superradiance Lattice

Coupling the momenta of timed Dicke states [460] can form superradiance lattices in the momentum space [461]. There has been many developments on the superradiance lattice platform, which is in many ways a versatile, room-temperature complement to the ultracold synthetic momentum state lattice platform. A number of lattice structures have been configured in the superradiance lattice to explore the novel phenomena associated with the Aharonov-Bohm phase. Wang et al. implemented a sawtooth superradiance lattice in the BEC, and the chiral edge currents were experimentally observed by designing the relative spatial phase between the two standing-wave field to induce artificial magnetic field [462]. He et al. constructed a Creutz superradiance lattice in room-temperature cesium atoms, where the interplay between flat-band localization and the Aharonov-Bohm phases were experimentally investigated through independently tuning magnitudes and phases of the hopping coefficients [see Fig. 22(a)] [463]. Besides, they also experimentally synthesized a zigzag lattice to manifest the chiral edge currents in the superradiance lattice, which was observed for the first time in an ensemble of atoms at room temperature [464]. Wang et al. unveiled the topological phase transition by inserting modulation phases of the coupling fields in a honeycomb superradiance lattice of timed Dicked states [see Fig. 22(b)] [465]. Mao et al. realized the extraction of the topological invariant of the 1D room-temperature superradiance lattices based on the spectroscopic method, which shows possible implication in the optical devices with topological physics [466]. Wang et al. implemented a velocity scanning tomography technique to distinguish the response of atoms with difference velocities, so that it enables the quantum simulation of 2D Chern insulators in ambient condition [467].

Other degrees of freedom of light can be incorporated to expand the dimension and design even richer physical effects. Xu et al. applied the Floquet engineering on the superradiance lattice and extended the 1D superradiance lattice to 2D synthetic lattice with an extra frequency dimension [see Fig. 22(c)]. The dynamic localization, delocalization, as well as the Floquet chiral currents were experimentally observed by implementing the Doppler shifts and the Floquet modulation to engineer effective force with arbitrary direction, and utilizing the phase delay between two driving fields to induce artificial magnetic fluxes [468]. The characteristic of superradiant emission can also be explored in synthetic superradiance lattices of ultracold atoms. Chen et al. experimentally constructed a 1D superradiance lattice in BEC of $^{87}$Rb ultracold atoms, where the measured directional emission spectra show the dependence on the band structure [469]. Mi et al. experimentally revealed the superradiant emission with subnanosecond resolution, as well as the long-lasting oscillation in the superradiant emission [470]. Li et al. theoretically analyzed the interplay between the synthetic magnetic flux and many-body atomic interaction, obtaining exotic sliding and chiral superfluid phases in superradiance lattices [471].

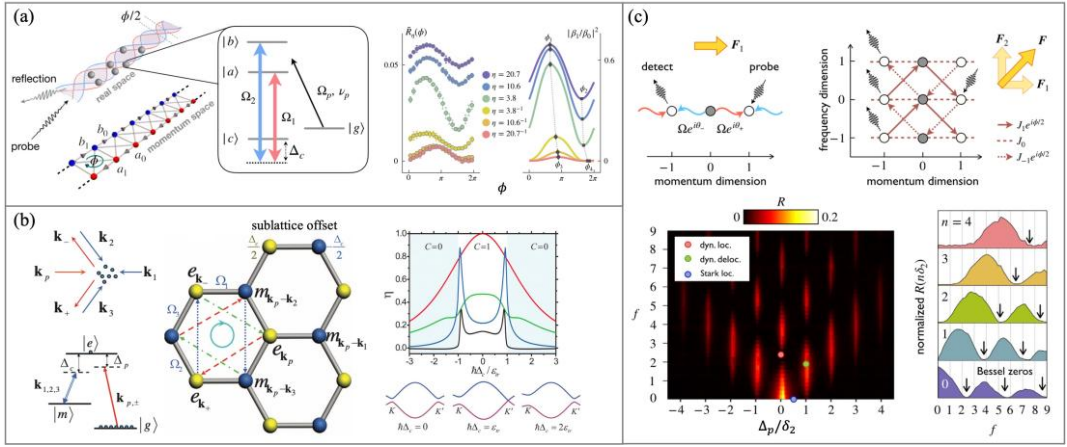

Fig. 22 (a) The averaged reflectivity and the normalized probability versus $\phi$ by selectively probing the flat or chiral band. Adapted from Ref. [463]. (b) The phase transition in the honeycomb superradiance lattice. Adapted from Ref. [465]. (c) The dynamic localization and delocalization in Floquet superradiance lattice. Adapted from Ref. [468].

## C. Electronic Circuit

With the recent development on the manufacturing technology of the integrated circuit, the platform of electric circuits shows the potential for fulfilling the exploration of inaccessible topological phase due to its flexibility and controllability [472, 473]. The dimension of the lattice in the electronic circuit is not strictly defined by the geometric space but rather by the coupling provided by the electric components such as capacitors [474]. While quite different from synthetic dimensions in its approach, analog experiments with electronic circuit lattices (and other platforms such as mechanical lattices [475-481]) are motivated by similar goals of engineering and exploring exotic Hamiltonians. We introduce several achievements that are made in this area.

Wang et al. used the electric circuits to fabricate the 4D topological insulator with the discrete elements and the interconnections built by capacitors and inductors, where the 3D surface state of the 4D topological insulator was experimentally demonstrated with the use of the impedance measurements [see Fig. 23(a)] [482]. The 3D topological feature associated with the Weyl physics was also experimentally investigated by Lu et al., where Berry curvature, Berry flux, as well as surface states were shown [see Fig. 23(b)] [483]. Moreover, Luo et al. reported the topological nodal line state and Weyl state in a 3D inductor-capacitor (LC) circuit lattice [484]. Ningyuan et al. experimentally demonstrated the topological band structure, site-resolved edge-mode distribution, and the Möbius global topology in the 2D Hofstadter model [485]. Hofmann et al. theoretically studied the possibility of utilizing negative impedance converters with the current inversion to access the so-called topolectrical Chern circuit, where the chiral edge mode is manifested in simulations [486]. The possibility of simulating non-Abelian Aharonov-Bohm effect in the linear circuit was further discussed by Albert et al. [487]. The nontrivial edge state of 2D SSH model was also theoretically and experimentally demonstrated by Liu et al. in the electric circuit [488].

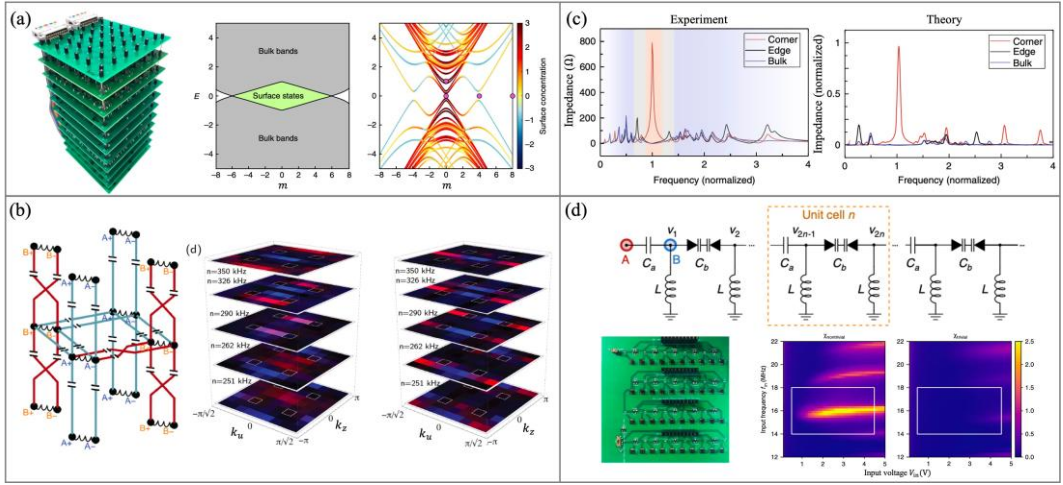

Fig. 23 (a) The 3D surface states for the circuit implementation of a 4D topological insulator. Adapted from Ref. [482]. (b) The surface states of the 3D Weyl circuit. Adapted from Ref. [483]. (c) The corner states of the quadrupole insulators. Adapted from Ref. [489]. (d) The enhanced third-order harmonic generation in the nonlinear SSH lattice. Adapted from Ref. [490].

Higher-order topological phenomena in the electric circuit can also be studied in the platform of electronic circuits. For example, topological corner states were observed in the experiment conducted by Imhof et al. with the quadrupole insulators [see Fig. 23(c)] [489]. The high-order topological transitions of the quadrupole insulators were experimentally observed by Serra-Garcia et al. in an LC circuit [491]. In the electric circuit framework, Ezawa et al. found huge resonance peak in the breathing kagome and pyrochlore lattices due to the corner states of the topological phase [492]. The topological effect can have prominent influence on the harmonic generation process, where Wang et al. used a nonlinear SSH circuit to display the enhanced third-order harmonic generation [see Fig. 23(d)] [490]. On the other hand, nonlinear effects can also assist the topological transition. Hadad et al. constructed a 1D SSH lattice in the nonlinear circuit arrays, and self-induced topological protection was emerged in the experiment [474]. The few-body interacting system can be explored in the electric circuit, where Olekhno et al. uncovered the topological edge states of inter interacting photon pairs in the experiments [493]. The exploration of the electronic circuits has further been extended into the non-Hermicity realm. In particular, the Floquet parity-time phase diagrams were uncovered in the numerical analysis performed by Huerta-Morales et al. through implementing gain and loss in a time-modulated chain of inductively coupled RLC C (where R stands for resistance, L for inductance, and C for capacitance) electronic circuits [494]. Zheng et al. studied the topological properties of a 5D non-Hermitian model including Yang monospheres, Fermi cylinder surfaces, and the 5D skin effect [495]. Riechert et al. demonstrated that electrical circuits could be useful for exploring lattice gauge theories in their classical (large-field) limit [496]. Finally, the electronic circuit shows its possibility of being utilized to construct hyperbolic lattices to study the flat band [35] and physics of negatively curved spaces [497].

### D. Other Synthetic Dimensions

Although we have seen many methods in constructing the synthetic dimension using various discrete physical states, there are also other approaches to build the artificial dimensionality which may be not able to be easily categorized in any sections above. Here we briefly introduce some relevant forefront research. In an early pioneering proposal to construct the synthetic dimension, Jukić et al. proposed to combine evanescent coupling and waveguide coupling between resonators in a 3D photonic crystal to construct the 4D photonic lattice with three spatial dimension and one synthetic dimension [498]. In their work, the additional dimension is built by using the spatial degree of freedom in a super unit cell, i.e., several resonators in each unit cell, and applying additional couplings to introduce the connectivity for the fourth synthetic dimension. Baum et al. proposed to form the Floquet synthetic dimension by implementing Floquet driving, where they manifested the topological boundary states in the 2D synthetic space [499]. The quantum states of light and matter can also be connected via the interaction to form synthetic dimension. For example, Rahmani et al. proposed to build a synthetic dimension composed of photon and exciton modes by implementing the light-matter coupling, where they demonstrated the transition between flat band and dispersive band in a synthetic Stub lattice [500].

The concept of synthetic dimension has also been developed in optomechanical systems, where the coupling between vibrational modes and optical modes are utilized in constructing the additional dimensionality. Therein, the optomechanical modulation phase can be used to create the effective magnetic field for photons in the synthetic dimension [501]. Poshakinskiy et al. further designed the exciton-mediated light-sound interaction in optomechanical system, which forms a synthetic dimension with three legs composed of Stokes and anti-Stokes photons [502].

Multiple bosons may also be utilized to create the synthetic dimension. For example, Wu et al. theoretically explored the 4D quantum Hall effect in a 1D quasicrystal by using two bosons [503] and Cheng et al. theoretically studied the multiboson dynamics on a 1D lattice that can be mapped to a single boson on an $N$-dimensional lattice [504]. Gräfa et al. theoretically and experimentally manifested that the evolution of correlated biphoton in nonlinear arrays of evanescently coupled waveguides which can be mapped to a model related to the single photon in 2D lattice [505].

In additions, Naumis et al. showed that an incommensurable system can map to a higher dimensional space, where they demonstrated that the incommensurable 1D cavity can be described by the Hamiltonian of a 2D triangular lattice [506]. Maczewsky et al. formulated a universal approach to fabricate higher dimensions by deliberately designing the local inhomogeneous coupling in the 1D lattices, where the dynamics in a 7D hypercubic lattice were experimentally displayed in a mapped 1D laser-written photonic lattice [507]. Following the similar mapping method, Edge et al. mapped the 4D Hamiltonian onto a 1D dynamical system, where the 4D phase diagram of the quantum Hall effect was demonstrated [508], building on earlier work related to embedding extra dimensions in quasiperodically driven kicked rotors [342-346, 509]. With the development in the field of synthetic dimensions, we envision ourselves to see more approaches arising for constructing synthetic dimensions in the near future.

## VI. Parameter Synthetic Dimension

We reviewed various ideas for constructing synthetic dimensions using different degrees of freedom of photons and atoms to connect discrete modes for lattice models in previous sections. There is also another idea for synthetic dimension using the system parameters. Such idea has fundamental difference from previously discussed discrete lattice models where the connectivity is the key in building the synthetic dimension. Here, the synthetic dimension is formed by leveraging the degrees of freedom inherent in parameters that emulates a dimension with the continuous variable in a Hamiltonian [8, 11, 12, 14]. The dependency of the system on the parameter can thus be articulated within a synthetic space, wherein the parameter axis functions as an additional synthetic dimension supplementing the conventional physical space. Usually, the static parameter can be viewed as an additional quantum momentum, which allows for an extended Brillouin zone with an additional dimension. In this way, higher-dimensional phenomena manifests via the parameter dependency of a lower-dimensional system. This method is usually termed as "*parametric synthetic dimensions*" or "*parameter spaces*", and is connected to the well-studied and complementary ideas of "dimensional reduction" and "dimensional extension".

Compared to other approaches forming the synthetic dimension (as discussed in previous sections), the complexity in constructing parameter spaces does not scale up with the dimensionality of the systems. Even simple two-level or few-level systems can effectively be used to explore band structures or phenomena associated with two-, three-, and higher-dimensional systems by parametric embedding [32, 510-514]. To be mentioned, if the parameter is constant or varied adiabatically, it lacks a kinetic term, which implies that no transport happens in this parametric synthetic dimension. However, the parametric synthetic dimension is still a powerful tool for demonstrating high-dimensional topological phenomena, as it can provide the projection of the systematic evolution under a certain quantum momentum (at a certain parameter) or when the quantum momentum is adiabatically varying (i.e., the parameter is slowly tuned).

In what follows, we first present typical theoretical models to illustrate the basic idea of parameter space in **Sec. VIA**. Then in **Sec. VIB** we discuss several representative physical phenomena including topological pumps, synthetic Weyl degeneracies and high-dimensional topology, and parameter spaces in non-Hermitian systems. **Sec. VIC** lists some useful applications of parameter synthetic dimension in designing photonic devices, followed by a brief introduction of constructing parameter spaces in other systems in **Sec. VID**.

### A. Theoretical Model

#### 1. A representative model for parameter synthetic dimension

The basic idea of parameter synthetic dimension is to treat the system parameters as the non-spatial dimensions. As a simple illustration, let us consider a Hamiltonian $H(\boldsymbol{k}_\alpha; \boldsymbol{k}_p)$ describing an extended "Brillouin zone" in the momentum space. Here $\boldsymbol{k}_\alpha$ is an $n$-dimensional Bloch lattice momentum ($\alpha = x, y, z$ for 3D systems) and $\boldsymbol{k}_p$ represent $m$-dimensional parameter synthetic dimensions. In general, the extended Brillouin zone has $n + m$ dimensions. Therefore, the eigenstates of $H(\boldsymbol{k}_\alpha; \boldsymbol{k}_p)$ can exhibit topological phenomena beyond $n$ spatial dimensions. The Hamiltonian satisfies the Schrödinger equation:

$$H(\widetilde{\boldsymbol{k}})|\psi(\widetilde{\boldsymbol{k}})\rangle = E(\widetilde{\boldsymbol{k}})|\psi(\widetilde{\boldsymbol{k}})\rangle, \tag{86}$$

where $\widetilde{\boldsymbol{k}}$ is the extended Brillouin zone, based on which the topological invariants can be obtained. For example, the Chern number in a synthetic 2D system (one spatial dimension $k_x$ and one synthetic space dimension $k_p$) writes:

$$C = \oint_{\widetilde{L}} \langle \psi(\widetilde{\boldsymbol{k}})|\nabla_{\widetilde{k}}|\psi(\widetilde{\boldsymbol{k}})\rangle \mathrm{d}\widetilde{\boldsymbol{k}}, \tag{87}$$

where $\widetilde{L}$ is a loop on the extended Brillouin zone of $\widetilde{\boldsymbol{k}} = (k_\alpha; k_p)$. The integral kernel here gives the gauge potential in synthetic space (i.e., Berry connection in the momentum space)

$$\boldsymbol{A}(\widetilde{\boldsymbol{k}}) = i\langle \psi(\widetilde{\boldsymbol{k}})|\nabla_{\widetilde{k}}|\psi(\widetilde{\boldsymbol{k}})\rangle. \tag{88}$$

The crucial idea here is that the topology frequently applied to band structures in momentum space can indeed be employed for any parameter dependence. Furthermore, if one adjusts the parameter as a function of time in an adiabatic manner, the dynamics of such a time-variant system display characteristics of higher-dimensional topological physics projected onto the low-dimensional real space. A typical model is known as the AAH model - an elegant example of the parameter space, with the following Hamiltonian [515]:

$$H(n) = \sum_n \lambda c_n^\dagger c_n \cos(2\pi b n + \phi) + \left[t c_n^\dagger c_{n+1} + \mathrm{h.c.}\right], \tag{89}$$

where $c_n(c_n^\dagger)$ is the annihilation (creation) operator of site $n$ in the $x$ direction, and $\lambda$, $t$, $\phi$, $b$ are parameters determining the on-site potential and coupling strength. Here, the parameter $\phi$ can be viewed as a synthetic momentum (i.e., $\phi \equiv k_m$), so that the Hamiltonian $H(k_n; k_m)$ represents a 2D square lattice $(n, m)$ with synthetic magnetic flux $b$:

$$H(n, m) = \sum_{n,m} \left(\lambda e^{i2\pi b n} c_{n,m}^\dagger c_{n,m+1} + t c_{n,m}^\dagger c_{n+1,m} + \mathrm{h.c.}\right). \tag{90}$$

Therefore, one can see that although the system is a 1D chain, this fictitious 2D lattice has the extended Brillouin zone in the coordinates $(k_n; k_m)$ and has a non-zero Chern due to the magnetic flux $b$. In this way, the topological properties of 2D systems can be probed in a 1D spatial structure.

## 2. Parameter space for synthetic Weyl points

Another typical exploration of high-dimensional physics using parameter space involves the synthetic Weyl points. Weyl points, characterized by linear intersections of two bands at distinct locations in 3D momentum or synthetic space, can be likened to monopoles possessing quantized monopole charge [516]. Within this context, they function as either sources or sinks for the Berry curvature. The polarity of the monopole charge is dictated by the chirality inherent to each Weyl point. Using the two degenerate modes at the Weyl points as the basis, the general effective Hamiltonian near each Weyl point can be expressed as [517]

$$H_{\mathrm{eff}} = N_x k_x \sigma_x + N_y k_y \sigma_y + N_z k_z \sigma_z + T_x k_x I, \tag{91}$$

where $\sigma_{x,y,z}$ are Pauli matrices, $I$ is a 2×2 identity matrix, $N_{x,y,z}$ are the Fermi velocities, and $T_x$ is the tilted velocity of the Weyl point along the $x$ direction. The tilting of the Weyl dispersion cone is determined by the Weyl parameter $\alpha_{\mathrm{Weyl}} = T_x/N_x$, with $\alpha_{\mathrm{Weyl}} < 1$ corresponding to a type I Weyl system, and $\alpha_{\mathrm{Weyl}} > 1$ corresponding to a type II Weyl system.

As a representative example, here we illustrate how to construct a 3D synthetic space holding Weyl points using system structural parameters in 1D optical waveguide arrays [518]. Consider a 1D binary waveguide array, in the tight-binding approximation, the system

Hamiltonian is written as

$$H = \beta_1(l) \sum_j a_{1,j}^\dagger a_{1,j} + \beta_2(l) \sum_j a_{2,j}^\dagger a_{2,j}$$

$$+ \kappa_1(m) \sum_j \left(a_{1,j}^\dagger a_{2,j} + a_{2,j}^\dagger a_{1,j}\right) + \kappa_2(m) \sum_j \left(a_{1,j+1}^\dagger a_{2,j} + a_{2,j}^\dagger a_{1,j+1}\right), \quad (92)$$

where $\beta_{1(2)}(l)$ and $\kappa_{1(2)}(m)$ represent propagation constant and coupling coefficients of the waveguides, $l$ and $m$ are two independent numbers within $(-1, 1)$. They are introduced through $w_1 = w_{1c}(1 + f_1 l)$, $w_2 = w_{2c}(1 - f_2 l)$, $d_1 = d_c(1 + m)$, $d_2 = d_c(1 - m)$, where $w_1$ and $w_2$ are the widths of two waveguides in a unit cell, $d_1$ and $d_2$ are the gaps of waveguides, and $f_i = (w_{1c} + w_{2c})/2w_{ic}$. The lattice constant is $\Lambda = 2d_c + w_{1c} + w_{2c}$. $l$ and $m$ construct a 3D synthetic space when incorporating the Bloch wave vector $k$ along the transverse direction $x$. We define three dimensionless coefficients $\delta l = l - l_c$, $\delta m = m - m_c$, and $\delta k = (k - k_c)/k_0$, where $k_0 = \pi/\Lambda$ and $(l_c, m_c, k_c) = (0,0,\pi/\Lambda)$ is the position of any possible degenerate point. Upon Fourier transformation of Eq. (92), we have

$$H = \begin{pmatrix} \beta_1(\delta l) & \kappa_1(\delta m) + \kappa_2(\delta m)e^{-ik\Lambda} \\ \kappa_1(\delta m) + \kappa_2(\delta m)e^{ik\Lambda} & \beta_2(\delta l) \end{pmatrix}. \quad (93)$$

Expanding $\beta_1(\delta l)$ and $\kappa_{1(2)}(\delta m)$ with respect to $\delta l$ and $\delta m$:

$$\begin{cases} \beta_1(\delta l) = \beta_{1c} + b_1 \delta l + O(\delta l^2), \\ \beta_2(\delta l) = \beta_{2c} - b_2 \delta l + O(\delta l^2), \end{cases} \quad (94)$$

and

$$\begin{cases} \kappa_1(\delta m) = \kappa_c + c\delta m + O(\delta m^2), \\ \kappa_2(\delta m) = \kappa_c - c\delta m + O(\delta m^2), \end{cases} \quad (95)$$

where

$$b_i = \frac{\partial \beta_i}{\partial \delta l}\bigg|_{\delta l = 0} = w_{1c} f_i \left(\frac{\partial \beta_i}{\partial w_i}\bigg|_{w_i = w_{ic}}\right) = (w_{1c} + w_{2c})/2 \left(\frac{\partial \beta_i}{\partial w_i}\bigg|_{w_i = w_{ic}}\right), \quad i = 1,2, \quad (96)$$

and

$$c = \frac{\partial \kappa}{\partial \delta m}\bigg|_{\delta m = 0} = d_c \left(\frac{\partial \kappa}{\partial d}\bigg|_{d = d_c}\right). \quad (97)$$

Substituting Eqs. (94) and (95) into Eq. (93) and expanding $H$ with respect to $(\delta l, \delta m, \delta k)$ up to the first order, we finally have

$$H = 2c\delta m \sigma_x + K_0 \delta k \sigma_y + b_+ \delta l(\sigma_z + \alpha_{\text{Weyl}}\sigma_0) + \beta_- \sigma_z + \beta_+ \sigma_0, \quad (98)$$

where $\sigma = (\sigma_x, \sigma_y, \sigma_z)$ are the Pauli matrix, $\sigma_0$ is a 2×2 identity matrix. $K_0 = -\kappa_c k_0 \Lambda$, $b_\pm = (b_1 \pm b_2)/2$, and $\beta_\pm = (\beta_1 \pm \beta_2)/2$. Here the parameter

$$\alpha_{\text{Weyl}} = \frac{b_-}{b_+} = \frac{(b_1 - b_2)}{(b_1 + b_2)}, \quad (99)$$

determines the types of the Weyl point, i.e., $\alpha_{\text{Weyl}} < 1 (> 1)$ corresponding to a type-I (-II) Weyl system. Based on this model, the synthetic Weyl point, including type-I and –II have been successfully observed in 1D photonic waveguides, in both Hermitian and non-Hermitian systems (will become Weyl exceptional ring). More complex topological degeneracies can also be realized in various systems following the similar principle, such as 5D Yang monopoles, high-dimensional exceptional contours in non-Hermitian systems, etc., as will be shown in **Sec. VIB** later.

### B. Physical Phenomena

#### 1. Adiabatic evolution and topological pumps

As has been demonstrated, the 1D AAH lattice model represents a 2D synthetic lattice by treating the parameters as the system momenta (see theoretical model in **Sec. VIA**). The synthetic dimension can be explored by considering the dynamics of a time-dependent parametric system, i.e., the configuration of the system is modified cyclically in time by varying the parameter $\phi$. In this way, scanning $\phi$ is equivalent to scanning the 2D extended Brillouin zone with a 1D line. In a finite system, the scanning process results in variations of the eigenfrequencies. These changes are directly related to the dispersion of the one-way edge state within the quantum Hall system. If we assume that at time $t = 0$, the system is in an edge state at one end, then by gradually changing $\phi$ as a function of time, the mode will evolve based on the edge-state dispersion. This evolution will lead to the mode becoming a bulk state, and eventually reappearing as an edge state localized at the opposite end. This process, termed as topological pump, reviews the topological insights of adiabatic mode evolutions, in which the adiabatic pumping process in a time-varying 1D system offers a straightforward method for investigating the characteristics of the associated 2D system, achieved by transforming a 1D lattice into a segment of a 2D model.

Photonic waveguides provide an excellent platform for exploring topological pumping effects, where the propagation direction (e.g., $z$) of light serve as the time dimension in a Schrödinger equation. By adiabatically varying $\phi$ along $z$, Kraus et al. experimentally observed the topological adiabatic pumping in waveguide lattices using the AAH model [see Fig. 24(a)-(b)] [519]. When light is injected into the edge state at one side of the system, it evolves into bulk modes propagating along the $z$-axis. Eventually, it reappears as an edge state on the opposite end, which aligns with the predications of the AAH model. Note that such an AAH model, in the real spatial dimension, can form a quasi-crystal, i.e., nonperiodic structures with long-range order. This implies that the arrangement of a quasi-crystal can be understood as stemming from periodic structures in a dimension that surpasses the physical one. In this vein, topological phase transitions in photonic quasicrystals and topological pump in a Fibonacci quasicrystal have also been demonstrated [520, 521].

Extending this idea to a 2D physical lattice, 4D quantum Hall physics can also be studied– making possible the probing of topological physics beyond the realistic 3D physics spaces. In Ref. [30], the topological pump is realized in a 2D "off-diagonal" AAH model, $H = \sum_{x,y} t_x(\phi_x)c^{\dagger}_{x,y}c_{x+1,y} + t_y(\phi_y)c^{\dagger}_{x,y}c_{x,y+1} + \text{h. c.}$, where $t_i(\phi_i) = t_i + \lambda_i \cos(2\pi b_i i + \phi_i)$ with $i = x, y$ are modulated hopping amplitudes in the $i$ direction. The pump parameters $\phi_x$ and $\phi_y$ correspond to synthetic momenta and $b_i$ is the synthetic magnetic fields. Therefore, the Hamiltonian maps to four dimensions—two real ($k_x$; $k_y$) and two synthetic ($\phi_x$; $\phi_y$). The bandgaps of the 2D pump with non-trivial second Chern numbers can be characterized as manifesting a quantized bulk response with 4D symmetry [see Fig. 24(c)]. This achievement further suggests the potential for studying 6D topological phases, which has not been experimentally verified yet [522, 523].

The realm of topological pumps continues to expand beyond serving as a tool for probing high-dimensional physics; it provides insights into adiabatic mode evolutions that can be further enriched by incorporating elements such as nonlinearity, nonparaxiality, and non-Abelian operations. In nonlinear optical systems, integer and fractional topological quantized transport of solitons have been demonstrated by considering Kerr-type nonlinearity, which quantizes transport via soliton formation [41, 524] [see Fig. 24(d)]. Beyond the paraxial condition commonly adopted in most quantum –classical analogy, Refs. [525, 526] indicate that breaking the paraxial assumption leads to an asymmetric topological pumping effect, arising from the destruction of chiral symmetry due to the emergence of long-range interactions [see Fig. 24(e)]. Unlike traditional Thouless pumps, which describe the adiabatic evolution of a physical system following a non-degenerate band, the non-Abelian pumps occur when degenerate bands are present. Such non-Abelian Thouless pump was theoretically proposed in a photonic Lieb lattice [527], and successfully realized in photonic [528, 529] [Fig. 24(f)] and acoustic [530] waveguides. The topological pump transforms the identical initial state into distinct final pumped states by switching the sequence of pumping operations – a signature of non-Abelian effects.

The pump process discussed above is restricted to the adiabatic condition, which inevitably increases the system's evolution duration and reduces the compactness of the system. Recently, Song et al. addressed the limitation of slow adiabatic evolution in topological pumps by leveraging the concept of quantum metric in quantum geometry (i.e., the real part of quantum geometry tensor that has been less explored compared to its imaginary counterpart, the Berry curvature) [531]. They discovered that the quantum metric tensor can serve as an important criterion to measure the adiabaticity of a topological pump process. Through experiments in bilayer silicon waveguides, they showed that judicious modification of the quantum metric can improve the evolution speed and even reach the adiabatic infimum of topological pumps [532].

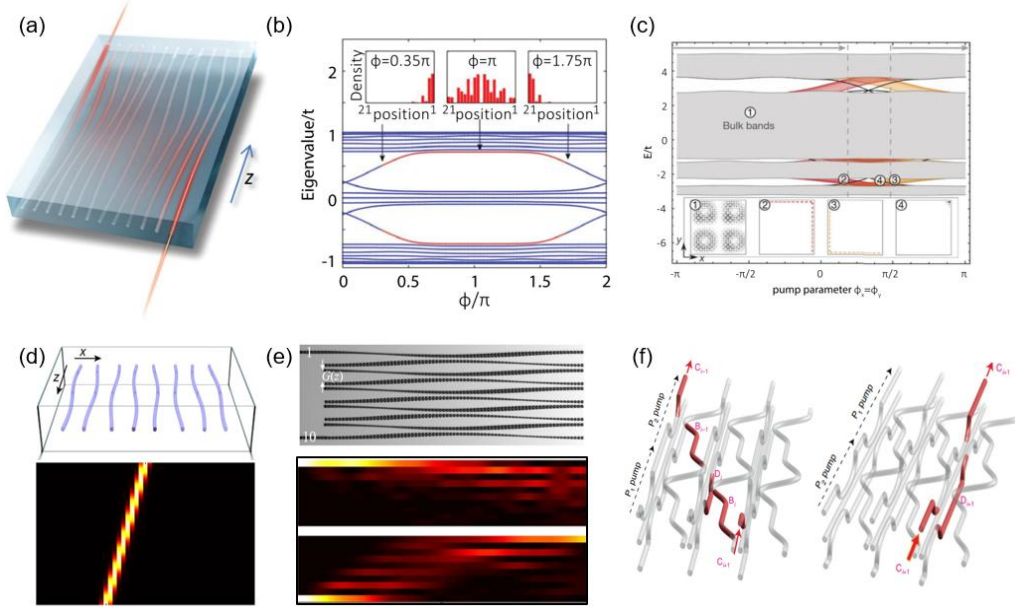

Fig. 24 (a) Schematics of topological adiabatic pumping in waveguide lattices. (b) Mode evolutions as the function of the pumping parameter $\phi$. The insert pictures illustrate the mode profile during the evolution process. (a)-(b) are adapted from Ref. [519]. (c) Band diagram of a topological pumping in 2D lattice manifesting the 4D integer quantum Hall effect. Adapted from Ref. [30]. (d) Nonlinear Thouless pumps with quantized soliton transport. Adapted from Ref. [41]. (e) Asymmatric topological pumps with nonparaxial conditions. Adapted from Ref.[525]. (f) Non-Abelian Thouless pumps. Adapted from Ref. [528].

## 2. Synthetic Weyl points and higher-dimensional nontrivial topology

In the preceding part, we discussed how the higher-dimensional physics of a system characterized by a parametric Hamiltonian is unveiled when the parameters are permitted to fluctuate as a function of time within a singular physical framework. In fact, the exploration of higher-dimensional physics can also be undertaken by examining the properties of a set of physical structures with varying parameters. In this way, the synthetic dimensions are explored statically, i.e., one point at a time, usually achieved by manual reconfiguration of the system. In this vein, viewing a static parameter as the synthetic momentum has been applied to experimentally observe Fermi arc and Weyl point in three dimensions, second Chern crystal in four dimensions, and even Yong monopole in five dimensions. Here we review the progress of parameter synthetic dimensions in a variety of optical systems ranging from photonic crystal, waveguide arrays, to metamaterials, specifically focusing on how to construct the parameter synthetic dimension in different systems.

Aspects of 3D Weyl-point physics can be explored using a simple 1D photonic crystal structure. In 2017, Wang et al. [533] considered a 1D photonic crystal, the unit cell of which consists of four layers with thicknesses of $(1 + p)d_a$, $(1 + q)d_b$, $(1 - p)d_a$ and $(1 - q)d_b$, where $p$ and $q$ are specially introduced structural parameters that forms a two-parameter space [see Fig. 25(a)]. When incorporated with $k$ (1D momentum space), it forms a 3D space containing the Weyl points. The manifestation of the physical signature for such a Weyl point is observed in the reflection phase when a wave, at the frequency of the Weyl point, is incident from air onto the photonic crystal along the direction of normal incidence. More specifically, by adjusting the parameters $p$ and $q$, the reflection phase wraps around the point where both $p$ and $q$ equal zero, i.e., around the synthetic Weyl point. In this spirit, a charge-2 Dirac point in a 1D optical superlattice system is also observed, with two parameters controlling the on-site potential and coupling coefficient forming two synthetic dimensions [534] [see Fig. 25(b)]. Recently, controllable photonic Weyl nodal line semimetals were also demonstrated in simple 1D photonic structures, where multiple phase transitions, e.g., from type-I to type-II Weyl points, were realized by flexibly modulating the structural parameters [535].

Compared to the Weyl point demonstrated in a real 3D spatial dimension, synthetic Weyl points realized in parameter synthetic dimensions are much easier to access and open new possibilities for studying unconventional physics associated with the Weyl point, which would be quite difficult for a real Weyl lattice. For example, the Weyl interface, the domain wall formed by two independent Weyl media, suggests novel physical effects across the interface. It is predicted that Fermi arcs would hybridize and alter their connectivity at the interface between two Weyl semimetals. However, the construction of Weyl interface remains quite challenging due to the difficulty in lattice matching and specific crystal orientation in higher dimensions. Recently, it has been theoretically proposed to construct the interface of synthetic Weyl semimetals and observe the Fermi arc reconstruction in a 1D dielectric trilayer grating, where the relative displacements between adjacent layers play the role of two synthetic momenta [536] [see Fig. 25(c)]. Moreover, experimental demonstrations have been conducted on arbitrary interfaces between two Weyl structures with orientations that can be freely rotated within the synthetic parameter space utilizing 1D optical waveguide arrays [537]. Besides the momentum space $k$ provided by the waveguide array, two structural parameters (waveguide widths and gaps) can be utilized to form two parameter spaces if engineered properly. Gapless topological interface states of the two Weyl structures have been observed. Note that due to the intrinsic limitation of waveguide systems, only type-I Weyl point can be realized in this system. Furthermore, in Ref. [518], a nanostructured subwavelength grating (SWG) waveguide is employed [see Fig. 25(d)], which allows flexible control of the waveguide dispersion and gives rise to continuous controllability for building not only a type-I but also a type-II Weyl points with conical-like Fermi surface and tilted dispersions (see theoretical model in **Sec. VIA**). As such, a type-II Weyl heterostructure can be constructed, which supports topologically protected interface states. More interestingly, the type-II Weyl point with tilted dispersion can set new rules for the emergence of topological bound state – which can disappear and become an extended mode even with the topological phase transition. This bound-extended mode transition can be controlled by tuning its rotational phases in the parameter spaces. More interestingly, in addition to the widely adopted structural parameters, the magnetic field strength can also be treated as a parameter synthetic dimension, thus enabling fully tunability to the Weyl point and the corresponding photonic Fermi arcs [538].

In principle, including more dimensions, whether spatial or non-spatial synthetic, enables the demonstration of physical effects in arbitrarily high-dimensional space, even beyond the real 3D dimensions. Lu et al. [539] considered the angle of helical modulation to the 3D Weyl photonic crystal fibers to construct a 4D synthetic spaces [see Fig. 25(e)]. They annihilated Weyl points by helical modulation of the photonic crystals, thus obtaining a 3D topological bandgap that separates forward and reverse propagating photonic channels. Spiral modulation pulls out a topological line defect in space, which can be used as the core layer of a unidirectional transmission fiber. Unlike traditional optical fibers, the optical signal in one direction can bypass arbitrarily shaped impurities or defects without scattering. Different from the topological principle of the unidirectional edge state of a 2D topological photonic crystal (the invariant is the first number of the 2D space), the topological invariant of the one-way fiber is the second number of the 4D parametric space. Such a second Chern crystal in a 4D parameter space can also be realized in 2D photonic crystals by introducing two extra synthetic translation dimensions [540] [see Fig. 25(f)]. Recently, Ma et al. utilized electromagnetic metamaterials as fundamental components to achieve a 5D generalization of a topological Weyl semimetal [42, 541] [see Fig. 25(g)]. They included two bi-anisotropy material parameters as synthetic dimensions in addition to the three real momentum dimensions, and demonstrated both linked Weyl surfaces and Yang monopoles. In summary, the exploration of higher-dimensional systems, which are inaccessible through experiments in three real-space dimensions, is made possible by synthetic parameter spaces, offering an intriguing avenue for studying exotic physics in high dimensions.

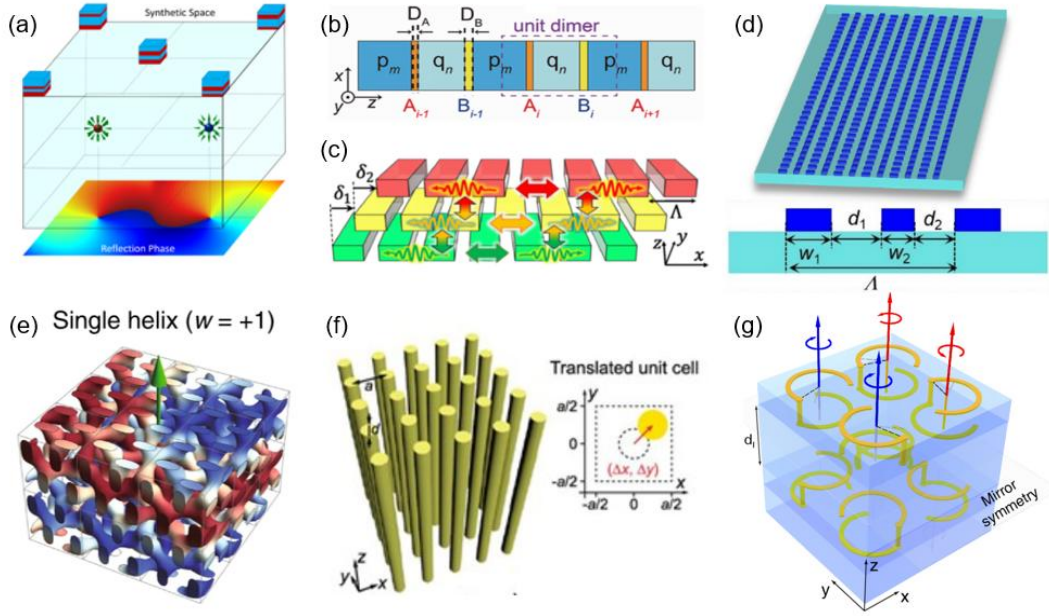

Fig. 25 (a) Optical Weyl points in 1D photonic crystals with two structural parameters as additional dimensions. Adapted from Ref. [533]. (b) Charge-2 Dirac points in a 1D optical superlattice with two structural parameters forming two synthetic dimensions. Adapted from Ref. [534]. (c) Fermi arc reconstruction in a 1D dielectric trilayer grating with the relative displacements between adjacent layers as two synthetic momenta. Adapted from Ref. [536]. (d) Type-II Weyl points and interface states in 1D optical SWG waveguide lattice, where two structural parameters controlling the coupling and on-site energy serve as parameter dimension. Adapted from Ref. [518]. (e) One-way fiber in 3D Weyl photonic crystal using the angle of helical modulation to construct a 4D synthetic spaces. Adapted from Ref. [539]. (f) 4D second Chern crystal realized in 2D photonic crystals with two extra synthetic translation dimensions. Adapted from Ref. [540]. (g) 5D Yang monopoles in 3D metamaterials, which includes two bi-anisotropy material parameters as synthetic dimensions. Adapted from Ref. [42].

## 3. Non-Hermitian parameter synthetic dimension

The exploration of synthetic topological systems reveals a richer realm of physics, especially when delving into nonconservative systems that engage in energy exchange with their surroundings. From a mathematical standpoint, these systems can be characterized as non-Hermitian. A distinctive attribute of such systems is the potential existence of specific conditions, termed exceptional points (EPs), where unconventional behavior is anticipated. These exceptional points are spectral singularities within the system's parameter space, marked by the concurrent convergence of two or more eigenvalues and their associated eigenvectors [146] [see Fig. 26(a)].

The realization of non-Hermitian photonic systems has attained great success, largely due to the fact that the complex potential can be easily controlled by gain and loss in an optical system. The spontaneous PT breaking process in the parameter space has been demonstrated in coupled waveguides [542-544] and whispering gallery micro-ring resonators [274, 545] [see Fig. 26(b)]. The intriguing features of EP can result in non-trivial light behaviors. For example, asymmetric light propagations were proposed by operating the system precisely at the EP [546] or utilizing the topological feature of EP by encircling it in the parameter space [24, 25, 547] [see Fig. 26(c)]. Besides, sensing in the vicinity of an EP is proposed [27, 28], based on the fact that the energy splitting of two coalescing levels subjected to a parameter perturbation is enhanced by the EP. The lasing system is naturally a non-Hermitian case, therefore it is straightforward to explore the lasing properties under the gain-loss arrangement [548, 549], which is utilized to generate single-mode lasing and suppress other unwanted modes. The EP encircling process has also been utilized to modify the laser profile, and a spatial evolving mode that faithfully settles into a pair of bi-orthogonal states at the two opposing facets of a laser cavity is realized [46].

The interaction between EPs and topology has become a focal point in recent research, as evidenced by extensive studies across various systems [550]. These investigations offer valuable insights for achieving novel symmetry-protected non-Hermitian phases that are entirely absent in the Hermitian domain. For instance, it has been demonstrated that a Dirac point can evolve into continuous rings of EPs when non-Hermiticity is introduced [21]. Furthermore, a new configuration involving isolated pairs of EPs has been realized, leading to a unique double–Riemann sheet topology. This configuration results in a bulk Fermi arc that connects the two EPs [33]. In addition to these developments in 2D, a set of NH degeneracies known as Weyl exceptional rings have been theoretically predicted [551], and was later confirmed experimentally in a 3D Weyl photonic lattice composed of helical waveguides etched into silica [552].

The Weyl exceptional ring demonstrated in 3D spatial space with deliberate and complicated structures lacks tenability. The realization of a synthetic Weyl exceptional ring opens doors to new possibilities by treating the non-Hermitian parameters as new synthetic dimensions [553] [see Fig. 26(d)]. Recently, non-Hermitian Weyl interface physics was experimentally demonstrated in complex synthetic parameter space within a loss-controlled silicon waveguide array, which overcomes the difficulty of realizing the non-Hermitian Weyl heterostructures in 3D spaces [554] [see Fig. 26(e)]. The researchers treated the non-Hermitian parameter as the additional dimension to construct non-Hermitian synthetic dimensions and explore the concept of *non-Hermitian order*, which refers to the spatial arrangement of non-Hermitian components, rather than the magnitude of non-Hermiticity (i.e., PT modulation), for advanced topological light manipulations. New types of Weyl interfaces hidden in the non-Hermitian dimension were revealed, explained by sign-reversed imaginary mass (reversed non-Hermitian order) and phase transition crossing the Weyl exceptional ring.

In general, the degeneracy points of Hermitian topological bands transform to non-Hermitian EP contour with the dimensionality being $d - 2$ in spatial dimension $d$ (because two constraints need to be satisfied to create an EP). For example, the diabolic points can be changed to isolated points (zero-dimensional, 0D) [33] and Weyl points can be changed to Weyl exceptional rings (1D) [552], as has been demonstrated. Therefore, the realization of exceptional surfaces (ESs) requires more degrees of freedom and dimensions, posing a significant challenge for the experimental demonstration. Recently, Zhang et al. successfully constructed a 3D exceptional surface based on parameter synthetic dimensions with magnon polaritons [555] [see Fig. 26(f)]. Due to the flexibility in manipulating the structure in parameter spaces, they showed that the ES can be conveniently adjusted across multiple dimensions to form an exceptional saddle point (ESP). This discovery of the ES will pave the way for advanced control over non-Hermitian systems in high dimensions. For instance, beyond ESs, it is possible to create exceptional volumes or even more complex EP structures by incorporating higher-order synthetic dimensions.

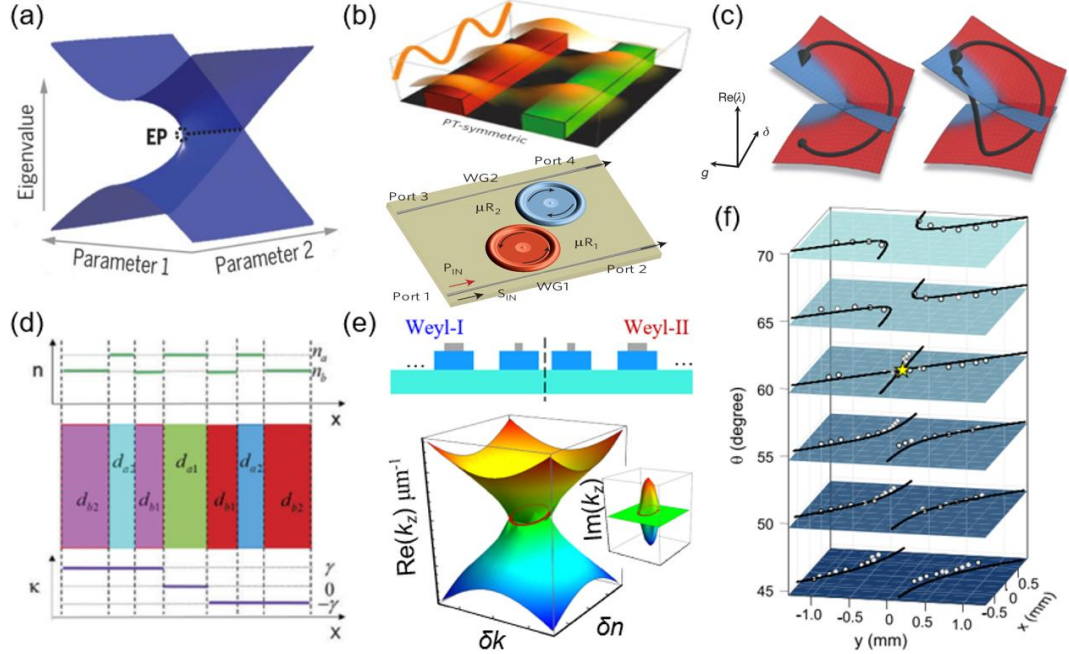

Fig. 26 (a) Exceptional points formed in parameter spaces. Adapted from Ref. [146]. (b) Realization of EP in different photonic systems. Top panel: in coupled waveguides. Adapted from Ref. [542]. Bottom panel: in micro resonators. Adapted from Ref. [274]. (c) Encircling the EP with different directions gives rise to asymmetric mode switching. Adapted from Ref. [24]. (d) Parameter synthetic dimension realized in PT symmetric photonic crystal systems. Adapted from Ref. [553]. (e) Realization of Weyl exceptional ring and non-Hermitian Weyl interface modes in synthetic parameter waveguide lattices. Adapted from Ref. [554]. (f) Exceptional surfaces in parameter synthetic dimensions with magnon polaritons. Adapted from Ref. [555].

## C. Applications

Besides offering a powerful method for exploring higher-dimensional topological physics, the parameter synthetic dimension is also empowered for useful photonic devices. In Refs. [556, 557], a topological rainbow concentrator that leverages topological photonic states is proposed by using the concept of synthetic dimensions. This synthetic dimension is achieved by utilizing the translational degree of freedom inherent in the nanostructures within a 2D photonic crystal's unit cell [see Fig. 27(a)]. The translational deformation triggers a non-trivial topology in this synthetic dimension, leading to robust interface states at varying frequencies. This topological rainbow has the ability to confine states of differing frequencies, which can be precisely controlled by adjusting the spatial modulation of the interface state group velocities. Both the operational frequency and bandwidth of the topological rainbow can be finely tuned by manipulating the band gap of the photonic crystal. Further researches have demonstrated a nanophotonic topological rainbow working in the near-infrared regime on a silicon chip [558] and a gigahertz surface acoustic wave topological rainbow in nanoscale phononic crystals [559], paving the way for realizing topological slow light, topological router, and topological temporary storage in integrated chips [see Fig. 27(b)]. To be mentioned, these prior researches have solely concentrated on the interactions between a stationary photonic crystal and a translating photonic crystal. It is physically unfeasible to achieve complete gap dynamic tuning using just one translation parameter. More recently, by incorporating two translation parameters, Guan et al. illustrated the phase diagram within the 2D translation parameter space and examined the domain wall separating two translating photonic crystals [560]. This enables the creation of dynamically adjustable bandpass filters, a capability that cannot be achieved with only one translation parameter, and will pave the way for the development of tunable topological devices based on parameter synthetic dimension.

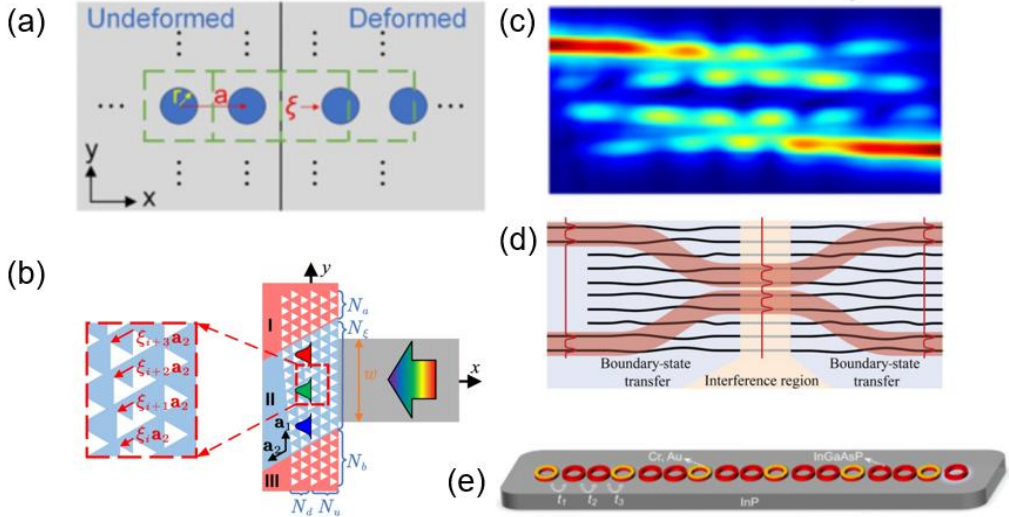

Fig. 27 (a) Forming synthetic dimension utilizing the translational degree of freedom of a 2D photonic crystal's unit cell. Adapted from Ref. [556]. (b) A topological rainbow concentrator thus can be realized and demonstrated in integrated nanophotonic chips. Adapted from Ref. [558]. (c) Robust and broadband optical coupler using topological pump. Adapted from Ref. [561]. (d) A topological splitter for high-visibility quantum interference of single-photon states. Adapted from Ref. [562]. (e) A topological lasing in a 1D coupled ring resonator array that can be mapped to a 2D non-Hermitian Chern insulator with synthetic dimension. Adapted from Ref. [563].

On the other hand, topological pumps have potential applications in on-chip optical routing and light steering. For example, Sun et al demonstrated broadband and fabrication tolerant power coupling and mode-order conversion using Thouless pumping mechanism [561] [see Fig. 27(c)]. They considered a finite Rice-Mele (RM) modeled silicon photonic waveguide array and exploited edge-to-edge pumping process to build the topological photonic directional couplers and mode-order converters. These topological devices exhibit an ultrabroad bandwidth of 120 nm and are quite tolerant to significant structural deviations ($-50 \sim 150$ nm) compared to the conventional counterpart. However, these topological devices typically have a large footprint, indicating a tradeoff between device size and topological robustness. Recently, these challenges have been addressed with the guidance of quantum metric [531], adiabatic infimum [532], and shortcut to adiabaticity concepts [529]. These advancements indicate viable pathways to achieve both robustness and a compact footprint for topological devices, thereby facilitating the development of practical topological integrated devices on photonic chips. Beyond the classical optics, topological pumps also show capability in quantum information processing. For example, Tambasco et al. harnessed the topological pumps to realize a high-visibility quantum interference of single-photon states in an integrated photonic circuit [562] [see Fig. 27(d)]. Their device implements the off-diagonal AAH model in a time-varying fashion, and the resulting topological pumping effects can be utilized to implement a 50:50 beamsplitter. Using this "topological beamsplitter", the Hong-Ou-Mandel interference is measured with $93.1 \pm 2.8\%$ visibility, demonstrating nonclassical behavior of topological states. More recently, a topological lasing with a well-defined non-Hermitian bulk topology in a 1D coupled ring resonator array is demonstrated [563] [see Fig. 27(e)]. This 1D structure is equivalent to a 2D non-Hermitian Chern insulator using the synthetic dimension. All these progress in parameter synthetic dimensions can certainly boost progress in on-chip information processing and lasing, high-speed optical communications, and integrated quantum photonics.

## D. Other Systems for Parameter Synthetic Dimension

The concept of parameter synthetic dimensions expands largely, demonstrating its effectiveness in various physics systems beyond optics, ranging from cold atoms and acoustics to elastic and mechanics. Now, we briefly summarize the specialties involved in different systems when forming parameter synthetic dimensions.

The ultracold atoms in optical superlattices have become a perfect platform for implementing quantized topological charge pumps through synthetic dimensions of parameters [564-566]. For example, the realization of a dynamical version of the 4D integer quantum Hall effect was achieved through the implementation of a 2D topological pump with ultracold bosonic atoms in a 2D angled optical superlattice [29] [see Fig. 28(a)]. The corresponding 2D model is represented by a square superlattice, which is composed of two 1D superlattices along the $x$ and $y$ directions, each created by overlaying two lattices. This pumping method is equivalent to threading the flux in the 4D model. In this way, researchers successfully observe a bulk response with intrinsic 4D topology and demonstrate its quantization by measuring the associated second Chern number.

In addition to the photonics and cold atoms, acoustic and elastic waveguide lattices have also emerged as a distinct and fruitful ground for constructing parameter synthetic dimensions. For instance, emulating an on-site modulating Harper model enables the intertwining of edge states within a finite acoustic waveguide lattice, and the Landau-Zener transition in a topological pumping process is demonstrated [567]. Moreover, the off-diagonal terms—representing the hopping effect—can be made positive or negative depending on their orientation, which is advantageous for achieving higher Chern numbers in a hopping-tuned Harper model [568]. Most importantly, by tailoring the parameters of the coupling tubes to conserve chiral symmetry, the braiding of degenerate acoustic modes, a crucial step toward realizing logic operations with sound, is demonstrated [569]. Almost at the same time, a non-Abelian Thouless process has also been successfully experimentally realized in the acoustic waveguides [530] [see Fig. 28(b)]. Using the structural parameters as extra system momenta, the 3D Weyl physics can also be explored in acoustic systems [570-572] [see Fig. 28(c)].

For elastic waves, in addition to discrete lattice systems, continuous elastic waveguides with elaborately decorated mass density or stiffness have emerged as promising platforms for realizing topologically protected wave evolution. For example, a 2D elastic plate with spatial stiffness modulation in one direction mimics a coupling elastic waveguide array supporting edge states in bandgap with non-zero Chern number. When smooth modulation is introduced along another direction with adiabaticity satisfied, a robust edge-to-edge topological pumping process has been proposed and experimentally demonstrated [573, 574] [see Fig. 28(d)]. Parameter synthetic dimensions have also been implemented in mechanical systems [575] for pursuing stable exceptional chains in non-Hermitian systems. Additionally, the electric circuits also can serve as an excellent platform for exploring exotic topological effects and offer a promising alternative to the synthetic dimensions approach for realizing higher-dimensional lattices [482, 576-578].

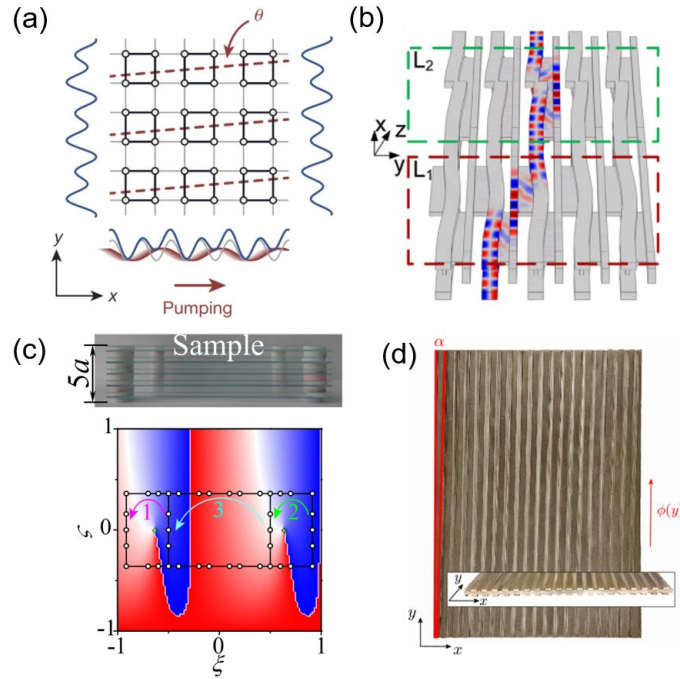

Fig. 28 (a) Topological pump with ultracold bosonic atoms in a 2D angled optical superlattice. Adapted from Ref. [29]. (b) Non-Abelian Thouless pumping with structural modulated acoustic waveguides. Adapted from Ref. [530]. (c) Weyl points in 1D sonic crystal. Adapted from Ref. [570]. (d) Realizing topological pumping in elastic waves. Adapted from Ref. [574].

We briefly summarize this section that the idea of parameter synthetic dimension serves similar purpose as that of synthetic dimensions using discrete states to simulate higher-dimensional physics in low-dimensional physical structures. Although the strategies are distinct and there are different advantages in both ideas, the unique physical phenomena in parameter synthetic dimensions also provide many potential applications. A combination between different ideas of synthetic dimensions may provide the unique opportunity for simulating complex physical problems and seeking novel wave manipulations.

## VII. Summary and Discussions

In this comprehensive review article, we discussed previous works focusing on the concept of synthetic dimensions, expanding the discussion from optical and photonic systems to atomic and molecular systems. The key idea for creating a synthetic dimension is to utilize various degrees of freedom of waves, materials, or systems to add additional dimensionality to spatial ones. As we summarized in this article, the frequency, orbital angular momentum, polarization of light has been used to construct discrete photonic lattices; the pulse arrival times have been used to build time-multiplexed networks; and intrinsic atomic states or momenta have been used to form discrete synthetic dimensions with atoms. Other important proposals with classical waves in electronic circuits and acoustic platforms as well as using photon numbers and superradiant states in quantum optics are also briefly discussed. All of these ideas require suitable design of the connectivity to induce the coupling between discrete lattice sites such that the model can mimic a tight-binding Hamiltonian in condensed matter physics. On the other hand, the use of system parameters was also discussed, where a continuous tunable parameter is usually taken to emulate the momentum in the Bouillon zone reciprocal to a Hamiltonian of a discrete lattice model. In both approaches for building synthetic dimensions, one or more additional dimensions are added to the geometric dimensionality of a physical structure.

An immediate advantage brought forth by synthetic dimensions is the potential for exploring high-dimensional physics in lower-dimensional geometric structures. For example, as we discussed in **Sec. VIB** and **Sec. VID**, one can study 4D topological physics in a 3D photonic structure (the 2D waveguide array) [30] or a 2D atomic array [29] once the system parameters are included to form the parameter synthetic space. This advantage can also be viewed from the opposite perspective, namely to study 1D or 2D physics in a 0D physical structure. For example, we have seen demonstrations of quantum Hall physics in a single ring resonator [37] and Weyl physics in 1D or 2D photonic crystals [537, 554]. The simplification of the spatial geometry may provide important on-chip applications where the spatial footprint is limited and complicated spatial structuring is difficult to construct [102, 105, 106, 108, 133, 179] but the wave dynamics can still be expanded to higher dimensions.

Another well-acknowledged advantage for synthetic dimensions is the unprecedented flexibility in designing the connectivity and the resulting model Hamiltonian. We have seen several examples in previous sections to study models with effective gauge fields [22, 23, 37], non-Hermiticity [39, 40], and long-range couplings [93, 126]. Different from their spatial counterparts, the model built with synthetic dimensions can provide unique ingredients to these models as such connectivity is often difficult to directly realize by the spatial geometry.

With the above advantages in mind, probably the most important application for synthetic dimensions is towards quantum simulations of physics in high dimensions and complicated models, yet in simple physical geometries that are possible to implement with current technology. Although most current works still focus on quantum simulations with linear Hamiltonians, the potential to demonstrate Hamiltonians with particle-particle interactions is rapidly growing. Recent developments with Rydberg atom synthetic dimensions [16] and ultracold atoms [382] show promise for the simulation of many-body physics with synthetic dimensions. Quantum simulation for many-body physics with synthetic dimensions in photonics is indeed challenging [159] due to the weakness of direct photon-photon interactions, but is still promising with recent experimental achievements [108, 150].

The last important point for synthetic dimensions we want to emphasize is the potential for technological applications in manipulating quantum states or optical fields in unconventional ways using ideals from fundamental physics. This point becomes particularly significant when developing on-chip photonic devices with tunability and reconfigurability, as light propagation and conversion may be controlled by varying the Hamiltonian in synthetic space [518]. Photonic calculations have also been proposed using additional degrees of freedom of light with synthetic dimensions [97, 107, 168-171, 239]. Moreover, the quantum operations such as manipulation of the state of photons, the generation of quantum entanglement and quantum-gate operations may also be manipulated in the synthetic space based on recent theoretical proposals [157, 226].

The field of synthetic dimensions still holds several challenges if one wants to fully take all of the advantages listed above. For photonics, loss must be decreased to ensure long evolution time for a Hamiltonian. Disorders and perturbations should be diminished so a precise model is built that can be extended to high dimensions, although several synthetic dimensions such as frequency are inherently less disordered than real-space photonic lattices. Proposals with table-top photonic structures require further optimization or alternative approaches to be realized in the on-chip footprint size [137]. Large photon-photon interactions, or all-optical nonlinearity are also desired to explore many-body physics with synthetic dimensions in photonics, and alternative approaches that incorporate natural or artificial atoms within photonic synthetic dimensions show promise in this regard. On the other hand, for atoms, there remain platform-specific challenges to be overcome for the exploration of novel many-body phases and phenomena in synthetic dimensions. Advancing these atomic synthetic dimensions experiments may further help to advance general quantum simulation and computation techniques.

The field of synthetic dimensions continues to grow rapidly with numerous significant contributions from the broader physical sciences community still emerging as this review is being written. We hope this comprehensive review not only serves the purpose of summarizing representative examples of recent developments in synthetic dimensions but also elicits broad interest from scientists from a wider variety of fields, catalyzing new ideas that bridge synthetic dimensions with their areas of study.

*Acknowledgments*

The research is supported by National Natural Science Foundation of China (Nos. 12122407, 12204304, 12104297, 12192252, and 12204233), and National Key Research and Development Program of China (Nos. 2023YFA1407200 and 2021YFA1400900). B. G. acknowledges support by the National Science Foundation under grant No. 1945031 and by the AFOSR MURI program under agreement number FA9550-22-1-0339.